%% file: janis_newman_algorithm.tex
\documentclass[10pt, a4paper]{article}
\pdfoutput=1

\usepackage[utf8]{inputenc}
\usepackage[T1]{fontenc}
\usepackage{lmodern, textcomp}
\usepackage[british]{babel}
\usepackage{csquotes}

\usepackage[heightrounded, top=4cm, bottom=4cm, left=3.5cm, right=3.5cm]{geometry}

\usepackage{eurosym, xspace}
\usepackage[nottoc]{tocbibind}
\usepackage{graphicx, subfig, float}
\usepackage[multiple]{footmisc}
\usepackage{authblk}
\usepackage{fancyhdr}

\usepackage[usenames, dvipsnames]{xcolor}

\usepackage[backend=bibtex8, citestyle=numeric-comp, maxbibnames=99,
			sorting=none, sortcites=true, firstinits=true]{biblatex}
\input{arxiv.def}

\usepackage{amsmath, amsfonts, amssymb}
\usepackage{cancel}

\usepackage{maths_min}

\usepackage[pdftex, breaklinks=true, linktocpage,
	colorlinks=true, urlcolor=blue, linkcolor=blue, citecolor=red,
	pdfauthor={Harold Erbin}
]{hyperref}
\usepackage[all]{hypcap}

\graphicspath{{images/}}

\numberwithin{equation}{section}

\DeclareUnicodeCharacter{00A0}{~}
\DeclareUnicodeCharacter{202F}{~}

\newcommand{\email}[1]{\thanks{\href{mailto:#1}{\nolinkurl{#1}}}}

\usepackage[sort&compress, english]{cleveref}

\hypersetup{
	pdftitle={Janis-Newman algorithm: generating rotating and NUT charged black holes},
}

\title{Janis--Newman algorithm: generating rotating and NUT charged black holes}

\author[1]{Harold Erbin\email{erbin@lpt.ens.fr}}
\affil[1]{\textsc{Cnrs}, \textsc{Lptens}, École Normale Supérieure, F-75231 Paris, France}

\begin{document}

\maketitle

\begin{abstract} 
In this review we present the most general form of the Janis--Newman algorithm.
This extension allows to generate configurations which contain all bosonic fields with spin less than or equal to two (real and complex scalar fields, gauge fields, metric field) and with five of the six parameters of the Plebański--Demiański metric (mass, electric charge, magnetic charge, NUT charge and angular momentum).
Several examples are included to illustrate the algorithm.
We also discuss the extension of the algorithm to other dimensions.
\end{abstract}

\newpage

\hrule
\pdfbookmark[1]{\contentsname}{toc}
\tableofcontents
\bigskip
\hrule

\newpage

\input{sections/introduction}
\input{sections/algorithm}
\input{sections/extension}
\input{sections/general}
\input{sections/derivation}

\input{sections/examples}
\input{sections/five_dimensions}

\input{sections/higher_dimensions}

\section*{Acknowledgments}

I am particularly grateful and indebted to Lucien Heurtier for our collaboration and our many discussions on this project.
I thank also Nick Halmagyi and Dietmar Klemm for interesting discussions, and I am grateful to the latter and Marco Rabbiosi for allowing me to reproduce an unpublished example of application.
Finally I wish to thank the members of the Harish--Chandra Research Institute (Allahabad, India) for organizing the set of lectures that helped me to transform my thesis in the current review.

\appendix

\input{sections/coordinates}
\input{sections/sugra}
\input{sections/properties}

\printbibliography[heading=bibintoc]

\end{document}

%% file: sections/introduction.tex
\section{Introduction}
\label{sec:intro}

\subsection{Motivations}

General relativity is the theory of gravitational phenomena.
It describes the dynamical evolution of spacetime through the Einstein--Hilbert action that leads to Einstein equations.
The latter are highly non-linear differential equations and finding exact solutions is a notoriously difficult problem.

There are different types of solutions but this review will cover only black-hole-like solutions (type-D in the Petrov classification) which can be described as particle-like objects that carry some charges, such as a mass or an electric charge.

Black holes are important objects in any theory of gravity for the insight they provide into the quantum gravity realm.
For this reason it is a key step, in any theory, to obtain all possible black holes solutions.
Rotating black holes are the most relevant subcases for astrophysics as it is believed that most astrophysical black holes are rotating.
These solutions may also provide exterior metric for rotating stars.

The most general solution of this type in pure Einstein--Maxwell gravity with a cosmological constant $\Lambda$ is the Plebański--Demiański metric~\cite{Plebanski:1975:ClassSolutionsEinsteinMaxwell, Plebanski:1976:RotatingChargedUniformly}: it possesses six charges: mass $m$, NUT charge $n$, electric charge $q$, magnetic charge $p$, spin $a$ and acceleration $\alpha$.
A challenging work is to generalize this solution to more complex Lagrangians, involving scalar fields and other gauge fields with non-minimal interactions, as is typically the case in supergravity.
As the complexity of the equations of motion increase, it is harder to find exact analytical solutions, and one often consider specific types of solutions (extremal, BPS), truncations (some fields are constant, equal or vanishing) or solutions with restricted number of charges.
For this reason it is interesting to find solution generating algorithms -- procedures which transform a seed configuration to another configuration with a greater complexity (for example with a higher number of charges).

An \emph{on-shell} algorithm is very precious because one is sure to obtain a solution if one starts with a seed configuration which solves the equations of motion.
On the other hand \emph{off-shell} algorithms do not necessarily preserve the equations of motion but they are nonetheless very useful: they provide a motivated ansatz, and it is always easier to check if an ansatz satisfy the equations than solving them from scratch.
Even if in practice this kind of solution generating technique does not provide so many new solutions, it can help to understand better the underlying theory (which can be general relativity, modified gravities or even supergravity) and it may shed light on the structure of gravitational solutions.

\subsection{The Janis--Newman algorithm}

The Janis--Newman (JN) algorithm is one of these (off-shell) solution generating techniques, which -- in its original formulation -- generates rotating metrics from static ones.
It was found by Janis and Newman as an alternative derivation of the Kerr metric~\cite{Newman:1965:NoteKerrSpinningParticle}, while shortly after it has been used again to discover the Kerr--Newman metric~\cite{Newman:1965:MetricRotatingCharged}.

This algorithm provides a way to generate axisymmetric metrics from a spherically symmetric seed metric through a particular complexification of radial and (null) time coordinates, followed by a complex coordinate transformation.
Often one performs eventually a change of coordinates to write the result in Boyer--Lindquist coordinates.

The original prescription uses the Newman--Penrose tetrad formalism, which appears to be very tedious since it requires to invert the metric, to find a null tetrad basis where the transformation can be applied, and lastly to invert again the metric.
In~\cite{Giampieri:1990:IntroducingAngularMomentum} Giampieri introduced another formulation of the JN algorithm which avoids gymnastics with null tetrads and which appears to be very useful for extending the procedure to more complicated solutions (such as higher dimensional ones).
However it has been so far totally ignored in the literature.
We stress that \emph{all} results are totally equivalent in both approaches, and every computation that can be done with the Giampieri prescription can be done with the other.
Finally~\cite{Nawarajan:2016:CartesianKerrSchildVariation} provides an alternative view on the algorithm.

In order for the metric to be still real, the radial functions inside the metric must be transformed such that reality is preserved.\footnotemark{}%
\footnotetext{%
	For simplifying, we will say that we complexify the functions inside the metric when we perform this transformation, even if in practice we "realify" them.
}
Despite that there is \emph{no} rigorous statement concerning the possible complexification of these functions, some general features have been worked out in the last decades and a set of rules has been established.
Note that this step is the same in both prescriptions.
In particular these rules can be obtained by solving the equations of motion for some examples and by identifying the terms in the solution~\cite{Demianski:1972:NewKerrlikeSpacetime}.
Another approach consists in expressing the metric functions in terms of the Boyer--Lindquist functions -- that appear in the change of coordinates and which are real --, the latter being then determined from the equations of motion~\cite{Drake:2000:UniquenessNewmanJanisAlgorithm, AzregAinou:2014:StaticRotatingConformal}.

It is widely believed that the JN algorithm is just a trick without any physical or mathematical basis, which is not accurate.
Indeed it was proved by Talbot~\cite{Talbot:1969:NewmanPenroseApproachTwisting} shortly after its discovery why this transformation was well-defined, and he characterizes under which conditions the algorithm is on-shell for a subclass of Kerr--Schild (KS) metrics (see also~\cite{Gurses:1975:LorentzCovariantTreatment}).\footnotemark{}%
\footnotetext{%
	It has not been proved that the KS condition is necessary, but all known examples seem to fit in this category.
}
KS metrics admit a very natural formulation in terms of complex functions for which (some) complex change of coordinates can be defined.
Note that KS metrics are physically interesting as they contain solutions of Petrov type II and D.
Another way to understand this algorithm has been provided by Schiffer et al.~\cite{Schiffer:1973:KerrGeometryComplexified} (see also~\cite{Finkelstein:1975:GeneralRelativisticFields}) who showed that some KS metrics can be written in terms of a unique complex generating function, from which other solutions can be obtained through a complex change of coordinates.
In various papers, Newman shows that the imaginary part of complex coordinates may be interpreted as an angular momentum, and there are similar correspondences for other charges (magnetic…)~\cite{Newman:1973:ComplexCoordinateTransformations, Newman:1974:CuriosityConcerningAngular, Newman:1976:HeavenItsProperties}.
More recently Ferraro shed a new light on the JN algorithm using Cartan formalism~\cite{Ferraro:2014:UntanglingNewmanJanisAlgorithm}.
Uniqueness results for the case of pure Einstein theory have been derived in~\cite{Drake:2000:UniquenessNewmanJanisAlgorithm}.
A recent account on these different points can be found in~\cite{Adamo:2014:KerrNewmanMetricReview}.

In its current form the algorithm is independent of the gravity theory under consideration since it operates independently at the level of each field in order to generate an ansatz, and the equations of motion are introduced only at the end to check if the configuration is a genuine solution.
We believe that a better understanding of the algorithm would lead to an on-shell formulation where the algorithm would be interpreted as some kind of symmetry or geometric property.
One intuition is that every configuration found with the JN algorithm and solving the equations of motion is derived from a seed that also solves the equations of motion (in particular no useful ansatz has been generated from an off-shell seed configuration).

Other solution generating algorithms rely on a complex formulation of general relativity which allows complex changes of coordinates.
This is the case of the Ernst potential formulation~\cite{Ernst:1968:NewFormulationAxially-1, Ernst:1968:NewFormulationAxially-2} or of Quevedo's formalism who decomposes the Riemann tensor in irreducible representations of $\group{SO}(3,\C) \sim \group{SO}(3,1)$ and then uses the symmetry group to generate new solutions~\cite{Quevedo:1992:ComplexTransformationsCurvature, Quevedo:1992:DeterminationMetricCurvature}.

Despite its long history the Janis--Newman algorithm has not produced any new rotating solution for non-fluid configurations (which excludes radiating and interior solutions) beside the Kerr--Newman metric~\cite{Newman:1965:MetricRotatingCharged}, and very few known examples have been reproduced~\cite{Newman:1965:NoteKerrSpinningParticle, Xu:1988:ExactSolutionsEinstein, Kim:1997:NotesSpinningAdS3, Kim:1999:SpinningBTZBlack, Yazadjiev:2000:NewmanJanisMethodRotating}.
Generically the application the Janis--Newman algorithm to interior and radiating systems~\cite{Herrera:1982:ComplexificationNonrotatingSphere, Drake:1997:ApplicationNewmanJanisAlgorithm, Glass:2004:KottlerLambdaKerrSpacetime, Ibohal:2005:RotatingMetricsAdmitting, AzregAinou:2014:GeneratingRotatingRegular, AzregAinou:2014:StaticRotatingConformal} consist in deriving a configuration that do not solve the equations of motion by itself and to interpret the mismatch as a fluid (whose properties can be studied) -- in this review we will not be interested by this kind of applications.
Moreover the only solutions that have been fully derived using the algorithm are the original Kerr metric~\cite{Newman:1965:NoteKerrSpinningParticle}, the $d = 3$ BTZ black hole~\cite{Kim:1997:NotesSpinningAdS3, Kim:1999:SpinningBTZBlack} and the $d$-dimensional Myers--Perry metric with a single angular momentum~\cite{Xu:1988:ExactSolutionsEinstein}: only the metric was found in the other cases~\cite{Newman:1965:MetricRotatingCharged, Yazadjiev:2000:NewmanJanisMethodRotating} and the other fields had to be obtained using the equations of motion.

A first explanation is that there is no real understanding of the algorithm in its most general form (as reviewed above it is understood in some cases): there is no geometric or symmetry-related interpretation.
Another reason is that the algorithm has been defined only for the metric (and real scalar fields) and no extension to the other types of fields was known until recently.
It has also been understood that the algorithm could not be applied in the presence of a cosmological constant~\cite{Demianski:1972:NewKerrlikeSpacetime}: in particular the (a)dS--Kerr(--Newman) metrics~\cite{Carter:1968:HamiltonJacobiSchrodingerSeparable} (see also~\cite{Plebanski:1976:RotatingChargedUniformly, Plebanski:1975:ClassSolutionsEinsteinMaxwell, Gibbons:1977:CosmologicalEventHorizons, Klemm:1997:RotatingTopologicalBlack}) cannot be derived in this way despite various erroneous claims~\cite{Ibohal:2005:RotatingMetricsAdmitting, deUrreta:2015:ExtendedNewmanJanisAlgorithm}.
Finally many works~\cite{Mallett:1988:MetricRotatingRadiating, Viaggiu:2006:InteriorKerrSolutions, Whisker:2008:BraneworldBlackHoles, Lessner:2008:ComplexTrickFivedimensional, Capozziello:2010:AxiallySymmetricSolutions, Caravelli:2010:SpinningLoopBlack, Dadhich:2013:RotatingBlackHole, Ghosh:2013:SpinningHigherDimensional, Ghosh:2015:RotatingBlackHole} (to cite only few) are (at least partly) incorrect or not reliable because they do not check the equations of motion or they perform non-integrable Boyer--Lindquist changes of coordinates~\cite{AzregAinou:2011:CommentSpinningLoop, AzregAinou:2014:GeneratingRotatingRegular, Xu:1998:RadiatingMetricRetarded}.

The algorithm was later shown to be generalizable by Demiański and Newman who demonstrated by writing a general ansatz and solving the equations of motion that other parameters can be added~\cite{Demianski:1966:CombinedKerrNUTSolution, Demianski:1972:NewKerrlikeSpacetime}, even in the presence of a cosmological constant.
While one parameter corresponds to the NUT charge, the other one did not receive any interpretation until now.\footnotemark{}%
\footnotetext{%
	Demiański's metric has been generalized in~\cite{Patel:1978:RadiatingDemianskitypeSpacetimes, Krori:1981:ChargedDemianskiMetric, Patel:1988:RadiatingDemianskitypeMetrics}.
}
Unfortunately Demiański did not express his result to a concrete algorithm (the normal prescription fails in the presence of the NUT charge and of the cosmological constant) which may explain why this work did not receive any further attention.
Note that the algorithm also failed in the presence of magnetic charges.

A way to avoid problems in defining the changes of coordinates to the Boyer--Lindquist system and to find the complexification of the functions has been proposed in~\cite{Drake:2000:UniquenessNewmanJanisAlgorithm} and extended in~\cite{AzregAinou:2014:GeneratingRotatingRegular}: the method consists in writing the unknown complexified function in terms of the functions of the coordinate transformation.
This philosophy is particularly well-suited for providing an ansatz which does not relies on a static seed solution.

More recently it has been investigated whether the JN algorithm can be applied in modified theories of gravity.
Pirogov put forward that rotating metrics obtained from the JN algorithm in Brans--Dicke theory are not solutions if $\alpha \neq 1$~\cite{Pirogov:2013:RotatingScalarvacuumBlack}.
Similarly Hansen and Yunes have shown a similar result in quadratic modified gravity (which includes Gauss--Bonnet)~\cite{Hansen:2013:ApplicabilityNewmanJanisAlgorithm}.\footnotemark{}%
\footnotetext{%
	There are some errors in the introduction of~\cite{Hansen:2013:ApplicabilityNewmanJanisAlgorithm}: they report incorrectly that the result from~\cite{Pirogov:2013:RotatingScalarvacuumBlack} implies that Sen's black hole cannot be derived from the JN algorithm, as was done by Yazadjiev~\cite{Yazadjiev:2000:NewmanJanisMethodRotating}.
	But this black hole corresponds to $\alpha = 1$ and as reported above there is no problem in this case (see~\cite{Horne:1992:RotatingDilatonBlack} for comparison).
	Moreover they argue that several works published before 2013 did not take into account the results of Pirogov~\cite{Pirogov:2013:RotatingScalarvacuumBlack}, published in 2013…
}
These do not include Sen's dilaton--axion black hole for which $\alpha = 1$ (\cref{sec:examples:rotating-T3}), nor the BBMB black hole from conformal gravity (\cref{sec:examples:bbmb}).
Finally it was proved in~\cite{CiriloLombardo:2006:NewmanJanisAlgorithmRotating} that it does not work either for Einstein--Born--Infled theories.\footnotemark{}%
\footnotetext{%
	It may be possible to circumvent the result of~\cite{CiriloLombardo:2006:NewmanJanisAlgorithmRotating} by using the results described in this review since several tools were not known by its author.
}
We note that all these no-go theorem have been found by assuming a transformation with only rotation.

Previous reviews of the JN algorithm can be found in~\cites{Adamo:2014:KerrNewmanMetricReview}[chap.~19]{DInverno:1992:IntroducingEinsteinsRelativity}{Drake:2000:UniquenessNewmanJanisAlgorithm}[sec.~5.4]{Whisker:2008:BraneworldBlackHoles} (see also~\cite{Reed:1974:ImaginaryTetradtransformationsEinstein}).

\subsection{Summary}

The goal of the current work is to review a series of recent papers~\cite{Erbin:2015:JanisNewmanAlgorithmSimplifications, Erbin:2015:FivedimensionalJanisNewmanAlgorithm, Erbin:2016:DecipheringGeneralizingDemianskiJanisNewman, Erbin:2015:SupergravityComplexParameters} in which the JN algorithm has been extended in several directions, opening the doors to many new applications.
This review evolved from the thesis of the author~\cite{Erbin:2015:BlackHolesN}, which presented the material from a slightly different perspective, and from lectures given at \textsc{Hri} (Allahabad, India).

As explained in the previous section, the JN algorithm was formulated only for the metric and all other fields had to be found using the equations of motion (with or without using an ansatz).
For example neither the Kerr--Newman gauge field or its associated field strength could be derived in~\cite{Newman:1965:MetricRotatingCharged}.
The solution to this problem is to perform a gauge transformation in order to remove the radial component of the gauge field in null coordinates~\cite{Erbin:2015:JanisNewmanAlgorithmSimplifications}.
It is then straightforward to apply the JN algorithm in either prescription.\footnotemark{}%
\footnotetext{%
	Another solution has been proposed by Keane~\cite{Keane:2014:ExtensionNewmanJanisAlgorithm} but it is applicable only to the Newman--Penrose coefficients of the field strength.
	Our proposal requires less computations and yields directly the gauge field from which all relevant quantities can easily be derived.
}
Another problem was exemplified by the derivation of Sen's axion--dilaton rotating black hole~\cite{Sen:1992:RotatingChargedBlack} by Yazadjiev~\cite{Yazadjiev:2000:NewmanJanisMethodRotating}, who could find the metric and the dilaton, but not the axion (nor the gauge field).
The reason is that while the JN algorithm applies directly to real scalar fields, it does not for complex scalar fields (or for a pair of real fields that can naturally be gathered into a complex scalar).
Then it is necessary to consider the complex scalar as a unique object and to perform the transformation without trying to keep it real~\cite{Erbin:2015:SupergravityComplexParameters}.
Hence this completes the JN algorithm for all bosonic fields with spin less than or equal to two.

A second aspect for which the original form of the algorithm was deficient is configuration with magnetic and NUT charges and in presence of a cosmological constant.
The issue corresponds to finding how one should complexify the functions: the usual rules do not work and if there were no way to obtain the functions by complexification then the JN algorithm would be of limited interest as it could not be exported to other cases (except if one is willing to solve equations of motion, which is not the goal of a solution generating technique).
We have found that to reproduce Demiański's result~\cite{Demianski:1972:NewKerrlikeSpacetime} it is necessary to complexify also the mass and to consider the complex parameter $m + i n$~\cite{Erbin:2016:DecipheringGeneralizingDemianskiJanisNewman, Erbin:2015:SupergravityComplexParameters} and to shift the curvature of the spherical horizon.
Similarly for configurations with magnetic charges one needs to consider the complex charge $q + i p$~\cite{Erbin:2015:SupergravityComplexParameters}.
Such complex combinations are quite natural from the point of view of the Plebański--Demiański solution~\cite{Plebanski:1975:ClassSolutionsEinsteinMaxwell, Plebanski:1976:RotatingChargedUniformly} described previously.
It is to notice that the appearance of complex coordinate transformations mixed with complex parameter transformations was a feature of Quevedo's solution generating technique~\cite{Quevedo:1992:ComplexTransformationsCurvature, Quevedo:1992:DeterminationMetricCurvature}, yet it is unclear what the link with our approach really is.
Hence the final metric obtained from the JN algorithm may contain (for vanishing cosmological constant) five of the six Plebański--Demiański parameters~\cite{Plebanski:1975:ClassSolutionsEinsteinMaxwell, Plebanski:1976:RotatingChargedUniformly} along with Demiański's parameter.

An interesting fact is that the previous argument works in the presence of the cosmological constant only if one considers the possibility of having a generic topological horizons (flat, hyperbolic or spherical) and for this reason we have provided an extension of the formalism to this case~\cite{Erbin:2016:DecipheringGeneralizingDemianskiJanisNewman}.

We also propose a generalization of the algorithm to any dimension~\cite{Erbin:2015:FivedimensionalJanisNewmanAlgorithm}, but while new examples could be found for $d = 5$ the program could not be carried to the end for $d > 5$.

All these results provide a complete framework for most of the theories of gravity that are commonly used.
As a conclusion we summarize the features of our new results:
\begin{itemize}
	\item all bosonic fields with spin $\le 2$;
	\item topological horizons;
	\item charges $m, n, q, p, a$ (with $a$ only for $\Lambda = 0$);
	\item extend to $d = 3, 5$ dimensions (and proposal for higher).
\end{itemize}
We have written a general \emph{Mathematica} package for the JN algorithm in Einstein--Maxwell theory.\footnotemark{}%
\footnotetext{
	Available at \url{http://www.lpthe.jussieu.fr/~erbin/}.
}
Here is a list of new examples that have been completely derived using the previous results (all in $4d$ except when said explicitly):
\begin{itemize}
	\item Kerr--Newman--NUT;
	\item dyonic Kerr--Newman;
	\item Yang--Mills Kerr--Newman black hole~\cite{Perry:1977:BlackHolesAre};
	\item adS--NUT Schwarzschild;
	\item Demiański's solution~\cite{Demianski:1972:NewKerrlikeSpacetime};
	\item ungauged $N = 2$ BPS solutions~\cite{Behrndt:1998:StationarySolutionsN2};
	\item non-extremal solution in $T^3$ model~\cite{Sen:1992:RotatingChargedBlack} (partly derived in~\cite{Yazadjiev:2000:NewmanJanisMethodRotating});
	\item SWIP solutions~\cite{Bergshoeff:1996:StationaryAxionDilatonSolutions};
	\item (a)dS--charged Taub--NUT--BBMB~\cite{Bardoux:2013:IntegrabilityConformallyCoupled};
	\item $5d$ Myers--Perry~\cite{Myers:1986:BlackHolesHigher};
	\item $5d$ BMPV~\cite{Breckenridge:1997:DbranesSpinningBlack};
	\item NUT charged black hole\footnotemark{} in gauged $N = 2$ sugra with $F = - i\, X^0 X^1$~\cite{Gnecchi:2014:RotatingBlackHoles}.
\end{itemize}
\footnotetext{%
	Derived by D.\ Klemm and M.\ Rabbiosi, unpublished work.
}
We also found a more direct derivation of the rotating BTZ black hole (derived in another way by Kim~\cite{Kim:1997:NotesSpinningAdS3, Kim:1999:SpinningBTZBlack}).

\subsection{Outlook}

A major playground for this modified Janis--Newman (JN) algorithm is (gauged) supergravity -- where many interesting solutions remain to be discovered -- since all the necessary ingredients are now present.
Moreover important solutions are still missing in higher-dimensional Einstein--Maxwell (in particular the charged Myers--Perry solution) and one can hope that understanding the JN algorithm in higher dimensions would shed light on this problem.
Another open case is whether black rings can also be derived using the algorithm.

A major question about the JN algorithm is whether it is possible to include rotation for non-vanishing cosmological constant.
A possible related problem concerns the addition of acceleration $\alpha$, which is the only missing parameter when $\Lambda = 0$.
It is indeed puzzling that one could get all Plebański--Demiański parameters but the acceleration, which appears in the combination $a + i \alpha$.
Both problems are linked to the fact that the JN algorithm -- in its current form -- does not take into account various couplings between the parameters (such as the spin with the cosmological constant or the acceleration with the mass in the simplest cases).
On the other hand it does not mean that it is impossible to find a generalization of the algorithm: philosophically the problem is identical to the ones of adding NUT and magnetic charges.

In any case the meaning and a rigorous derivation of the JN algorithm -- perhaps elevating it to the status of a true solution generating algorithm -- are still to be found.
It is also interesting to note that almost all of the examples quoted in the previous section can be embedded into $N = 2$ supergravity.
This calls for a possible interpretation of the algorithm in terms of some hidden symmetry of supergravity, or even of string theory.

We hope that our new extension of the algorithm will help to bring it outside the shadow where it stayed since its creation and to establish it as a standard tool for deriving new solutions in the various theories of gravity.

\subsection{Overview}

In \cref{sec:algo} we review the original Janis--Newman algorithm and its alternative form due to Giampieri before illustrating it with some examples.
\Cref{sec:extension} shows how to extend the algorithm to more complicated set of fields (complex scalars, gauge fields) and parameters (magnetic and NUT charges, topological horizon).
Then \cref{sec:general} provides a general description of the algorithm in its most general form.
The complex transformation described in the previous section are derived in \cref{sec:derivation}.
\Cref{sec:examples} describes several examples.
Finally \cref{sec:five} extends first the algorithm to five dimensions and \cref{sec:higher} generalizes these ideas to any dimension.

\Cref{app:coord} gathers useful formulas on coordinate systems in various numbers of dimensions.
\Cref{app:N=2-sugra} reviews briefly the main properties of $N = 2$ supergravity.
Finally \cref{app:technical-properties} discusses some additional properties of the JN algorithm.

In our conventions the spacetime signature is mostly plus.

%% file: sections/algorithm.tex
\section{Algorithm: main ideas}
\label{sec:algo}

In this section we summarize the original algorithm together with its extension to gauge fields.
We will see that the algorithm involves the transformations of two different objects (the tensor structure and the coordinate-dependent functions of the fields) which can be taken care of separately.
The transformation of the tensor structure is simple and no new idea (for $d = 4$) will be needed after this section since we will be dealing with the two most general tensor structures for bosonic fields of spin less than or equal to two (the metric and vector fields).
On the other hand the transformation of the functions is more involved and we will introduce new concepts through simple examples in the next section before giving the most general formulation in~\cref{sec:general}.
We review the two different prescriptions for the transformation and we illustrate the algorithm with two basic examples: the flat space and the Kerr--Newman metrics.

\subsection{Summary}
\label{sec:algo:summary}

The general procedure for the Janis--Newman algorithm can be summarized as follows:
\begin{enumerate}
	\item Perform a change of coordinates $(t, r)$ to $(u, r)$ and a gauge transformation such that $g_{rr} = 0$ and $A_r = 0$.
	
	\item Take $u, r \in \C$ and replace the functions $f_i(r)$ inside the real fields by new real-valued functions $\tilde f_i(r, \bar r)$ (there is a set of “empirical” rules).
	
	\item Perform a complex change of coordinates and transform accordingly:
	\begin{enumerate}
		\item the tensor structure, i.e.\ the $\dd x^\mu$ (two prescriptions: Janis--Newman~\cite{Newman:1965:NoteKerrSpinningParticle} and Giampieri~\cite{Giampieri:1990:IntroducingAngularMomentum});
		\label{algo:list:procedure-tensor}
		
		\item the functions $\tilde f_i(r, \bar r)$.
		\label{algo:list:procedure-functions}
	\end{enumerate}
	
	\item Perform a change of coordinates to simplify the metric (for example to Boyer--Lindquist system).
	If the transformation is infinitesimal then one should check that it is a valid diffeomorphism, i.e. that it is integrable.
\end{enumerate}
Note that in the last point the operations (a) and (b) are independent.
In practice one is performing the algorithm for a generic class of configurations with unspecified $f_i(r)$ in order to obtain general formulas.
One leaves point 2 and (3b) implicit since the other steps are independent of the form of the functions.
Then given a specific configuration one can perform 2 and (3b).

Throughout the review we will not be interested in showing that the examples discussed are indeed solutions but merely to explain how to extend the algorithm.
All examples we are discussing have been shown to be solutions of the theory under concerned and we refer the reader to the original literature for more details.
For this reason we will rarely mention the action or the equations of motion and just discussed the fields and their expressions.

One could add a fifth point to the list: checking the equations of motion.
We stress again that the algorithm is \emph{off-shell} and there is no guarantee (except in some specific cases~\cite{Adamo:2014:KerrNewmanMetricReview}) that a solution is mapped to a solution.

\subsection{Algorithm}
\label{sec:algo:algorithm}

We present the algorithm for a metric $g_{\mu\nu}$ and a gauge field $A_\mu$ associated with a $\group{U}(1)$ gauge symmetry.
This simple case is sufficient to illustrate the main features of the algorithm.

As already mentioned in the introduction, the authors of~\cite{Newman:1965:MetricRotatingCharged} failed to derive the field strength of the Kerr--Newman black hole from the Reissner--Nordström one.
In the null tetrad formalism it is natural to write the field strength in terms of its Newman--Penrose coefficients, but a problem arises when one tries to generate the rotating solution since one of the coefficients is zero in the case of Reissner--Nordström, but non-zero for Kerr--Newman.
Three different prescriptions have been proposed: two works in the Newman--Penrose formalism -- one with the field strength~\cite{Keane:2014:ExtensionNewmanJanisAlgorithm} and one with the gauge field~\cite{Erbin:2015:JanisNewmanAlgorithmSimplifications} -- while the third extends Giampieri's approach to the gauge field~\cite{Erbin:2015:JanisNewmanAlgorithmSimplifications}.
Since the proposals from~\cite{Erbin:2015:JanisNewmanAlgorithmSimplifications} fit more directly (and parallel each other) inside the prescriptions of Janis--Newman and Giampieri, we will focus on them.
It is also more convenient to work with the gauge fields since any other quantity can be easily computed from them.

\subsubsection{Seed metric and gauge fields}

The seed metric and gauge field take the form
\begin{subequations}
\label{algo:eq:static:tr}
\begin{gather}
	\label{algo:eq:static:metric:tr}
	\dd s^2 = - f(r)\, \dd t^2 + f(r)^{-1}\, \dd r^2 + r^2 \dd \Omega^2, \qquad
	\dd \Omega^2 = \dd\theta^2 + H(\theta)^2\, \dd \phi^2, \\
	\label{algo:eq:static:vector:tr}
	A = f_A(r)\, \dd t.
\end{gather}
\end{subequations}
The normalized curvature of the $(\theta, \phi)$ sections (or equivalently of the horizon) is denoted by $\kappa$
\begin{equation}
	\kappa =
	\begin{cases}
		+1 & S^2, \\
		-1 & H^2
	\end{cases}
\end{equation} 
where $S^2$ and $H^2$ are respectively the sphere and the hyperboloid,\footnotemark{} and one has%
\footnotetext{%
	We leave aside the case of the plane $\R^2$ with $\kappa = 0$.
	The formulas can easily be extended to this case.
}
\begin{equation}
	H(\theta) =
	\begin{cases}
		\sin \theta & \kappa = 1, \\
		\sinh \theta & \kappa = -1.
	\end{cases}
\end{equation} 
In all this section we will consider the case of spherical horizon with $\kappa = 1$.

Introduce Eddington--Finkelstein coordinates $(u, r)$
\begin{equation}
	\dd u = \dd t - f^{-1} \dd r
\end{equation} 
in order to remove the $g_{rr}$ term of the metric~\cite{Newman:1965:NoteKerrSpinningParticle}.
Under this transformation the gauge field becomes
\begin{equation}
	A = f_A\, (\dd u + f^{-1} \dd r).
\end{equation} 
The changes of coordinate has introduced an $A_r$ component but since it depends only on the radial coordinate $A_r = A_r(r)$ it can be removed by a gauge transformation.

At the end the metric and gauge fields are
\begin{subequations}
\label{algo:eq:static:ur}
\begin{gather}
	\label{algo:eq:static:metric:ur}
	\dd s^2 = - f\, \dd t^2 + 2 \dd u \dd r + r^2 \dd \Omega^2, \\
	\label{algo:eq:static:vector:ur}
	A = f_A\, \dd u.
\end{gather} 
\end{subequations}

This last step was missing in~\cite{Newman:1965:MetricRotatingCharged} and explains why they could not derive the full solution from the algorithm.
The lesson to draw is that the validity of the algorithm depends a lot on the coordinate basis\footnotemark{} and of the parametrization of the fields, although guiding principle founded on all known examples seems that one needs to have
\footnotetext{%
	The canonical example being that the Kerr metric in quasi-isotropic coordinates cannot be derived from the Schwarzschild metric in isotropic coordinates while it can be derived in the usual coordinates (see \cref{sec:algorithm:kerr-newman}).
}
\begin{equation}
	g_{rr} = 0, \qquad
	A_r = 0.
\end{equation}

\subsubsection{Janis--Newman prescription: Newman--Penrose formalism}

The Janis--Newman prescription for transforming the tensor structure relies on the Newman--Penrose formalism~\cite{Newman:1965:NoteKerrSpinningParticle, Newman:1965:MetricRotatingCharged, Adamo:2014:KerrNewmanMetricReview}.

First one needs to obtain the contravariant expressions of the metric and of the gauge field
\begin{subequations}
\begin{gather}
	\frac{\pd^2}{\pd s^2} = g^{\mu\nu} \pd_\mu \pd_\nu
		= f\, \pd r^2 - 2\, \pd u \pd r + \frac{1}{r^2} \left( \pd_\theta^2 + \frac{\pd_\phi^2}{\sin^2 \theta} \right), \\
	A = - f_A\, \pd_r.
\end{gather}
\end{subequations}
Then one introduces null complex tetrads
\begin{equation}
	Z_a^\mu = \{ \ell^\mu, n^\mu, m^\mu, \bar m^\mu \}
\end{equation} 
with flat metric
\begin{equation}
	\eta^{ab} =
		\begin{pmatrix}
			0 & -1 & 0 & 0 \\
			-1 & 0 & 0 & 0 \\
			0 &  0 & 0 & 1 \\
			0 &  0 & 1 & 0 \\
		\end{pmatrix}
\end{equation} 
such that
\begin{equation}
	g^{\mu\nu} = \eta^{ab} Z^\mu_a Z^\nu_b
		= - \ell^\mu n^\nu - \ell^\nu n^\mu + m^\mu \bar m^\nu + m^\nu \bar m^\mu.
\end{equation} 
The explicit tetrad expressions are
\begin{equation}
	\label{algo:eq:static:tetrads}
	\ell^\mu = \delta_r^\mu, \qquad
	n^\mu = \delta_u^\mu -\frac{f}{2}\; \delta_r^\mu, \qquad
	m^\mu = \frac{1}{\sqrt{2} \bar r} \left(\delta_\theta^\mu + \frac{i}{\sin \theta}\; \delta_\phi^\mu \right)
\end{equation}
and the gauge field is
\begin{equation}
	A^\mu = -f_A\, \ell^\mu.
\end{equation} 
Note that without the gauge transformation there would be an additional term and the expression of $A^\mu$ in terms of the tetrads would be ambiguous.

At this point $u$ and $r$ are allowed to take complex values but keeping $\ell^\mu$ and $n^\mu$ real and $\conj{(m^\mu)} = \bar m^\mu$ and replacing
\begin{equation}
	f(r) \longrightarrow \tilde f(r, \bar r) \in \R, \qquad
	f_A(r) \longrightarrow \tilde f_A(r, \bar r) \in \R.
\end{equation} 
Consistency implies that one recovers the seed for $\bar r = r$ and $\bar u = u$.

Finally one can perform a complex change of coordinates
\begin{equation}
	\label{algo:eq:change:complexification-ur}
	u = u' + i a \cos \theta, \qquad
	r = r' - i a \cos \theta
\end{equation} 
where $a$ is a parameter (to be interpreted as the angular momentum per unit of mass) and $r', u' \in \R$.
While this transformation seems arbitrary we will show later (\cref{sec:general,sec:derivation}) how to extend it and that general consistency limits severely the possibilities.
The tetrads transform as vectors
\begin{equation}
	Z'^\mu_a = \frac{\pd x'^\mu}{\pd x^\nu}\, Z^\nu_a
\end{equation} 
and this lead to the expressions
\begin{equation}
\begin{gathered}
	\label{algo:eq:rotating:tetrads}
	\ell'^\mu = \delta_r^\mu, \qquad
	n'^\mu = \delta_u^\mu - \frac{\tilde f}{2}\, \delta_r^\mu, \\
	m'^\mu = \frac{1}{\sqrt{2} (r' + i a \cos \theta)} \left(\delta_\theta^\mu + \frac{i}{\sin \theta}\, \delta_\phi^\mu - i a \sin \theta\, (\delta_u^\mu - \delta_r^\mu) \right).
\end{gathered}
\end{equation} 
After inverting the contravariant form of the metric and the gauge field one is lead to the final expressions
\begin{subequations}
\label{algo:eq:rotating:ur}
\begin{gather}
	\label{algo:metric:rotating:ur}
		\dd s'^2  = - \tilde f\, (\dd u' - a \sin^2 \theta\, \dd\phi)^2
			- 2\, (\dd u' - a \sin^2 \theta\, \dd\phi) (\dd r' + a \sin^2 \theta\, \dd\phi)
			+ \rho^2 \dd \Omega^2, \\
	A' = \tilde f_A\, (\dd u' - a \sin^2 \theta\, \dd \phi).
\end{gather}
\end{subequations}
where
\begin{equation}
	\label{algo:eq:rotating:metric:rho}
	\rho^2 = \abs{r}^2
		= r'^2 + a^2 \cos^2 \theta.
\end{equation} 
The coordinate dependence of the functions can be written as
\begin{equation}
	\tilde f = \tilde f(r, \bar r)
		= \tilde f(r', \theta)
\end{equation} 
in the new coordinates (and similarly for $\tilde f_A$), but note that the $\theta$ dependence is not arbitrary and comes solely from $\Im r$.

\subsubsection{Giampieri prescription}

The net effect of the transformation \eqref{algo:eq:change:complexification-ur} on the tensor structure amounts to the replacements
\begin{equation}
	\label{algo:eq:replacement-diff}
	\dd u \longrightarrow \dd u' - a \sin^2 \theta\, \dd \phi, \qquad
	\dd r \longrightarrow \dd r' + a \sin^2 \theta\, \dd \phi
\end{equation} 
by comparing \eqref{algo:eq:static:ur} and \eqref{algo:eq:rotating:ur}, up to the $r^2 \to \rho^2$ in front of $\dd\Omega^2$.
Is it possible to obtain the same effect by avoiding the Newman--Penrose formalism and all the computations associated to changing from covariant to contravariant expressions?
Inspecting the infinitesimal form of \eqref{algo:eq:change:complexification-ur}
\begin{equation}
	\dd u = \dd u' - i a \sin \theta\, \dd \theta, \qquad
	\dd r = \dd r' + i a \sin \theta\, \dd \theta,
\end{equation} 
one sees that \eqref{algo:eq:replacement-diff} can be recovered if one sets~\cite{Giampieri:1990:IntroducingAngularMomentum}
\begin{equation}
	\label{algo:eq:giampieri-ansatz}
	i \dd \theta = \sin \theta\, \dd\phi.
\end{equation} 
Note that it should be done only in the infinitesimal transformation and not elsewhere in the metric.
Although some authors~\cite{Ibohal:2005:RotatingMetricsAdmitting, Ferraro:2014:UntanglingNewmanJanisAlgorithm} mentioned the equivalence between the tetrad computation and \eqref{algo:eq:replacement-diff}, it is surprising that this direction has not been followed further.

While this new prescription is not rigorous and there is no known way to derive \eqref{algo:eq:giampieri-ansatz}, it continues to hold for the most general seed (\cref{sec:general}) and it gives systematically the same results as the Janis--Newman prescription, as can be seen by simple inspection.
In particular this approach is not adding nor removing any of the ambiguities due to the function transformations that are already present and well-known in JN algorithm.
Since this prescription is much simpler we will continue to use it throughout the rest of this review (we will show in \cref{sec:general} how it is modified for topological horizons).

Finally the comparison of the two prescriptions clearly shows that the $r^2$ factor in front of $\dd\Omega^2$ should be considered as a function instead of a part of the tensor structure: the replacement $r^2 \to \rho^2$ is dictated by the rules given in the next section.
We did not want to enter into these subtleties here but this will become evident in \cref{sec:general}.

\subsubsection{Transforming the functions}

The transformation of the functions is common to both the Janis--Newman and Giampieri prescriptions since they are independent of the tensor structure.
This step is the main weakness of the Janis--Newman algorithm because there is no unique way to perform the replacement and for this reason the final result contains some part of arbitrariness.
This provides another incentive for checking systematically if the equations of motion are satisfied.
Nonetheless examples have provided a small set of rules~\cite{Newman:1965:NoteKerrSpinningParticle, Newman:1965:MetricRotatingCharged, Drake:2000:UniquenessNewmanJanisAlgorithm, Erbin:2015:JanisNewmanAlgorithmSimplifications}
\begin{subequations}
\label{algo:eq:rules}
\begin{align}
	r & \longrightarrow \frac{1}{2} (r + \bar r) = \Re r, \\
	\frac{1}{r} & \longrightarrow \frac{1}{2} \left(\frac{1}{r} + \frac{1}{\bar r}\right) = \frac{\Re r}{\abs{r}^2}, \\
	r^2 & \longrightarrow \abs{r}^2.
\end{align}
\end{subequations}
The idea is to use geometric or arithmetic means.
All other functions can be reduced to a combination of them, for example $1 / r^2$ is complexified as $1 / \abs{r}^2$.

Every known configuration which does not involve a magnetic, a NUT charge, complex scalar fields or powers higher of $r$than quadratic can be derived with these rules (these cases will be dealt with in \cref{sec:extension,sec:general}).
Hence despite the fact that there is some arbitrariness, it is ultimately quite limited and very few options are possible in most cases.

\subsubsection{Boyer--Lindquist coordinates}

Boyer--Lindquist coordinates are defined to be those with the minimal number of non-zero off-diagonal components in the metric.
Performing the transformation (the primes in \eqref{algo:eq:rotating:ur} are now omitted)
\begin{equation}
	\label{algo:eq:change:bl}
	\dd u' = \dd t' - g(r) \dd r', \qquad
	\dd \phi = \dd \phi' - h(r) \dd r,
\end{equation} 
the conditions $g_{tr} = g_{r\phi'} = 0$ are solved for
\begin{equation}
	\label{algo:eq:change:bl:solution-gh}
	g(r) = \frac{r^2 + a^2}{\Delta}, \qquad
	h(r) = \frac{a}{\Delta}
\end{equation} 
where we have defined
\begin{equation}
	\label{algo:eq:rotating:tr-delta}
	\Delta(r) = \tilde f \rho^2 + a^2 \sin^2 \theta.
\end{equation} 
As indicated by the $r$-dependence this change of variables is integrable provided that $g$ and $h$ are functions of $r$ only.
However $\Delta$ as given in \eqref{algo:eq:rotating:tr-delta} for a generic configuration contains a $\theta$ dependence: one should check that this dependence cancels once restricted to the example of interest.
Otherwise one is not allowed to perform this change of coordinates (but other systems may still be found).

Given \eqref{algo:eq:change:bl:solution-gh} one gets the metric and gauge fields (deleting the prime)
\begin{subequations}
\label{algo:eq:rotating:tr}
\begin{gather}
	\label{algo:eq:rotating:metric:tr}
	\dd s^2 = - \tilde f\, \dd t^2
		+ \frac{\rho^2}{\Delta}\, \dd r^2
		+ \rho^2 \dd\theta^2
		+ \frac{\Sigma^2}{\rho^2} \sin^2 \theta\, \dd\phi^2
		+ 2 a (\tilde f - 1) \sin^2 \theta\, \dd t \dd\phi, \\
	\label{algo:eq:rotating:vector:tr}
	A = \tilde f_A\, \left(\dd t - \frac{\rho^2}{\Delta}\, \dd r - a \sin^2 \theta\, \dd \phi \right)
\end{gather}
\end{subequations}
with
\begin{equation}
	\label{algo:eq:rotating:tr-sigma}
	\frac{\Sigma^2}{\rho^2} = r^2 + a^2 + a g_{t\phi}.
\end{equation} 
The $rr$-term has been computed from
\begin{equation}
	\label{algo:eq:g-plus-a-sin-h}
	g - a \sin^2 \theta\, h = \frac{\rho^2}{\Delta}.
\end{equation} 
Generically the radial component of the gauge field depends only on radial coordinate $A_r = A_r(r)$ ($\theta$-dependence of the function $\tilde f_A$ sits in a factor $1 / \rho^2$ which cancels the one in front of $\dd r$) and one can perform a gauge transformation in order to set it to zero, leaving
\begin{equation}
	A = \tilde f_A\, \left(\dd t - a \sin^2 \theta\, \dd \phi \right).
\end{equation}

\subsection{Examples}
\label{sec:algo:examples}

\subsubsection{Flat space}
\label{sec:algorithm:flat}

It is straightforward to check that the algorithm applied to the Minkowski metric -- which has $f = 1$, leading to $\tilde f = 1$ -- in spherical coordinates
\begin{equation}
	\dd s^2 = - \dd t^2 + \dd r^2 + r^2 \big( \dd\theta^2 + \sin^2 \theta\; \dd \phi^2 \big)
\end{equation} 
gives again the Minkowski metric but in spheroidal coordinates \eqref{coord:metric:4d:spheroidal} (after a Boyer--Lindquist transformation)
\begin{equation}
	\dd s^2 = - \dd t^2 + \frac{\rho^2}{r^2 + a^2}\; \dd r^2 + \rho^2 \dd\theta^2 + (r^2 + a^2) \sin^2 \theta\; \dd \phi^2,
\end{equation} 
recalling that $\rho^2 = r^2 + a^2 \cos^2 \theta$.
The metric is exactly diagonal because $g_{t\phi} = 0$ for $\tilde f = 1$ from \eqref{algo:eq:rotating:metric:tr}.

Hence for flat space the JN algorithm reduces to a change of coordinates, from spherical to (oblate) spheroidal coordinates: the $2$-spheres foliating the space in the radial direction are deformed to ellipses with semi-major axis $a$.

This fact is an important consistency check that will be useful when extending the algorithm to higher dimensions (\cref{sec:higher}) or to other coordinate systems (such as one with direction cosines).
Moreover in this case one can forget about the time direction and consider only the transformation of the radial coordinate.

\subsubsection{Kerr--Newman}
\label{sec:algorithm:kerr-newman}

The seed function is the Reissner--Nordström for which the metric and gauge field are
\begin{equation}
	\label{algo:eq:rn:functions}
	f(r) = 1 - \frac{2m}{r} + \frac{q^2}{r^2}, \qquad
	f_A = \frac{q}{r}.
\end{equation} 
Applications of the rules \eqref{algo:eq:rules} leads to
\begin{subequations}
\begin{gather}
	\tilde f = 1 - \frac{2 m \Re r}{\abs{r}^2} + \frac{q^2}{\abs{r}^2}
		= 1 + \frac{q^2 - 2 m r'}{\rho^2}, \\
	\tilde f_A = \frac{q \Re r}{\abs{r}^2}
		= \frac{q r'}{\rho^2}.
\end{gather}
\end{subequations}
These functions together with \eqref{algo:eq:rotating:tr} describe correctly the Kerr--Newman solution~\cite{Visser:2009:KerrSpacetimeBrief, Adamo:2014:KerrNewmanMetricReview}.
For completeness we spell out the expressions of the quantities appearing in the metric
\begin{subequations}
\begin{gather}
	\frac{\Sigma^2}{\rho^2} = r^2 + a^2 - \frac{q^2 - 2 m r}{\rho^2}\, a^2 \sin^2 \theta, \\
	\Delta = r^2 - 2 m r + a^2 + q^2.
\end{gather}
\end{subequations}
In particular $\Delta$ does not contain any $\theta$ dependence and the BL transformation is well defined.
Moreover the radial component of the gauge field is
\begin{equation}
	A_r = - \frac{\tilde f_A \rho^2}{\Delta}
		= \frac{q r}{\Delta}
\end{equation} 
and it is independent of $\theta$.

%% file: sections/extension.tex
\section{Extension through simple examples}
\label{sec:extension}

In this section we motivate through simple examples modifications to the original prescription for the transformation of the functions.

\subsection{Magnetic charges: dyonic Kerr--Newman}
\label{sec:extension:dyonic}

The dyonic Reissner--Nordström metric is obtained from the electric one \eqref{algo:eq:rn:functions} by the replacement~\cite[sec.~6.6]{Carroll:2004:SpacetimeGeometryIntroduction}
\begin{equation}
	q^2 \longrightarrow \abs{Z}^2 = q^2 + p^2
\end{equation} 
where $Z$ corresponds to the central charge
\begin{equation}
	Z = q + i p.
\end{equation} 
Then the metric function reads
\begin{equation}
	f = 1 - \frac{2m}{r} + \frac{\abs{Z}^2}{r^2}.
\end{equation} 
The gauge field receives a new $\phi$-component
\begin{equation}
	\label{ext:eq:static:vector}
	A = f_A\, \dd t - p \cos \theta\, \dd\phi
		= f_A\, \dd u - p \cos \theta\, \dd\phi
\end{equation}
(the last equality being valid after a gauge transformation) and
\begin{equation}
	f_A = \frac{q}{r}.
\end{equation} 

The transformation of the function $f$ under \eqref{algo:eq:change:complexification-ur} is straightforward and yields
\begin{equation}
	\tilde f = 1 - \frac{2m r' - \abs{Z}^2}{\rho^2}.
\end{equation} 
On the other hand transforming directly the $r$ inside $f_A$ according to \eqref{algo:eq:rules} does not yield the correct result.
Instead one needs to first rewrite the gauge field function as
\begin{equation}
	f_A = \Re\left(\frac{Z}{r}\right)
\end{equation} 
from which the transformation proceeds to
\begin{equation}
	\tilde f_A = \frac{\Re(Z \bar r)}{\abs{r}^2}
		= \frac{q r' - p a \cos \theta}{\rho^2}.
\end{equation} 
Note that it not useful to replace $p$ by $\Im Z$ in \eqref{ext:eq:static:vector} since it is not accompanied by any $r$ dependence.
Moreover it is natural that the factor $\abs{Z}^2$ appears in the metric and this explains why the charges there do not mix with the coordinates.

The gauge field in BL coordinates is finally
\begin{subequations}
\begin{align}
	A &= \frac{q r - p a \cos \theta}{\rho^2}\, \dd t
			+ \left(- \frac{q r}{\rho^2}\, a \sin^2 \theta + \frac{p(r^2 + a^2)}{\rho^2}\, \cos\theta \right) \dd\phi \\
		&= \frac{q r}{\rho^2} (\dd t - a \sin^2 \theta \dd\phi)
			+ \frac{p \cos \theta}{\rho^2} \left(a\, \dd t + (r^2 + a^2)\, \dd\phi \right).
\end{align}
\end{subequations}
The radial component has been removed thanks to a gauge transformation since it depends only on $r$
\begin{equation}
	\Delta \times A_r = - \frac{q r - p a \cos \theta}{\rho^2}\, \rho^2 - p a \cos \theta
		= - q r.
\end{equation} 

There is a coupling between the parameters $a$ and $p$ which can be interpreted from the fact that a rotating magnetic charge has an electric quadrupole moment.
This coupling is taken into account from the product of the imaginary parts which yield a real term.
In view of the form of the algorithm such contribution could not arise from any other place.
Moreover the combination $Z = q + i p$ appears naturally in the Plebański--Demiański solution~\cite{Plebanski:1975:ClassSolutionsEinsteinMaxwell, Plebanski:1976:RotatingChargedUniformly}.

The Yang--Mills Kerr--Newman black hole found by Perry~\cite{Perry:1977:BlackHolesAre} can also be derived in this way, starting from the seed
\begin{equation}
	A^I = \frac{q^I}{r}\, \dd t + p^I \cos \theta\, \dd\phi, \qquad
	\abs{Z}^2 = q^I q^I + p^I p^I
\end{equation} 
where $q^I$ and $p^I$ are constant elements of the Lie algebra.

\subsection[NUT charge, cosmological constant and topological horizon: (anti-)de Sitter Schwarzschild--NUT]
{NUT charge and cosmological constant and topological horizon: (anti-)de Sitter Schwarzschild--NUT}
\label{sec:extension:nut}

In this subsection we consider general topological horizons
\begin{equation}
	\dd \Omega^2 = \dd\theta^2 + H(\theta)^2\, \dd \phi^2, \qquad
	H(\theta) =
	\begin{cases}
		\sin \theta & \kappa = 1 \quad (S^2), \\
		\sinh \theta & \kappa = -1 \quad (H^2).
	\end{cases}
\end{equation} 
The cosmological constant is denoted by $\Lambda$.
We give only the main formulas to motivate the modification of the algorithm, leaving the details of the transformation for \cref{sec:general}.

The complex transformation that adds a NUT charge is
\begin{subequations}
\label{ext:eq:change:jna-nut}
\begin{gather}
	u = u' - 2 \kappa \ln H(\theta), \qquad
	r = r' + i n, \\
	m = m' + i \kappa n, \qquad
	\kappa = \kappa' - \frac{4\Lambda}{3}\, n^2.
\end{gather}
\end{subequations}
Note that it is $\kappa$ and not $\kappa'$ that appears in $m$.
After having shown

The metric derived from the seed \eqref{algo:eq:static:metric:tr} is
\begin{equation}
	\dd s^2 = - \tilde f\, (\dd t - 2 \kappa n H'(\theta)\, \dd\phi)^2
		+ \tilde f^{-1}\, \dd r^2
		+ \rho^2\, \dd\Omega^2,
\end{equation}
see \eqref{gen:eq:rotating:tr-F-cst}, where
\begin{equation}
	\rho^2 = r'^2 + n^2.
\end{equation} 

The function corresponding to the (a)dS--Schwarzschild metric is
\begin{equation}
	f = \kappa - \frac{2m}{r} - \frac{\Lambda}{3}\, r^2
		= \kappa - 2 \Re\left(\frac{m}{r}\right) - \frac{\Lambda}{3}\, r^2.
\end{equation} 
The transformation is
\begin{equation}
	\tilde f = \kappa
			- \frac{2 \Re(m \bar r)}{\abs{r}^2}
			- \frac{\Lambda}{3}\, \abs{r}^2
		= \kappa' - \frac{4\Lambda}{3}\, n^2
			- \frac{2 \left[ m' r' + \left( \kappa' - \frac{4\Lambda}{3}\, n^2 \right) n^2 \right]}{\rho^2}
			- \frac{\Lambda}{3}\, \rho^2
\end{equation} 
which after simplification gives
\begin{equation}
	\label{ext:eq:nut-tilde-f}
	\tilde f = \kappa' - \frac{2 m' r' + 2 \kappa' n^2}{\rho^2}
		- \frac{\Lambda}{3} (r'^2 + 5 n^2)
		+ \frac{8\Lambda}{3}\, \frac{n^4}{\rho^2}
\end{equation} 
which corresponds correctly to the function of (a)dS--Schwarzschild--NUT~\cite{AlonsoAlberca:2000:SupersymmetryTopologicalKerrNewmannTaubNUTaDS}.

Note that it is necessary to consider the general case of massive black hole with topological horizon (if $\Lambda \neq 0$ for the latter) even if one is ultimately interested in the $m = 0$ or $\kappa = 1$ cases.

The transformation \eqref{ext:eq:change:jna-nut} can be interpreted as follows.
In similarity with the case of the magnetic charge, writing the mass as a complex parameter is needed to take into account some couplings between the parameters that would not be found otherwise.
Moreover the shift of $\kappa$ is required because the curvature of the $(\theta, \phi)$ section should be normalized to $\kappa = \pm 1$ but the coupling of the NUT charge with the cosmological constant modifies the curvature: the new shift is necessary to balance this effect and to normalize the $(\theta, \phi)$ curvature to $\kappa' = \pm 1$ in the new metric.
The NUT charge in the Plebański--Demiański solution~\cite{Plebanski:1975:ClassSolutionsEinsteinMaxwell, Plebanski:1976:RotatingChargedUniformly} is
\begin{equation}
	\ell = n \left( 1 - \frac{4\Lambda}{3}\, n^2 \right)
\end{equation} 
so the natural complex combination is $m + i \ell$ and not $m + i \kappa n$ from this point of view, and similarly for the curvature~\cite[sec.~5.3]{Griffiths:2006:NewLookPlebanskiDemianski} (such relations appear when taking limit of the Plebański--Demiański solution to recover subcases).

Finally we conclude this section with two remarks to quote different contexts where the above expression appear naturally :
\begin{itemize}
	\item Embedding Einstein--Maxwell into $N = 2$ supergravity with a negative cosmological constant $\Lambda = - 3 g^2$, the solution is BPS if~\cite{AlonsoAlberca:2000:SupersymmetryTopologicalKerrNewmannTaubNUTaDS}
	\begin{equation}
		\kappa' = -1, \qquad
		n = \pm \frac{1}{2g},
	\end{equation} 
	in which case $\kappa' = \kappa$.
	
	\item The Euclidean NUT solution is obtained from the Wick rotation
	\begin{equation}
		t = - i \tau, \qquad
		n = i \nu.
	\end{equation}
	The condition for regularity is~\cite{Chamblin:1999:LargeNPhases, Johnson:2014:ThermodynamicVolumesAdSTaubNUT}
	\begin{equation}
		m = m' - \nu \left( \kappa + \frac{4\Lambda}{3}\, \nu^2 \right)
			= 0.
	\end{equation} 
\end{itemize}

\subsection{Complex scalar fields}

For a complex scalar field, or any pair of real fields that can be naturally gathered as a complex field, one should treat the full field as a single entity instead of looking at the real and imaginary parts independently.
In particular one should not impose any reality condition.
A typical case of such system is the axion--dilaton pair
\begin{equation}
	\tau = \e^{-2\phi} + i \sigma.
\end{equation} 

In order to demonstrate this principle consider the seed (for a complete example see \cref{sec:examples:rotating-T3})
\begin{equation}
	\tau = 1 + \frac{\mu}{r}
\end{equation} 
where only the dilaton is non-zero.
Then the transformation \eqref{algo:eq:change:complexification-ur} gives
\begin{equation}
	\tau' = 1 + \frac{\mu}{r}
		= 1 + \frac{\mu}{r' - i a \cos\theta}
		= 1 + \frac{\mu r'}{\rho^2} + i\, \frac{\mu a \cos\theta}{\rho^2}.
\end{equation} 
The transformation generates an imaginary part which cannot be obtained if $\Im \tau$ is treated separately: the algorithm does not change fields that vanish except if they are components of a larger field.
Note that both $\tau$ and $\tau'$ are harmonic functions.

%% file: sections/general.tex
\section{Complete algorithm}
\label{sec:general}

In this section we gather all the facts on the Janis--Newman algorithm and we explain how to apply it to a general setting.
We write the formulas corresponding to the most general configurations that can be obtained.
We insist again on the fact that all these results can also be derived from the tetrad formalism.

\subsection{Seed configuration}
\label{sec:general:seed}

We consider a general configuration with a metric $g_{\mu\nu}$, gauge fields $A_\mu^I$, complex scalar fields $\tau^i$ and real scalar fields $q^u$.
The initial parameters of the seed configuration are the mass $m$, electric charges $q^I$, magnetic charges $p^i$ and some other parameters $\lambda^A$ (such as the scalar charges).
The electric and magnetic charges are grouped in complex parameters
\begin{equation}
	Z^I = q^I + i p^I.
\end{equation} 
All indices run over some arbitrary ranges.

The seed configuration is spherically symmetric and in particular all the fields depend only on the radial direction $r$
\begin{subequations}
\label{gen:eq:static:tr}
\begin{gather}
	\label{gen:eq:static:metric:tr}
	\dd s^2 = - f_t(r)\, \dd t^2 + f_r(r)\, \dd r^2 + f_\Omega(r)\, \dd\Omega^2, \\
	A^I = f^I(r)\, \dd t + p^I H'(\theta)\, \dd\phi, \\
	\tau^i = \tau^i(r), \qquad
	q^u = q^u(r)
\end{gather}
\end{subequations}
where
\begin{equation}
	\dd \Omega^2 = \dd\theta^2 + H(\theta)^2\, \dd \phi^2, \qquad
	H(\theta) =
	\begin{cases}
		\sin \theta & \kappa = 1 \quad (S^2), \\
		\sinh \theta & \kappa = -1 \quad (H^2).
	\end{cases}
\end{equation} 
Note that only two functions in the metric are relevant since the last one can be fixed through a diffeomorphism.
All the real functions are denoted collectively by
\begin{equation}
	f_i = \{ f_t, f_r, f_\Omega, f^I, q^u \}.
\end{equation} 

The transformation to null coordinates is
\begin{equation}
	\label{gen:eq:change:null}
	\dd t = \dd u - \sqrt{\frac{f_r}{f_t}}\, \dd r
\end{equation} 
and yields
\begin{subequations}
\label{gen:eq:static:ur}
\begin{gather}
	\label{gen:eq:static:metric:ur}
	\dd s^2 = - f_t\, \dd u^2 - 2 \sqrt{f_t f_r}\, \dd r^2 + f_\Omega\, \dd\Omega^2, \\
	A^I = f^I\, \dd u + p^I H'\, \dd\phi
\end{gather}
\end{subequations}
where the radial component of the gauge field
\begin{equation}
	A^I_r = f^I \sqrt{\frac{f_r}{f_t}}
\end{equation} 
has been set to zero through a gauge transformation.

\subsection{Janis--Newman algorithm}
\label{sec:general:jna}

\subsubsection{Complex transformation}

One performs the complex change of coordinates
\begin{equation}
	\label{gen:eq:change:jna}
	r = r' + i\, F(\theta), \qquad
	u = u' + i\, G(\theta).
\end{equation}
In the case of topological horizons the Giampieri ansatz \eqref{algo:eq:giampieri-ansatz} generalizes to
\begin{equation}
	\label{gen:eq:giampieri-ansatz}
	i\, \dd \theta = H(\theta)\, \dd \phi
\end{equation} 
leading to the differentials
\begin{equation}
	\dd r = \dd r' + F'(\theta) H(\theta)\, \dd \phi, \qquad
	\dd u = \dd u' + G'(\theta) H(\theta)\, \dd \phi.
\end{equation} 
The ansatz \eqref{gen:eq:giampieri-ansatz} is a direct consequence of the fact that the $2$-dimensional slice $(\theta, \phi)$ is given by $\dd \Omega^2 = \dd\theta^2 + H(\theta)^2\, \dd \phi^2$, such that the function in the RHS of \eqref{gen:eq:giampieri-ansatz} corresponds to $\sqrt{g^\Omega_{\phi\phi}}$ (where $g$ is the static metric), as can be seen by doing the computation with $i\, \dd \theta = \mc H(\theta) \dd\phi$ and identifying $\mc H = H$ at the end.

The most general known transformation is
\begin{subequations}
\begin{gather}
	\label{gen:eq:change:jna-functions-FG}
	F(\theta) = n - a\, H'(\theta) + c \left( 1 + H'(\theta)\, \ln \frac{H(\theta/2)}{H'(\theta/2)} \right), \\
	G(\theta) = \kappa a\, H'(\theta)
		- 2 \kappa n \ln H(\theta)
		- \kappa c\, H'(\theta)\, \ln \frac{H(\theta/2)}{H'(\theta/2)}, \\
	m = m' + i \kappa n, \\
	\kappa = \kappa' - \frac{4\Lambda}{3}\, n^2,
\end{gather}
\end{subequations}
where $a, c \neq 0$ only if $\Lambda = 0$ (see \cref{sec:derivation} for the derivation).
The mass that is transformed is the physical mass: even if it written in terms of other parameters one should identify it and transform it.

The parameters $a$ and $n$ correspond respectively to the angular momentum and to the NUT charge.
On the other hand the constant $c$ did not receive any clear interpretation (see for example~\cites{Demianski:1972:NewKerrlikeSpacetime, Adamo:2014:KerrNewmanMetricReview}[sec.~5.3]{Krasinski:2006:InhomogeneousCosmologicalModels}).
It can be noted that the solution is of type II in Petrov classification (and thus the JN algorithm \emph{can} change the Petrov type) and it corresponds to a wire singularity on the rotation axis.
Moreover the BL transformation is not well-defined.

\subsubsection{Function transformation}
\label{sec:general:jna:functions}

All the real functions $f_i = f_i(r)$ must be modified to be kept real once $r \in \C$
\begin{equation}
	\label{gen:eq:complexification-functions}
	\tilde f_i = \tilde f_i(r, \bar r)
		= \tilde f_i\big(r', F(\theta) \big) \in \R.
\end{equation} 
The last equality means that $\tilde f_i$ can depend on $\theta$ only through $\Im r = F(\theta)$.
The condition that one recovers the seed for $\bar r = r = r'$ is
\begin{equation}
	\tilde f_i(r', 0) = f_i(r').
\end{equation} 

If all magnetic charges are vanishing or in terms without electromagnetic charges the rules for finding the $\tilde f_i$ are
\begin{subequations}
\label{gen:eq:rules}
\begin{align}
	\label{gen:eq:rules:r}
	r & \longrightarrow \frac{1}{2} (r + \bar r) = \Re r, \\
	\label{gen:eq:rules:1/r}
	\frac{1}{r} & \longrightarrow \frac{1}{2} \left(\frac{1}{r} + \frac{1}{\bar r}\right) = \frac{\Re r}{\abs{r}^2}, \\
	\label{gen:eq:rules:r2}
	r^2 & \longrightarrow \abs{r}^2.
\end{align}
\end{subequations}
Up to quadratic powers of $r$ and $r^{-1}$ these rules determine almost uniquely the result.
This is not anymore the case when the configurations involve higher power.
These can be dealt with by splitting it in lower powers: generically one should try to factorize the expression into at most quadratic pieces.
Some examples of this with natural guesses are
\begin{equation}
	r^4 - b^2 = (r^2 + b) (r^2 - b), \qquad
	r^4 + b = r^2 \left( r^2 + \frac{b}{r^2} \right).
\end{equation} 
Moreover the same power of $r$ can be transformed differently, for example
\begin{equation}
	\frac{1}{r^n} \longrightarrow \frac{1}{r^{n-2}}\, \frac{1}{\abs{r}^2}.
\end{equation} 

Denoting by $Q(r)$ and $P(r)$ collectively all functions that multiply $q^I$ and $p^I$ respectively, all such terms should be rewritten as
\begin{equation}
	\Big( q^I Q(r), p^I P(r) \Big) = \Big( \Re\big(Z^I Q(r)\big), \Im\big(Z^I P(r)\big) \Big)
\end{equation} 
before performing the transformation \eqref{gen:eq:change:jna}.
Note that in this case one does not use the rules \eqref{gen:eq:rules}.

Finally the transformed complex scalars are obtained by simply plugging \eqref{gen:eq:change:jna}
\begin{equation}
	\tau'^i(r', \theta) = \tau^i\big(r + i F(\theta)\big).
\end{equation}

\subsubsection{Null coordinates}

Plugging the transformation \eqref{gen:eq:change:jna} inside the seed metric and gauge fields \eqref{gen:eq:static:ur} leads to\footnotemark{}%
\footnotetext{%
	We stress that at this stage these formula do not satisfy Einstein equations, they are just proxies to simplify later computations.
}
\begin{subequations}
\label{gen:eq:rotating:ur}
\begin{gather}
	\dd s^2 = - \tilde f_t\, (\dd u' + \alpha\, \dd r' + \omega H\, \dd\phi )^2
		+ 2 \beta\, \dd r' \dd \phi
		+ \tilde f_\Omega\, \big(\dd\theta^2 + \sigma^2 H^2\, \dd\phi^2 \big), \\
	A^I = \tilde f^I\, (\dd u' + G' H\, \dd \phi) + p^I H'\, \dd\phi
\end{gather}
\end{subequations}
where (one should not confuse the primes to indicate derivatives from the primes on the coordinates)
\begin{equation}
	\omega = G' + \sqrt{\frac{\tilde f_r}{\tilde f_t}}\, F', \qquad
	\sigma^2 = 1 + \frac{\tilde f_r}{\tilde f_\Omega}\, F'^2, \qquad
	\alpha = \sqrt{\frac{\tilde f_r}{\tilde f_t}}, \qquad
	\beta = \tilde f_r\, F' H.
\end{equation}

\subsubsection{Boyer--Lindquist coordinates}

The Boyer--Lindquist transformation
\begin{equation}
	\label{gen:eq:change:bl}
	\dd u' = \dd t' - g(r') \dd r', \qquad
	\dd \phi = \dd \phi' - h(r') \dd r',
\end{equation} 
can be used to remove the off-diagonal $tr$ and $r\phi$ components of the metric
\begin{equation}
	g_{t'r'} = g_{r'\phi'} = 0.
\end{equation} 
The solution to these equations is
\begin{equation}
	\label{gen:eq:change:bl:solution-gh}
	g(r') = \frac{\sqrt{\big(\tilde f_t \tilde f_r \big)^{-1}}\, \tilde f_\Omega - F' G'}{\Delta}, \qquad
	h(r') = \frac{F'}{H \Delta}
\end{equation} 
where
\begin{equation}
	\label{gen:eq:change:bl:delta}
	\Delta = \frac{\tilde f_\Omega}{\tilde f_r}\, \sigma^2
		= \frac{\tilde f_\Omega}{\tilde f_r} + F'^2.
\end{equation} 
Remember that the changes of coordinate is valid only if $g$ and $h$ are functions of $r'$ only.

Inserting \eqref{gen:eq:change:bl:solution-gh} into \eqref{gen:eq:rotating:ur} yields
\begin{subequations}
\label{gen:eq:rotating:tr}
\begin{gather}
	\dd s^2 = - \tilde f_t\, (\dd t' + \omega H\, \dd\phi' )^2
		+ \frac{\tilde f_\Omega}{\Delta}\, \dd r'^2
		+ \tilde f_\Omega\, \big(\dd\theta^2 + \sigma^2 H^2\, \dd\phi'^2 \big), \\
	A^I = \tilde f^I\, \left(\dd t' - \frac{\tilde f_\Omega}{\Delta \sqrt{\tilde f_t \tilde f_r}}\, \dd r' + G' H\, \dd \phi' \right) + p^I H'\, \dd\phi'
\end{gather}
\end{subequations}
where we recall that
\begin{equation}
	\omega = G' + \sqrt{\frac{\tilde f_r}{\tilde f_t}}\, F', \qquad
	\sigma^2 = 1 + \frac{\tilde f_r}{\tilde f_\Omega}\, F'^2.
\end{equation} 
Generically one finds $A_r = A_r(r)$ which can be set to zero thanks to a gauge transformation.

Before closing this section we simplify the above formulas for few simple cases that are often used.

\paragraph{Degenerate Schwarzschild seed}

A degenerate seed (one unknown function) in Schwarzschild coordinates has
\begin{equation}
	f_r = f_t^{-1}, \qquad
	f_\Omega = r^2.
\end{equation} 
The above formulas for this case can be found in \cref{sec:derivation:ansatz}.

\paragraph{Degenerate isotropic seed}

A degenerate seed in isotropic coordinates has
\begin{equation}
	f_t = f^{-1}, \qquad
	f_r = f, \qquad
	f_\Omega = r^2 f.
\end{equation} 
In this case the above formulas reduced to
\begin{subequations}
\label{gen:eq:rotating:tr-degenerate-isotropic}
\begin{gather}
	\dd s^2 = - \tilde f^{-1}\, (\dd t + \omega H\, \dd\phi )^2
		+ \tilde f \rho^2 \left( \frac{\dd r^2}{\Delta}
			+ \dd\theta^2 + \sigma^2 H^2\, \dd\phi^2 \right), \\
	A^I = \tilde f^I\, \left(\dd t - \frac{\tilde f \rho^2}{\Delta}\, \dd r + G' H\, \dd \phi \right) + p^I H'\, \dd\phi
\end{gather}
\end{subequations}
where we recall that
\begin{equation}
	\omega = G' + \tilde f\, F', \qquad
	\sigma^2 = 1 + \frac{F'^2}{\rho^2}, \qquad
	\Delta = \tilde f \rho^2 + F'^2.
\end{equation} 

\paragraph{Constant $F$}

The expressions simplify greatly if $F' = 0$ (for example when $\Lambda \neq 0$).
First all functions depend only on $r$ since $F(\theta) = \cst$
\begin{equation}
	\tilde f_i(r, \theta) = \tilde f_i(r, 0).
\end{equation} 
As a consequence the Boyer--Lindquist transformation \eqref{gen:eq:change:bl:solution-gh}
\begin{equation}
	g(r') = \sqrt{\frac{\tilde f_r}{\tilde f_t}}, \qquad
	h(r') = 0
\end{equation} 
is always well-defined.
For the same reason it is always possible to perform a gauge transformation.
Finally the metric and gauge fields \eqref{gen:eq:rotating:tr} becomes
\begin{subequations}
\label{gen:eq:rotating:tr-F-cst}
\begin{gather}
	\dd s^2 = - \tilde f_t \big(\dd t + G' H\, \dd\phi \big)^2
		+ \tilde f_r\, \dd r^2
		+ \tilde f_\Omega\, \dd\Omega^2, \\
	A^I = \tilde f^I\, \left(\dd t' + G' H\, \dd \phi' \right) + p^I H'\, \dd\phi'.
\end{gather}
\end{subequations}

\subsection{Open questions}

The algorithm we have described help to work with five (four if $\Lambda \neq 0$) of the six parameters of the Plebański--Demiański (PD) solution.
It is tempting to conjecture that it can be extended to the full set of parameters by generalizing the ideas described in \cref{sec:extension:nut} (shifting $\kappa$, writing $a + i \alpha$…).
Indeed we have found that these operations were quite natural in the context of the  PD solution and inspiration could be found in~\cite{Griffiths:2006:NewLookPlebanskiDemianski}.

%% file: sections/derivation.tex
\section{Derivation of the transformations}
\label{sec:derivation}

The goal of this section is to derive the form \eqref{gen:eq:change:jna-functions-FG} of the possible complex transformations.
This method was first used by Demiański~\cite{Demianski:1972:NewKerrlikeSpacetime} and then generalized in~\cite{Erbin:2016:DecipheringGeneralizingDemianskiJanisNewman}.
The idea is to perform the algorithm in a simple setting (metric with one unknown function and one gauge field), leaving arbitrary the functions $F(\theta)$ and $G(\theta)$ in \eqref{gen:eq:change:jna} and $\tilde f_i$ before solving the equations of motion to determine them.
Then the result can be reinterpreted in terms of rules to get the functions $\tilde f_i$ from $f_i$ (this last part was not discussed in~\cite{Demianski:1972:NewKerrlikeSpacetime}).
This selects the possible complex transformations.
Then one can hope that these transformations will be the most general ones (under the assumptions that are made), and one can use these transformations in other cases without having to solve the equations.
The latter claim can be justified by looking at the equations of motions for more complex examples: even if one cannot find directly a solution, one finds that the same structure persists~\cite{Erbin:2016:DecipheringGeneralizingDemianskiJanisNewman} (this is also motivated by the solutions in~\cite{Krori:1981:ChargedDemianskiMetric, Patel:1988:RadiatingDemianskitypeMetrics}).
Another strength of this approach is to remove the ambiguity of the algorithm since the functions are found from the equations of motion, and this may help when one does not know how to perform precisely the algorithm (for example in higher dimensions, see \cref{sec:higher}).

Another goal of this section is to expose the full technical details of the computations: Demiański's paper~\cite{Demianski:1972:NewKerrlikeSpacetime} is short and results are extremely condensed.
In particular we uncover an underlying assumption on the form of the metric function and we show how this lead to an error an in his formula (14) (already pointed out in~\cite{Quevedo:1992:ComplexTransformationsCurvature}).
A generalization of this hypothesis leads to other equations that we could not solve analytically and which may lead to other complex transformations.

Finally this analysis shows the impossibility to derive the (a)dS--Kerr(--Newman) solutions from the JN algorithm.
As discussed in the previous section generalization of the ansatz may help to avoid this no-go theorem.

\subsection{Setting up the ansatz}
\label{sec:derivation:ansatz}

We first recall the action and equations of motion before describing the ansatz for the metric and gauge fields.
We refer to \cref{sec:general} for the general formulas from which the expressions in this section are derived.

\subsubsection{Action and equations of motion}

The action for Einstein--Maxwell gravity with cosmological constant $\Lambda$ reads
\begin{equation}
	\label{deriv:eq:einstein-maxwell-action}
	S = \int \dd^4 x\; \sqrt{- g} \left( \frac{1}{2 \varkappa^2} (R - 2 \Lambda) - \frac{1}{4}\, F^2 \right),
\end{equation} 
where $\varkappa^2 = 8 \pi G$ is the Einstein coupling constant, $g_{\mu\nu}$ is the metric with Ricci scalar $R$ and $F = \dd A$ is the field strength of the Maxwell field $A_\mu$.
In the rest of this section we will set $\varkappa = 1$.
The corresponding equations of motion (respectively Einstein and Maxwell) are
\begin{equation}
	\label{deriv:eq:einstein-maxwell-eom}
	G_{\mu\nu} + \Lambda g_{\mu\nu} = 2\, T_{\mu\nu}, \qquad
	\grad_\mu F^{\mu\nu} = 0,
\end{equation} 
where energy--momentum tensor for the electromagnetic gauge field $A_\mu$ is
\begin{equation}
	T_{\mu\nu} = F_{\mu\rho} \tens{F}{_\nu^\rho} - \frac{1}{4}\, g_{\mu\nu} F^2.
\end{equation}

\subsubsection{Seed configuration}

We are interested in the subcase of \eqref{gen:eq:static:metric:tr} where
\begin{equation}
	\label{eq:static-ansatz-one-unknown}
	f_t = f, \qquad
	f_r = f^{-1}, \qquad
	f_\Omega = r^2.
\end{equation} 

The seed configuration is
\begin{subequations}
\label{deriv:eq:static:tr}
\begin{gather}
	\label{deriv:eq:static:metric:tr}
	\dd s^2 = - f(r)\, \dd t^2 + f(r)^{-1}\, \dd r^2 + r^2\, \dd\Omega^2, \\
	\label{deriv:eq:static:vector:tr}
	A = f_A(r)\, \dd t
\end{gather}
\end{subequations}
where we consider spherical and hyperbolic horizons
\begin{equation}
	\dd \Omega^2 = \dd\theta^2 + H(\theta)^2\, \dd \phi^2, \qquad
	H(\theta) =
	\begin{cases}
		\sin \theta & \kappa = 1, \\
		\sinh \theta & \kappa = -1.
	\end{cases}
\end{equation} 
In terms of null coordinates \eqref{gen:eq:change:null} the configuration reads
\begin{subequations}
\label{deriv:eq:static:ur}
\begin{gather}
	\label{deriv:eq:static:metric:ur}
	\dd s^2 = - f\, \dd u^2 - 2\, \dd u \dd r + r^2\, \dd\Omega^2, \\
	\label{deriv:eq:static:vector:ur}
	A = f_A\, \dd u.
\end{gather}
\end{subequations}

\subsubsection{Janis--Newman configuration}

The configuration obtained from the Janis--Newman algorithm with a general transformation \eqref{gen:eq:change:jna}
\begin{equation}
	r = r' + i\, F(\theta), \qquad
	u = u' + i\, G(\theta)
\end{equation}
corresponds to (we omit the primes on the coordinates)
\begin{subequations}
\label{deriv:eq:rotating:ur}
\begin{gather}
	\dd s^2 = - \tilde f\, (\dd u + \alpha\, \dd r + \omega H\, \dd\phi )^2
		+ 2 \beta\, \dd r \dd \phi
		+ \rho^2\, \big(\dd\theta^2 + \sigma^2 H^2\, \dd\phi^2 \big), \\
	A = \tilde f_A\, (\dd u + G' H\, \dd \phi)
\end{gather}
\end{subequations}
where
\begin{equation}
	\rho^2 = r^2 + F^2, \quad
	\omega = G' + \tilde f^{-1}\, F', \quad
	\sigma^2 = 1 + \frac{F'^2}{\tilde f \rho^2}, \quad
	\alpha = \tilde f^{-1}, \quad
	\beta = \tilde f^{-1}\, F' H.
\end{equation} 

The Boyer--Lindquist transformation \eqref{gen:eq:change:bl}
\begin{equation}
	\dd u = \dd t' - g(r) \dd r, \qquad
	\dd \phi = \dd \phi' - h(r) \dd r
\end{equation} 
with functions
\begin{equation}
	g(r) = \frac{\rho^2 - F' G'}{\Delta}, \qquad
	h(r) = \frac{F'}{H \Delta}, \qquad
	\Delta = \tilde f \rho^2\, \sigma^2
\end{equation} 
leads to (omitting the primes on the coordinates)
\begin{subequations}
\label{deriv:eq:rotating:tr}
\begin{gather}
	\dd s^2 = - \tilde f_t\, (\dd t + \omega H\, \dd\phi )^2
		+ \frac{\rho^2}{\Delta}\, \dd r^2
		+ \rho^2\, \big(\dd\theta^2 + \sigma^2 H^2\, \dd\phi^2 \big), \\
	A = \tilde f_A\, \left(\dd t - \frac{\rho^2}{\Delta}\, \dd r + G' H\, \dd \phi \right).
\end{gather}
\end{subequations}

\subsection{Static solution}

It is straightforward to solve the equations \eqref{deriv:eq:einstein-maxwell-eom} for the static configuration \eqref{deriv:eq:static:tr}.

Only the $(t)$ component of Maxwell equations is non trivial
\begin{equation}
	2 f'_A + r f''_A = 0,
\end{equation} 
the prime being a derivative with respect to $r$, and its solution is
\begin{equation}
	f_A(r) = \alpha + \frac{q}{r}
\end{equation} 
where $q$ is a constant of integration that is interpreted as the charge and $\alpha$ is an additional constant that can be removed by a gauge transformation.

The only relevant Einstein equation is
\begin{equation}
	\frac{q^2}{r^2} - \kappa + r^2 \Lambda + f + r f' = 0
\end{equation} 
whose solution reads
\begin{equation}
	\label{eq:topdown-1:static-f}
	f(r) = \kappa - \frac{2m}{r} + \frac{q^2}{r^2} - \frac{\Lambda}{3}\, r^2,
\end{equation} 
$m$ being a constant of integration that is identified to the mass.

We stress that we are just looking for solutions of Einstein equations and we are not concerned with regularity (in particular it is well-known that only $\kappa = 1$ is well-defined for $\Lambda = 0$).

The solution we will find in the next section should reduce to this one upon setting $F, G = 0$.

\subsection{Stationary solution}

Since Boyer--Lindquist imposes additional restrictions on the solutions we will solve the equations of motion \eqref{deriv:eq:einstein-maxwell-eom} for the configuration in null coordinates \eqref{deriv:eq:rotating:ur}.

\subsubsection{Simplifying the equations}
\label{sec:derivation:stationary:simplifying}

The components $(rr)$ and $(r\theta)$ give respectively the equation
\begin{subequations}
\begin{align}
	G'' + \frac{H'}{H}\, G' &= \pm 2 F, \\
	F' \left( G'' + \frac{H'}{H}\, G' \right) &= 2 F F'.
\end{align}
\end{subequations}
If $F' = 0$ then $F$ is an arbitrary constant and the sign of the first equation can be absorbed into its definition.\footnotemark{}%
\footnotetext{%
	In particular all expressions are quadratic in $F$, but only linear in $F'$.
}
On the other hand if $F' \neq 0$ one can simplify by the latter in the second equation and this fixes the sign of the first equation.
Then in both cases the relevant equation reduces to
\begin{equation}
	\label{eq:topdown-1-F-Gd-bis}
	G'' + \frac{H'}{H}\, G' = 2 F,
\end{equation} 
which depends only on $\theta$ and allows to solve for $G$ in terms of $F$.

Integrating the $r$-component of the Maxwell equation gives
\begin{equation}
	\tilde f_A = \frac{q\, r}{r^2 + F^2} + \alpha\, \frac{r^2 - F^2}{r^2 + F^2}.
\end{equation}
The $\theta$-equation reads
\begin{equation}
	\alpha\, F' = 0
\end{equation}
which implies $\alpha = 0$ if $F' \neq 0$.
The $\phi$- and $t$-equations follow from these two equations.
As seen above, $\alpha$ can be removed in the static limit $F \to 0$ and in the rest of this section we consider only the case\footnotemark{}%
\footnotetext{%
	We relax this assumption in \cref{sec:derivation:relaxing:gauge-fields}.
}
\begin{equation}
	\alpha = 0.
\end{equation} 

The $(tr)$ equation contains only $r$-derivatives of $\tilde f$ and can be integrated, giving\footnotemark{}%
\footnotetext{%
	In~\cite{Demianski:1972:NewKerrlikeSpacetime} the last term of $\tilde f$ is missing as pointed out in~\cite{Quevedo:1992:ComplexTransformationsCurvature}.
}
\begin{equation}
	\tilde f = \kappa - \frac{2m r - q^2 + 2 F (\kappa\, F + K)}{r^2 + F^2} - \frac{\Lambda}{3}\, (r^2 + F^2) - \frac{4 \Lambda}{3}\, F^2 + \frac{8 \Lambda}{3}\, \frac{F^4}{r^2 + F^2}
\end{equation} 
where again $m$ is a constant of integration interpreted as the mass and the function $K$ is defined by
\begin{equation}
	2 K = F'' + \frac{H'}{H}\, F'.
\end{equation} 
This implies the equations $(r\phi)$ and $(\theta\theta)$.

As explained below \eqref{gen:eq:complexification-functions} the $\theta$-dependence should be contain in $F(\theta)$ only.
The second term of the function $\tilde f$ contains some lonely $\theta$ from the $H(\theta)$ in the function $K$: this means that they should be compensated by the $F$, and we therefore ask that the sum $\kappa F + K$ be constant\footnotemark{}%
\footnotetext{%
	In \cref{sec:derivation:relaxing:metric-function} we relax this last assumption by allowing non-constant $\kappa F + K$.
	In this context the equations and the function $\tilde f$ are modified and this provides an explanation for the Demiański's error in $\tilde f$ in~\cite{Demianski:1972:NewKerrlikeSpacetime}.
}
\begin{equation}
	\kappa\, F' + K' = 0
	\quad \Longrightarrow \quad
	\kappa\, F + K = \kappa n.
\end{equation} 
The parameter $n$ is interpreted as the NUT charge.

The components $(t\theta)$ and $(\theta\phi)$ give the same equation
\begin{equation}
	\Lambda\, F' = 0.
\end{equation} 

Finally one can check that the last three equations $(tt), (t\phi)$ and $(\phi\phi)$ are satisfied.

\subsubsection{Summary of the equations}

The equations to be solved are
\begin{subequations}
\label{eq:topdown-1}
\begin{align}
	\label{eq:topdown-1-F-Gd}
	2 F &= G'' + \frac{H'}{H}\; G', \\
	\label{eq:topdown-1-Fd-Kd}
	\kappa\, n &= \kappa\, F + K, \\
	\label{eq:topdown-1-lambda}
	0 &= \Lambda F'
\end{align}
and the function $\tilde f$ is
\begin{equation}
	\label{eq:topdown-1-tilde-f}
	\tilde f = \kappa - \frac{2m r - q^2 + 2 F (\kappa\, F + K)}{r^2 + F^2} - \frac{\Lambda}{3}\, (r^2 + F^2) - \frac{4 \Lambda}{3}\, F^2 + \frac{8 \Lambda}{3}\, \frac{F^4}{r^2 + F^2}.
\end{equation}
We also defined
\begin{equation}
	\label{eq:topdown-1-K-Fd}
	2 K = F'' + \frac{H'}{H}\, F'.
\end{equation} 
\end{subequations}

As explained in the introduction the second step will be to explain \eqref{eq:topdown-1-tilde-f} in terms of new rules for the algorithm: they have been found in~\cite{Erbin:2016:DecipheringGeneralizingDemianskiJanisNewman} and this was the topic of \cref{sec:general:jna}.

In the next subsections we solve explicitly the equations \eqref{eq:topdown-1} in both cases $\Lambda \neq 0$ and $\Lambda = 0$.

\subsubsection{Solution for \texorpdfstring{$\Lambda \neq 0$}{non-vanishing cosmological constant}}

Equation \eqref{eq:topdown-1-lambda} implies that $F' = 0$, from which $K = 0$ follows by definition; then one obtains
\begin{equation}
	F(\theta) = n
\end{equation} 
by compatibility with \eqref{eq:topdown-1-Fd-Kd} and since $K(\theta) = 0$.

Solution to \eqref{eq:topdown-1-F-Gd} is
\begin{equation}
	G(\theta) = c_1 - 2 \kappa\, n \ln H(\theta) + c_2 \ln \frac{H(\theta/2)}{H'(\theta/2)}
\end{equation} 
where $c_1$ and $c_2$ are two constants of integration.
Since only $G'$ appears in the metric we can set $c_1 = 0$.
On the other hand the constant $c_2$ can be removed by the transformation
\begin{equation}
	\dd u = \dd u' - c_2\, \dd\phi
\end{equation} 
since one has
\begin{equation}
	\left( \ln \frac{H(\theta/2)}{H'(\theta/2)} \right)' = \frac{1}{H(\theta)}.
\end{equation} 

The solution to the system \eqref{eq:topdown-1} is thus
\begin{equation}
	F(\theta) = n, \qquad
	G(\theta) = - 2 \kappa\, n \ln H(\theta).
\end{equation} 
The function $\tilde f$ then takes the form
\begin{equation}
	\label{eq:topdown-1:tilde-f-lambda}
	\tilde f = \kappa - \frac{2m r - q^2 + 2 \kappa n^2}{r^2 + n^2} - \frac{\Lambda}{3}\,\frac{r^4 + 6 n^2 r^2 - 3 n^4}{r^2 + n^2}.
	% \kappa - \frac{2m r - q^2 + 2 \kappa n^2}{r^2 + n^2} - \frac{\Lambda}{3} (r^2 + 5 n^2) + \frac{8 \Lambda}{3}\, \frac{n^4}{r^2 + n^2}
\end{equation} 
This corresponds to the (a)dS--Schwarzschild--NUT solution: compare with \eqref{ext:eq:nut-tilde-f} and \eqref{gen:eq:rotating:tr-F-cst}.

The parameter $\Delta$ in the BL transformation \eqref{gen:eq:change:bl:delta} is
\begin{equation}
	\Delta = \kappa r^2 - 2 m r + q^2 + \Lambda n^4 - \frac{\Lambda}{3}\, r^4 - n^2 (\kappa + 2 \Lambda r^2 ).
\end{equation} 

As noted by Demiański the only parameters that appear are the mass and the NUT charge, and it is not possible to add angular momentum for non-vanishing cosmological constant.\footnotemark{}%
\footnotetext{%
	In~\cite{Leigh:2014:GerochGroupEinstein} Leigh et al.\ generalized Geroch's solution generating technique and also found that only the mass and the NUT charge appear when $\Lambda \neq 0$. We would like to thank D.\ Klemm for this remark.
}
As a consequence the JN algorithm cannot provide a derivation of the (a)dS--Kerr--Newman solution.

\subsubsection{Solution for \texorpdfstring{$\Lambda = 0$}{vanishing cosmological constant}}
\label{sec:derivation:stationary:solution-no-cosmo}

The solution to the differential equation \eqref{eq:topdown-1-Fd-Kd} is
\begin{equation}
	F(\theta) = n - a\, H'(\theta) + c \left( 1 + H'(\theta)\, \ln \frac{H(\theta/2)}{H'(\theta/2)} \right)
\end{equation}
where $a$ and $c$ denote two constants of integration.

We solve the equation \eqref{eq:topdown-1-F-Gd} for $G$
\begin{equation}
	\begin{aligned}
		G(\theta) = c_1 &+ \kappa\, a\, H'(\theta)
			- \kappa\, c\, H'(\theta)\, \ln \frac{H(\theta/2)}{H'(\theta/2)}
			- 2 \kappa\, n \ln H(\theta) \\
			&+ (a + c_2) \ln \frac{H(\theta/2)}{H'(\theta/2)}
	\end{aligned}
\end{equation} 
and $c_1, c_2$ are constants of integration.
Again since only $G'$ appears in the metric we can set $c_1 = 0$.
We can also remove the last term with the transformation
\begin{equation}
	\dd u = \dd u' - (c_2 + a) \dd\phi.
\end{equation} 
One finally gets
\begin{subequations}
\begin{align}
	F(\theta) &= n - a\, H'(\theta) + c \left( 1 + H'(\theta)\, \ln \frac{H(\theta/2)}{H'(\theta/2)} \right), \\
	G(\theta) &= \kappa\, a\, H'(\theta)
		- \kappa\, c\, H'(\theta)\, \ln \frac{H(\theta/2)}{H'(\theta/2)}
		- 2 \kappa\, n \ln H(\theta).
\end{align}
\end{subequations}

This solution was already found in~\cite{Krori:1981:ChargedDemianskiMetric} for the case $\kappa = 1$ by solving directly Einstein--Maxwell equations, starting with a metric ansatz of the form \eqref{deriv:eq:rotating:ur}.
Our aim was to show that the same solution can be obtained by applying Demiański's method to all the quantities, including the gauge field.

The BL transformation is well defined only for $c = 0$, in which case
\begin{equation}
	g = \frac{r^2 + a^2 + n^2}{\Delta}, \qquad
	h = \frac{\kappa a}{\Delta}, \qquad
	\Delta = \kappa r^2 - 2 m r + q^2 - \kappa n^2 + \kappa a^2.
\end{equation} 
The function $\tilde f$ reads
\begin{equation}
	\label{eq:topdown-1:tilde-f-no-Lambda-no-c}
	\tilde f = \kappa - \frac{2 m r - q^2}{\rho^2} + \frac{\kappa\, n (n - a H')}{\rho^2}, \qquad
	\rho^2 = r^2 + (n - a\, H')^2
\end{equation} 
and this corresponds to the Kerr--Newman--NUT solution~\cite[sec.~2.2]{AlonsoAlberca:2000:SupersymmetryTopologicalKerrNewmannTaubNUTaDS}.

\subsection{Relaxing assumptions}
\label{sec:derivation:relaxing}

In the derivation of \cref{sec:derivation:stationary:simplifying} we have made two assumptions in order to recover the simplest case.
The goal of this section is to show how these assumptions can be lifted, even if this does not lead to useful results: one cannot solve the equations in one case while in the other it is not clear how to recast the result in terms of a complex transformation.

\subsubsection{Metric function \texorpdfstring{$F$}{F}-dependence}
\label{sec:derivation:relaxing:metric-function}

In \cref{sec:derivation:stationary:simplifying} we obtained the equation \eqref{eq:topdown-1-Fd-Kd}
\begin{equation}
	\kappa\, F + K = \kappa\, n, \qquad
	2 K = F'' + \frac{H'}{H}\, F'
\end{equation}
by requiring that the function \eqref{eq:topdown-1-tilde-f}
\begin{equation}
	\tilde f = \kappa - \frac{2m r - q^2 + 2 F (\kappa\, F + K)}{r^2 + F^2} - \frac{\Lambda}{3}\, (r^2 + F^2) - \frac{4 \Lambda}{3}\, F^2 + \frac{8 \Lambda}{3}\, \frac{F^4}{r^2 + F^2}
\end{equation} 
depends on $\theta$ only through $F(\theta)$.
A more general assumption would be that $\kappa F + K$ is some function $\chi = \chi(F)$
\begin{equation}
	\label{eq:topdown-1-Fd-Kd-chi}
	\kappa\, F + K = \kappa\, \chi(F).
\end{equation} 
First if $F' = 0$ then $K = 0$ and the definition of $K$ implies
\begin{equation}
	\chi = F = n.
\end{equation} 
The $(t\theta)$- and $(\theta\phi)$-components give the equation
\begin{equation}
	4 \Lambda\, F^2 F' = F'\, \pd_F \chi.
\end{equation} 

If $\Lambda = 0$ we find that
\begin{equation}
	\pd_F \chi = 0
	\Longrightarrow
	\chi = n
\end{equation} 
which reduces to the case studied in \cref{sec:derivation:stationary:simplifying}, while if $F' = 0$ this equation does not provide anything.

On the other hand if $F' \neq 0$ and $\Lambda \neq 0$ then the previous equation becomes
\begin{equation}
	\pd_F \chi = 4 \Lambda F^2
\end{equation} 
which can be integrated to
\begin{equation}
	\label{top-down:eq:chi-F-solution}
	\chi(F) = n + \frac{4}{3}\, \Lambda F^3
\end{equation} 
(notice that the limit $\Lambda \to 0$ is coherent).
Plugging this function into equation \eqref{eq:topdown-1-Fd-Kd-chi} one obtains
\begin{equation}
	\label{eq:topdown-1-Fd-Kd-chi-replaced}
	\kappa\, F + K = \kappa \left(n + \frac{4}{3}\, \Lambda F^3 \right)
\end{equation} 
(remember that $F' \neq 0$).
This differential equation is non-linear and we were not able to find an analytical solution.
Despite that this provides a generalization of the algorithm with non-constant $F$ in the presence of a cosmological constant this is not sufficient for obtaining (a)dS--Kerr: the form of $g_{\theta\theta}$ given in \eqref{deriv:eq:rotating:tr} is not the required one.

Nonetheless by inserting the expression of $\chi$ in $\tilde f$ we see that the last term is killed
\begin{equation}
	\tilde f = \kappa - \frac{2m r - q^2 + 2 \kappa\, n\, F}{r^2 + F^2} - \frac{\Lambda}{3}\, (r^2 + F^2) - \frac{4 \Lambda}{3}\, F^2.
\end{equation} 
One can recognize the function given by Demiański~\cite{Demianski:1972:NewKerrlikeSpacetime} and may explain his error.

\subsubsection{Gauge field integration constant}
\label{sec:derivation:relaxing:gauge-fields}

In \cref{sec:derivation:stationary:simplifying} we obtained a second integration constant $\alpha$ in the expression of the gauge field
\begin{equation}
	\tilde f_A = \frac{q\, r}{r^2 + F^2} + \alpha\, \frac{r^2 - F^2}{r^2 + F^2}.
\end{equation}
One of the Maxwell equation gives $\alpha = 0$ if $F' \neq 0$, but otherwise no equation fixes its value.
For this reason we focus on the case $F' = 0$ or equivalently $\Lambda \neq 0$ through equation \eqref{eq:topdown-1-lambda}.

In this case the function $\tilde f$ is modified to
\begin{equation}
	\tilde f = \kappa - \frac{2m r - q^2 + 2 F (\kappa\, F + K) + 4 \alpha^2 F^2}{r^2 + F^2} - \frac{\Lambda}{3}\, (r^2 + F^2) - \frac{4 \Lambda}{3}\, F^2 + \frac{8 \Lambda}{3}\, \frac{F^4}{r^2 + F^2}.
\end{equation} 
Equation \eqref{eq:topdown-1-lambda} is modified but it is still solved by $F' = 0$ and all other equations are left unchanged (in particular $\kappa F + K$ is still given by the function $\chi(F)$ \eqref{top-down:eq:chi-F-solution}).
For $\chi(F) = n$ the configuration with $\alpha \neq 0$ provides another solution when $\Lambda \neq 0$ but it is not clear how to get it from a complexification of the function.

%% file: sections/examples.tex
\section{Examples}
\label{sec:examples}

In this section we list several examples that can be derived from the JN algorithm described in \cref{sec:general}.
Other examples were described previously: Kerr--Newman in \cref{sec:algorithm:kerr-newman}, dyonic Kerr--Newman and Yang--Mills Kerr--Newman in \cref{sec:extension:dyonic}.
For simplicity we will always consider the case $\kappa = 1$ except when $\Lambda \neq 0$.

The first two examples are the Kerr--Newmann--NUT solution (already derived by another path in \cref{sec:derivation:stationary:solution-no-cosmo}) and the charged (a)dS--BBMB--NUT solution in conformal gravity.
We will also give examples from ungauged $N = 2$ supergravity coupled to $n_v = 0, 1, 3$ vector multiplets (pure supergravity, T$^3$ model and STU model): this theory is reviewed in \cref{app:N=2-sugra}.

\subsection{Kerr--Newman--NUT}

The Reissner--Nordström metric and gauge fields are given by
\begin{subequations}
\label{matter:eq:reissner-nordstrom}
\begin{gather}
	\label{matter:metric:reissner-nordstrom:tr}
	\dd s^2 = - f\, \dd t^2 + f^{-1}\, \dd r^2 + r^2 \dd \Omega^2, \qquad
	f = 1 - \frac{2m}{r} + \frac{q^2}{r^2}, \\
	A = f_A\, \dd t, \qquad
	f_A = \frac{q}{r},
\end{gather} 
\end{subequations}
$m$ and $q$ being the mass and the electric charge.

The two functions are complexified as
\begin{equation}
	\tilde f = 1 - \frac{2 \Re(m \bar r) + q^2}{\abs{r}^2}, \qquad
	\tilde f_A = \frac{q \Re r}{\abs{r}^2}.
\end{equation} 
Performing the transformation
\begin{equation}
	u = u' + \big( a \cos\theta - 2 n \ln \sin\theta \big), \qquad
	r = r' + i \big( n - a \cos\theta \big), \qquad
	m = m' + i n
\end{equation} 
gives (omitting the primes)
\begin{equation}
	\tilde f = 1 - \frac{2 m r + 2 n ( n - a \cos\theta) - q^2}{\rho^2}, \qquad
	\rho^2 = r^2 + (n - a \cos\theta)^2.
\end{equation}
The metric and the gauge fields in BL coordinates are
\begin{subequations}
\begin{gather}
	\dd s^2 = - \tilde f\, (\dd t + \Omega\, \dd\phi )^2
		+ \frac{\rho^2}{\Delta}\, \dd r^2
		+ \rho^2 (\dd\theta^2 + \sigma^2 H^2 \dd\phi^2), \\
	A = \frac{q r}{\rho^2}\, \Big( \dd t - (a \sin^2 \theta + 2 n \cos \theta) \dd \phi \Big) + A_r\, \dd r
\end{gather}
\end{subequations}
where
\begin{equation}
	\begin{gathered}
		\Omega = - 2 n \cos \theta - (1 - \tilde f^{-1})\, a \sin^2 \theta, \qquad
		\sigma^2 = \frac{\Delta}{\tilde f \rho^2}, \\
		\Delta = r^2 - 2 m r + a^2 + q^2 - n^2.
	\end{gathered}
\end{equation} 
This corresponds to the Kerr--Newman--NUT solution~\cite{AlonsoAlberca:2000:SupersymmetryTopologicalKerrNewmannTaubNUTaDS}.

One can check that $A_r$ is a function of $r$ only
\begin{equation}
	A_r = - \frac{q r}{\Delta}
\end{equation} 
and it can be removed by a gauge transformation.

\subsection{Charged (a)dS--BBMB--NUT}
\label{sec:examples:bbmb}

The action of Einstein--Maxwell theory with cosmological constant conformally coupled to a scalar field is~\cite{Bardoux:2013:IntegrabilityConformallyCoupled}
\begin{equation}
	S = \frac{1}{2} \int \dd^4 x\, \sqrt{-g} \left( R - 2 \Lambda - \frac{1}{6}\, R \phi^2 - (\pd \phi)^2 - 2 \alpha \phi^4 - F^2 \right),
\end{equation} 
where $\alpha$ is a coupling constant, and we have set $8\pi G = 1$.

For $F, \alpha, \Lambda = 0$, the Bocharova--Bronnikov--Melnikov--Bekenstein (BBMB) solution~\cite{Bekenstein:1974:ExactSolutionsEinsteinconformal, Bocharova:1970:ExactSolutionSystem} is static and spherically symmetric -- it can be seen as the equivalent of the Schwarzschild black hole in conformal gravity.

The general static charged solution with cosmological constant and quartic coupling reads
\begin{subequations}
\begin{gather}
	\dd s^2 = - f\, \dd t^2 + f^{-1}\, \dd r^2 + r^2\, \dd\Omega^2, \\
	A = \frac{q}{r}\, \dd t, \qquad
	\phi = \sqrt{- \frac{\Lambda}{6 \alpha}}\; \frac{m}{r - m}, \\
	f = - \frac{\Lambda}{3}\, r^2 + \kappa\, \frac{(r - m)^2}{r^2},
\end{gather}
\end{subequations}
where the horizon can be spherical or hyperbolic.
There is one constraint among the parameters
\begin{equation}
	\label{matter:eq:bbmb:constraint}
	q^2 = \kappa m^2 \left( 1 + \frac{\Lambda}{36 \alpha} \right)
\end{equation} 
and one has $\alpha \Lambda < 0$ in order for $\phi$ to be real.

In order to add a NUT charge one performs the JN transformation\footnotemark{}%
\footnotetext{%
	Due to the convention of~\cite{Bardoux:2013:IntegrabilityConformallyCoupled} there is no $\kappa$ in the transformations.
}
\begin{equation}
	u = u' - 2 n \ln H(\theta), \qquad
	r = r' + i n, \qquad
	m = m' + i n, \qquad
	\kappa = \kappa' - \frac{4\Lambda}{3}\, n^2.
\end{equation} 
One obtains the metric (omitting the primes)
\begin{equation}
	\dd s^2 = - \tilde f \big(\dd t - 2 n H'\, \dd\phi \big)^2
		+ \tilde f^{-1}\, \dd r^2
		+ (r^2 + n^2)\, \dd\Omega^2
\end{equation}
where the function $\tilde f$ is
\begin{equation}
	\tilde f = - \frac{\Lambda}{3}\, (r^2 + n^2) + \left( \kappa - \frac{4\Lambda}{3}\, n^2 \right)\, \frac{(r - m)^2}{r^2 + n^2}.
\end{equation} 
Note that the term $(r - m)$ is invariant.
Similarly one obtains the scalar field
\begin{equation}
	\phi = \sqrt{- \frac{\Lambda}{6 \alpha}}\; \frac{\sqrt{m^2 + n^2}}{r - m}
\end{equation} 
where the $m$ in the numerator as been complexified as $\abs{m}$.
Finally it is trivial to find the gauge field
\begin{equation}
	A = \frac{q}{r^2 + n^2}\, \big( \dd t - 2 n \cos \theta\, \dd \phi \big)
\end{equation} 
and the constraint \eqref{matter:eq:bbmb:constraint} becomes
\begin{equation}
	q^2 = \left( \kappa - \frac{4\Lambda}{3}\, n^2 \right) (m^2 + n^2) \left( 1 + \frac{\Lambda}{36 \alpha} \right).
\end{equation} 

An interesting point is that the radial coordinate is redefined in~\cite{Bardoux:2013:IntegrabilityConformallyCoupled} when obtaining the stationary solution from the static one.

Note that the BBMB solution and its NUT version are obtained from the limit
\begin{equation}
	\Lambda, \alpha \longrightarrow 0, \qquad
	\text{with} \qquad
	- \frac{\Lambda}{36 \alpha} \longrightarrow 1,
\end{equation} 
which also implies $q = 0$ from the constraint \eqref{matter:eq:bbmb:constraint}.
Since no other modifications are needed, the derivation from the JN algorithm also holds in this case.

\subsection{Ungauged \texorpdfstring{$N = 2$}{N = 2} BPS solutions}
\label{sec:examples:N=2-bps}

A BPS solution is a classical solution which preserves a part of the supersymmetry.
The BPS equations are obtained by setting to zero the variations of the fermionic partners under a supersymmetric transformation.
These equations are first order and under some conditions their solutions also solve the equations of motion.

In~\cite[sec.~3.1]{Behrndt:1998:StationarySolutionsN2} (see also~\cite[sec.~2.2]{Hristov:2010:BPSBlackHoles} for a summary), Behrndt, Lüst and Sabra obtained the most general stationary BPS solution for $N = 2$ ungauged supergravity.
The metric for this class of solutions reads
\begin{equation}
	\label{matter:metric:sugra:static-N=2}
	\dd s^2 = f^{-1} (\dd t + \Omega\, \dd\phi)^2 + f\, \dd \Sigma^2,
\end{equation} 
with the $3$-dimensional spatial metric given in spherical or spheroidal coordinates
\begin{subequations}
\label{matter:metric:flat-spatial}
\begin{align}
	\dd \Sigma^2 &= h_{ij}\, \dd x^i \dd x^j
		= \dd r^2 + r^2 (\dd\theta^2 + \sin^2 \theta\, \dd\phi^2) \\
		&= \frac{\rho^2}{r^2 + a^2}\; \dd r^2 + \rho^2 \dd\theta^2 + (r^2 + a^2) \sin^2 \theta\; \dd \phi^2,
\end{align}
\end{subequations}
where $i, j, k$ are flat spatial indices (which should not be confused with the indices of the scalar fields).
The functions $f$ and $\Omega$ depend on $r$ and $\theta$ only.

Then the solution is entirely given in terms of two sets of (real) harmonic functions\footnotemark{} $\{ H^\Lambda, H_\Lambda \}$%
\footnotetext{%
	We omit the tilde that is present in~\cite{Behrndt:1998:StationarySolutionsN2} to avoid the confusion with the quantities that are transformed by the JNA.
	No confusion is possible since the index position will always indicate which function we are using.
}
\begin{subequations}
\label{matter:eq:N=2-bps-equations}
\begin{gather}
	f = \e^{-K} = i (\bar X^\Lambda F_\Lambda - X^\Lambda \bar F_\Lambda), \\
	\levi{_{ijk}} \pd_j \Omega_k = 2 \e^{-K} \mc A_i = (H_\Lambda \pd_i H^\Lambda - H^\Lambda \pd_i H_\Lambda), \\
	F^\Lambda_{ij} = \frac{1}{2}\, \levi{_{ijk}} \pd_k H^\Lambda, \qquad
	G_{\Lambda\,ij} = \frac{1}{2}\, \levi{_{ijk}} \pd_k H_\Lambda, \\
	i (X^\Lambda - \bar X^\Lambda) = H^\Lambda, \qquad
	i (F_\Lambda - \bar F_\Lambda) = H_\Lambda.
\end{gather}
\end{subequations}
The object $\Omega_i$ is the connection of the line bundle corresponding to the fibration of time over the spatial manifold (its curl is related to the Kähler connection).
Its only non-vanishing component is $\Omega_\phi \equiv \Omega = \omega H$.

Starting from the metric \eqref{matter:metric:sugra:static-N=2} in spherical coordinates with $\Omega = 0$, one can use the JN algorithm of \cref{sec:general} with
\begin{equation}
	f_t = f^{-1}, \qquad
	f_r = f, \qquad
	f_\Omega = r^2 f,
\end{equation} 
leading to the formula \eqref{gen:eq:rotating:tr-degenerate-isotropic}.
The function $\Omega$ reads
\begin{equation}
	\Omega = \omega H
		= a (1 - \tilde f) \sin^2 \theta + 2 n \cos\theta.
\end{equation} 

Then one needs only to find the complexification of $f$ and to check that it gives the correct $\omega$, as would be found from the equations \eqref{matter:eq:N=2-bps-equations}.
However it appears that one cannot complexify directly $f$ since it should be viewed as a composite object made of complex functions.
Therefore one needs to complexify first the harmonic functions $H_\Lambda$ and $H^\Lambda$ (or equivalently $X^\Lambda$), and then to reconstruct the other quantities.
Nonetheless, equations \eqref{matter:eq:N=2-bps-equations} ensure that finding the correct harmonic functions gives a solution, thus it is not necessary to check these equations for all the other quantities.

In the next subsections we provide two examples,\footnotemark{} one for pure supergravity as an appetizer, and then one with $n_v = 3$ multiplets (STU model).%
\footnotetext{%
	They correspond to singular solutions, but we are not concerned with regularity here.
}

\subsubsection{Pure supergravity}
\label{sec:examples:N=2-bps:pure}

As a first example we consider pure (or minimal) supergravity, i.e. $n_v = 0$~\cite[sec.~4.2]{Behrndt:1998:StationarySolutionsN2}.
The prepotential reads
\begin{equation}
	F = - \frac{i}{4}\, (X^0)^2.
\end{equation} 
The function $H_0$ and $H^0$ are related to the real and imaginary parts of the scalar $X^0$
\begin{equation}
	H_0 = \frac{1}{2} (X^0 + \bar X^0) = \Re X^0, \qquad
	\bar H^0 = i (X^0 - \bar X^0) = - 2 \Im X^0,
\end{equation} 
while the Kähler potential is given by
\begin{equation}
	f = \e^{-K} = X^0 \bar X^0.
\end{equation} 
The static solution corresponds to
\begin{equation}
	\label{matter:eq:pure-sugra-static-X0}
	H_0 = X^0 = 1 + \frac{m}{r}
\end{equation} 

Performing the JN transformation for the angular momentum gives
\begin{equation}
	\tilde X^0 = 1 + \frac{m (r + i a \cos\theta)}{\rho^2}.
\end{equation}
This corresponds to the second solution of which is stationary with
\begin{equation}
	\omega = \frac{m (2r + m)}{\rho^2}\, a \sin^2 \theta.
\end{equation} 

Alternatively one can use the JN algorithm to add a NUT charge.
In this case using the rule 
\begin{equation}
	r \longrightarrow \frac{1}{2}\, (r + \bar r) = \Re r = r'
\end{equation} 
must be use for transforming $f$ and $r^2$ (in front of $\dd\Omega$), leading to
\begin{equation}
	\label{matter:eq:pure-sugra-static-X0-nut}
	X^0 = 1 + \frac{m + i n}{r}.
\end{equation} 
Note that it gives
\begin{equation}
	\tilde f = \left(1 + \frac{m}{r}\right)^2 + \frac{n^2}{r^2}.
\end{equation} 
It is slightly puzzling that the above rule should be used instead of the two others in \eqref{gen:eq:rules}.
One possible explanation is the following: in the seed solution shift the radial coordinate such that $r = R - m$ and apply the JN transformation in this coordinate system.
It is clear that every function of $r$ is left unchanged while the tensor structure transforms identically since $\dd r = \dd R$.
After the transformation one can undo the coordinate transformation.
As we mentioned earlier the algorithm is very sensible to the coordinate system and to the parametrization (but it is still not clear why the $R$-coordinate is the natural one).
This kind of difficulty will reappear in the SWIP solution (\cref{sec:examples:swip}).

\subsubsection{STU model}

We now consider the STU model $n_v = 3$ with prepotential~\cite[sec.~3]{Behrndt:1998:StationarySolutionsN2}
\begin{equation}
	F = - \frac{X^1 X^2 X^3}{X^0}.
\end{equation} 
The expressions for the Kähler potential and the scalar fields in terms of the harmonic functions are complicated and will not be needed (see ~\cite[sec.~3]{Behrndt:1998:StationarySolutionsN2} for the expressions).
Various choices for the functions will give different solutions.

A class of static black hole-like solutions are given by the harmonic functions~\cite[sec.~4.4]{Behrndt:1998:StationarySolutionsN2}
\begin{equation}
	\label{matter:eq:stu-static-functions}
	H_0 = h_0 + \frac{q_0}{r}, \qquad
	H^i = h^i + \frac{p^i}{r}, \qquad
	H^0 = H_i = 0.
\end{equation} 
These solutions carry three magnetic $p^i$ and one electric $q_0$ charges.

Let's form the complex harmonic functions
\begin{equation}
	\mc H_0 = H_0 + i\, H^0, \qquad
	\mc H_i = H^i + i\, H_i.
\end{equation} 
Then the rule for complex function leads to
\begin{equation}
	\mc H_0 = h_0 + \frac{q_0 (r + i a \cos\theta)}{\rho^2}, \qquad
	\mc H_i = h^i + \frac{p^i (r + i a \cos\theta)}{\rho^2},
\end{equation} 
for which the various harmonic functions read explicitly
\begin{equation}
	H_0 = h_0 + \frac{q_0 r}{\rho^2}, \qquad
	H^i = h^i + \frac{p^i r}{\rho^2}, \qquad
	H^0 = \frac{q_0 a \cos\theta}{\rho^2}, \qquad
	H_i = \frac{p^i a \cos\theta}{\rho^2}.
\end{equation}
This set of functions corresponds to the stationary solution of~\cite[sec.~4.4]{Behrndt:1998:StationarySolutionsN2} where the magnetic and electric dipole momenta are not independent parameters but obtained from the magnetic and electric charges instead.

\subsection{Non-extremal rotating solution in \texorpdfstring{$T^3$}{T3} model}
\label{sec:examples:rotating-T3}

The $T^3$ model under consideration corresponds to Einstein--Maxwell gravity coupled to an axion $\sigma$ and a dilaton $\phi$ (with specific coupling constants) and the action is given by \eqref{ex:eq:action:swip} with $M = 1$.
This model can be embedded in $N = 2$ ungauged supergravity with $n_v = 1$, equal gauge fields $A \equiv A^0 = A^1$ and prepotential\footnotemark{}%
\footnotetext{%
	This model can be obtained from the STU model by setting the sections pairwise equal $X^2 = X^0$ and $X^3 = X^1$~\cite{Chow:2014:BlackHolesN8}.
	It is also a truncation of pure $N = 4$ supergravity.
}
\begin{equation}
	F = - i\, X^0 X^1,
\end{equation} 
The dilaton and the axion corresponds to the complex scalar field
\begin{equation}
	\tau = \e^{-2\phi} + i\, \sigma.
\end{equation} 
Sen derived the rotating black hole for this theory using the fact that it can be embedded in heterotic string theory~\cite{Sen:1992:RotatingChargedBlack}.

The static metric, gauge field and the complex field read respectively
\begin{subequations}
\begin{align}
	\dd s^2 &= - \frac{f_1}{f_2}\, \dd t^2 + f_2 \Big(f_1^{-1}\, \dd r^2 + r^2\, \dd\Omega^2 \Big), \\
	A &= \frac{f_A}{f_2}\, \dd t, \\
	\tau &= \e^{-2\phi} = f_2
\end{align}
\end{subequations}
where
\begin{equation}
	f_1 = 1 - \frac{r_1}{r}, \qquad
	f_2 = 1 + \frac{r_2}{r}, \qquad
	f_A = \frac{q}{r}.
\end{equation} 
The radii $r_1$ and $r_2$ are related to the mass $m$ and the charge $q$ by
\begin{equation}
	r_1 + r_2 = 2 m, \qquad
	r_2 = \frac{q^2}{m}.
\end{equation} 

Applying the Janis--Newman algorithm with rotation, the two functions $f_1$ and $f_2$ are complexified as
\begin{equation}
	\tilde f_1 = 1 - \frac{r_1 r}{\rho^2}, \qquad
	\tilde f_2 = 1 + \frac{r_2 r}{\rho^2}.
\end{equation} 
The final metric in BL coordinates is given by
\begin{equation}
	\dd s^2 = - \frac{\tilde f_1}{\tilde f_2} \left[ \dd t - a \left(1 - \frac{\tilde f_2}{\tilde f_1}\right) \sin^2 \theta\, \dd\phi \right]^2
		+ \tilde f_2 \left( \frac{\rho^2 \dd r^2}{\Delta} + \rho^2 \dd\theta^2 + \frac{\Delta}{\tilde f_1}\, \sin^2 \theta\, \dd\phi^2 \right)
\end{equation}
for which the BL functions are
\begin{equation}
	g(r) = \frac{\hat \Delta}{\Delta}, \qquad
	h(r) = \frac{a}{\Delta}
\end{equation} 
with
\begin{equation}
	\label{matter:eq:sen-bh-delta}
	\Delta = \tilde f_1 \rho^2 + a^2 \sin^2 \theta, \qquad
	\hat \Delta = \tilde f_2 \rho^2 + a^2 \sin^2 \theta.
\end{equation} 

Once $f_A$ has been complexified as
\begin{equation}
	\tilde f_A = \frac{q r}{\rho^2}
\end{equation} 
the transformation of the gauge field is straightforward
\begin{equation}
	A = \frac{\tilde f_A}{\tilde f_2}\, (\dd t - a \sin^2 \theta\, \dd\phi )
		- \frac{q r}{\Delta}\, \dd r.
\end{equation} 
The $A_r$ depending solely on $r$ can again be removed thanks to a gauge transformation.

Finally the scalar field is complex and is transformed as
\begin{equation}
	\tau = 1 + \frac{r_2 \bar r}{\rho^2}.
\end{equation} 
The explicit values for the dilaton and axion are then
\begin{equation}
	\e^{-2\phi} = \tilde f_2, \qquad
	\sigma = \frac{r_2 a \cos \theta}{\rho^2}.
\end{equation} 

This reproduces Sen's solution and it completes the computation from~\cite{Yazadjiev:2000:NewmanJanisMethodRotating} which could not derive the gauge field nor the axion.
It is interesting to note that for another value of the dilaton coupling we cannot use the transformation~\cite{Horne:1992:RotatingDilatonBlack, Pirogov:2013:RotatingScalarvacuumBlack}.\footnotemark{}%
\footnotetext{%
	The authors of~\cite{Hansen:2013:ApplicabilityNewmanJanisAlgorithm} report incorrectly that~\cite{Pirogov:2013:RotatingScalarvacuumBlack} is excluding all dilatonic solutions.
}

\subsection{SWIP solutions}
\label{sec:examples:swip}

Let's consider the action~\cites{Bergshoeff:1996:StationaryAxionDilatonSolutions}[sec.~12.2]{Ortin:2004:GravityStrings}
\begin{equation}
	\label{ex:eq:action:swip}
	S = \frac{1}{16 \pi} \int \dd^4 x\, \sqrt{\abs{g}}\, \left( R
		- 2 (\pd \phi)^2 - \frac{1}{2}\, \e^{4\phi}\, (\pd \sigma)^2
		- \e^{-2\phi} F^i_{\mu\nu} F^{i\mu\nu} + \sigma\, F^i_{\mu\nu} \tilde{F}^{i\mu\nu} \right)
\end{equation} 
where $i = 1, \ldots, M$.
When $M = 2$ and $M = 6$ this action corresponds respectively to $N = 2$ supergravity with one vector multiplet and to $N = 4$ pure supergravity, but we keep $M$ arbitrary.
The axion $\sigma$ and the dilaton $\phi$ are naturally paired into a complex scalar
\begin{equation}
	\tau = \sigma + i \e^{-2\phi}.
\end{equation} 

In order to avoid redundancy we first provide the general metric with $a, n \neq 0$, and we explain how to find it from the restricted case $a = n = 0$.
The stationary Israel--Wilson--Perjés (SWIP) solutions correspond to
\begin{subequations}
\label{matter:eq:swip:solution}
\begin{gather}
	\dd s^2 = - \e^{2U} W (\dd t + A_\phi\, \dd \phi)^2 + \e^{-2U} W^{-1} \dd \Sigma^2, \\
	A^i_t = 2 \e^{2U} \Re(k^i H_2), \qquad
	\tilde A^i_t = 2 \e^{2U} \Re(k^i H_1), \qquad
	\tau = \frac{H_1}{H_2}, \\
	A_\phi = 2 n \cos \theta - a \sin^2 \theta (\e^{-2U} W^{-1} - 1), \\
	\e^{-2U} = 2 \Im(H_1 \bar H_2), \qquad
	W = 1 - \frac{r_0^2}{\rho^2}.
\end{gather}
\end{subequations}
This solution is entirely determined by the two harmonic functions
\begin{equation}
	\label{matter:eq:swip:harmonic-functions}
	H_1 = \frac{1}{\sqrt{2}}\, \e^{\phi_0} \left( \tau_0 + \frac{\tau_0 \mc M + \bar \tau_0 \Upsilon}{r - i a \cos \theta} \right), \qquad
	H_2 = \frac{1}{\sqrt{2}}\, \e^{\phi_0} \left( 1 + \frac{\mc M + \Upsilon}{r - i a \cos \theta} \right).
\end{equation} 
The spatial $3$-dimensional metric $\dd \Sigma^2$ reads
\begin{equation}
	\label{matter:metric:swip:flat-spatial}
	\dd\Sigma^2 = h_{ij}\, \dd x^i \dd x^j
		= \frac{\rho^2 - r_0^2}{r^2 + a^2 - r_0^2}\; \dd r^2 + (\rho^2 - r_0^2) \dd\theta^2 + (r^2 + a^2 - r_0^2) \sin^2 \theta\; \dd \phi^2.
\end{equation} 

Finally, $r_0$ corresponds to
\begin{equation}
	r_0^2 = \abs{\mc M}^2 + \abs{\Upsilon}^2 - \sum_i \abs{\Gamma^i}^2
\end{equation} 
where the complex parameters are
\begin{equation}
	\mc M = m + i n, \qquad
	\Gamma^i = q^i + i p^i,
\end{equation} 
$m$ being the mass, $n$ the NUT charge, $q^i$ the electric charges and $p^i$ the magnetic charges, while the axion--dilaton charge $\Upsilon$ takes the form
\begin{equation}
	\Upsilon = - \frac{1}{2} \sum_i \frac{(\bar \Gamma^i)^2}{\mc M}.
\end{equation} 
The latter together with the asymptotic values $\tau_0$ are defined by
\begin{equation}
	\tau \sim \tau_0 - i \e^{-2\phi_0} \frac{2 \Upsilon}{r}.
\end{equation} 
The complex constant $k^i$ are determined by
\begin{equation}
	k^i = - \frac{1}{\sqrt{2}}\, \frac{\mc M \Gamma^i + \bar \Upsilon \bar \Gamma^i}{\abs{\mc M}^2 - \abs{\Upsilon}^2}.
\end{equation} 

As discussed in the previous section, the transformation of scalar fields is different depending on one is turning on a NUT charge or an angular momentum.
For this reason, starting from the case $a = n = 0$, one needs to perform the two successive transformations
\begin{subequations}
\begin{gather}
	\label{matter:eq:swip:djn-nut}
	u = u' - 2 i n \ln \sin \theta, \qquad
	r = r' + i n, \qquad
	m = m' + i n, \\
	\label{matter:eq:swip:djn-rot}
	u = u' + i a \cos \theta, \qquad
	r = r' - i a \cos \theta,
\end{gather}
\end{subequations}
the order being irrelevant (for definiteness we choose to add the NUT charge first), the reason being that the transformations of the functions are different in both cases (as in \cref{sec:examples:N=2-bps:pure}).
As explained in \cref{app:group-properties}, group properties of the JN algorithm ensure that the metric will be transformed as if only one transformation was performed.
Then the metric and the gauge fields are directly obtained, which ensures that the general form of the solution \eqref{matter:eq:swip:solution} is correct.
For that one needs to shift $r^2$ by $r_0^2$ in order to bring the metric \eqref{matter:metric:swip:flat-spatial} to the form \eqref{matter:metric:flat-spatial}.
This modifies the function but one does not need this fact to obtain the general form. Then one can shift by $- r_0^2$ before dealing with the complexification of the functions.
See~\cite[p.~17]{Bergshoeff:1996:StationaryAxionDilatonSolutions} and \cref{sec:examples:N=2-bps:pure} for discussions about the changes of coordinates.
Since all the functions and the parameters depend only on $\mc M$, $H_1$ and $H_2$, it is sufficient to explain their complexification.

The function $W$ is transformed as a real function.
On the other hand $H_1$ and $H_2$ are complex harmonic functions and should be transformed accordingly.
For the NUT charge one should use the rule
\begin{equation}
	r \longrightarrow \Re r.
\end{equation} 
Then one can perform the second transformation \eqref{matter:eq:swip:djn-rot} in order to add the angular momentum by applying the usual rules \eqref{gen:eq:rules}.
On can see that it yields the correct result.

Finally let's note that it seems possible to also start from $p^i = 0$ and to turn them on using the transformation
\begin{equation}
	q^i = q'^i = q^i + i p^i,
\end{equation} 
using different rules for complexifying the various terms (depending whether one is dealing with a real or a complex function/parameter).

\subsection{Gauged \texorpdfstring{$N = 2$}{N = 2} non-extremal solution}
\label{sec:examples:gauged-N=2}

The simplest deformation of $N = 2$ supergravity with $n_v$ vector multiplets consists in the so-called Fayet--Iliopoulos (FI) gauging.
It amounts to gauging $(n_v + 1)$ times the diagonal $\group{U}(1)$ group of the $\group{SU}(2)$ part of the R-symmetry group (automorphism of the supersymmetry algebra).
The potential can be entirely written in terms of the quantities defined in \cref{app:N=2-sugra} and of the $(n_v + 1)$ coupling constants $g_I$, where $I = 0, \ldots, n_v$.

We consider the model with prepotential (see also \cref{sec:examples:rotating-T3})
\begin{equation}
	F = - i\, X^0 X^1.
\end{equation} 
for which the potential generated by the FI gauging is
\begin{equation}
	V(\tau, \bar \tau) = - \frac{4}{\tau + \bar \tau} \big(g_0^2 + g_0 g_1 (\tau + \bar\tau) + g_1^2 \abs{\tau}^2 \big).
\end{equation} 
The goal of this section is to derive the NUT charged black hole from~\cite{Gnecchi:2014:RotatingBlackHoles} using the JN algorithm.\footnotemark{}%
\footnotetext{%
	The original derivation is due to D.\ Klemm and M.\ Rabbiosi and has not been published.
	I am grateful to them for allowing me to reproduce it here.
}

The seed solution is taken to be eq.\ (4.22) from~\cite{Gnecchi:2014:RotatingBlackHoles} with $j = N = 0$
\begin{subequations}
\begin{gather}
	f_t = \kappa
		- \frac{2 m r - 2 \ell^2 \sum_I g_I \abs{Z^I}^2}{f_\Omega}
		+ \frac{f_\Omega}{\ell^2}, \\
	f_\Omega = r^2 - \Delta^2 - \delta^2, \\
	f^I = \frac{(r - \Delta) Q^I - \delta\, P^I}{f_\Omega}, \\
	\tau = \frac{g_0}{g_1}\, \frac{r + \Delta - i \delta}{r - \Delta + i \delta}.
\end{gather}
\end{subequations}
where the following quantities have been defined
\begin{subequations}
\begin{gather}
	m = \frac{\ell^2 P^0}{\Delta}\; \frac{g_1^2 \big[- (P^1)^2 P^0 + (Q^1)^2 P^0 - 2 Q^0 Q^1 P^1 \big] + g_0^2 P^0 \abs{Z^0}^2}{\abs{Z^0}^2}, \\
	\delta = - \Delta\, \frac{Q^0}{P^0}.
\end{gather}
\end{subequations}
The independent parameters are given by $Q^I$ (electric charges), $P^I$ (magnetic charges), $g_\Lambda$ (FI gaugings), $\Delta$ (scalar charge) and $\Lambda = - 3 / \ell^2$ (the cosmological constant).

In order to perform the complexification the functions are first rewritten as
\begin{subequations}
\begin{gather}
	f_t = \kappa
		- \frac{2 \Re(m \bar r) - 2 \ell^2 \sum_I g_I \abs{Z^I}^2}{f_\Omega}
		+ \frac{f_\Omega}{\ell^2}, \\
	f_\Omega = \abs{r}^2 - \Delta^2 - \delta^2
		= \abs{r}^2 - \frac{\Delta^2 \abs{Z^1}^2}{\Im(Z^1)^2}, \\
	f^I
		= \frac{\Re(Q^I \bar r) \Im Z^1 - \Delta \Im(Z^I Z^1)}{\Im Z^1\, f_\Omega}, \\
	%\tau = \frac{g_0}{g_1}\, \frac{P^1\, r + \Delta (P^1 + i Q^1)}{P^1\, r - \Delta (P^1 + i Q^1)}
	%	= \frac{g_0}{g_1}\, \frac{\Im(Z^1 \bar r) + i \Delta \bar Z^1}{\Im(Z^1 \bar r) - i \Delta \bar Z^1}.
	\tau = \frac{g_0}{g_1}\, \frac{\bar r + \Delta - i \delta}{\bar r - \Delta + i \delta}.
\end{gather}
\end{subequations}
Applying the transformations \eqref{gen:eq:change:jna} with \eqref{gen:eq:change:jna-functions-FG} gives (omitting the primes)
\begin{subequations}
\begin{gather}
	\tilde f_t = \kappa + \frac{4 n^2}{\ell^2}
		- \frac{2 m r + 2 \left(\kappa + 4 n^2 / \ell^2 \right) n^2 - 2 \ell^2 \sum_I g_I \abs{Z^I}^2}{\tilde f_\Omega}
		+ \frac{\tilde f_\Omega}{\ell^2}, \\
	\tilde f_\Omega = r^2 + n^2 - \Delta^2 - \delta^2, \\
	\tilde f^I = \frac{(Q^I r + P^I n) \Im Z^1 - \Delta \Im(Z^I Z^1)}{\Im Z^1\, \tilde f_\Omega}, \\
	\tilde \tau = \frac{g_0}{g_1}\, \frac{r + \Delta - i (\delta + n)}{r - \Delta + i (\delta - n)}.
\end{gather}
\end{subequations}
The last step is to simplify these expressions
\begin{subequations}
\begin{gather}
	\tilde f_t = \kappa + \frac{4 n^2}{\ell^2}
		- \frac{2 m r + 2 \kappa n^2 + 8 n^4 / \ell^2 - 2 \ell^2 \sum_I g_I \abs{Z^I}^2}{\tilde f_\Omega}
		+ \frac{\tilde f_\Omega}{\ell^2}, \\
	\tilde f_\Omega = r^2 + n^2 - \Delta^2 - \delta^2, \\
	\tilde f^I = \frac{Q^I (r - \Delta) + P^I (n - \delta)}{\tilde f_\Omega}, \\
	\tilde \tau = \frac{g_0}{g_1}\, \frac{r + \Delta - i (\delta + n)}{r - \Delta + i (\delta - n)}.
\end{gather}
\end{subequations}
It is straightforward to check that the form of the metric and gauge fields are correctly reproduced by the algorithm given in \cref{sec:general} for the tensor structure.
In total this reproduces the eq.\ (4.22) and formulas below in~\cite{Gnecchi:2014:RotatingBlackHoles} with $j = 0$.

An important thing that we learn here is that the mass parameter needs to be transformed as if it was not composed of other parameters.

%% file: sections/five_dimensions.tex
\section{Five dimensional algorithm}
\label{sec:five}

While in four dimensions we have at our disposal many theorems on the classification of solutions, this is not the case for higher dimensions and the bestiary for solutions is much wider and less understood~\cite{Emparan:2008:BlackHolesHigher, Adamo:2014:KerrNewmanMetricReview}.
Rotating solutions in higher dimensions are characterized by several angular momenta.
Important solutions have not yet been discovered, even in the simplest theories such as the charged rotating black holes with several angular momenta in pure Einstein--Maxwell gravity.

Generalizing the JN algorithm in other dimensions is challenging and only small steps have been taken in this direction.
For instance Xu recovered Myers--Perry solution with one angular momentum~\cite{Myers:1986:BlackHolesHigher} from the Schwarzschild--Tangherlini solution~\cite{Xu:1988:ExactSolutionsEinstein} (see also~\cite{Aliev:2006:RotatingBlackHoles}), and Kim showed how the rotating BTZ black hole~\cite{Banados:1992:BlackHoleThree} can be obtained from its static limit~\cite{Kim:1997:NotesSpinningAdS3, Kim:1999:SpinningBTZBlack}.
One of the difficulty is to be able to perform several successive transformations in order to introduce all the allowed angular momenta.

In this section we report the successful generalization of the JN algorithm to five dimensions where we recover two examples~\cite{Erbin:2015:FivedimensionalJanisNewmanAlgorithm}: the complete Myers--Perry black hole~\cite{Myers:1986:BlackHolesHigher} and the Breckenridge--Myers--Peet--Vafa (BMPV) extremal black hole~\cite{Breckenridge:1997:DbranesSpinningBlack}.
We give of proposal for extending this method to higher dimensions in the next section.

It appears that the two angular momenta can be added one after the other by performing two successive transformations, each using different rules for complexifying the functions.
These rules can be understood as transforming only the functions that appear in the part of the metric which describes the rotation plane associated to the angular momentum.
Our method makes use of the Giampieri prescription and we did not succeed in expressing it in terms of the Janis--Newman prescription.

A major application of our work would be to find the charged solution with two angular momenta of the $5d$ Einstein--Maxwell gravity.
This problem is highly non-trivial and there is few chances that this technique would work directly~\cite{Aliev:2006:RotatingBlackHoles}, but one can imagine that a generalization of Demiański's approach~\cite{Demianski:1972:NewKerrlikeSpacetime} (see \cref{sec:derivation}) could lead to new interesting solutions in five dimensions.
An intermediate step is represented by the CCLP metric~\cite{Chong:2005:GeneralNonExtremalRotating} which is a solution of the Einstein--Maxwell theory with a Chern--Simons term, but it cannot be derived from the JN algorithm and we give some intuition about this fact in the last subsection.

Finally one could seek for an extension of the algorithm to the derivation of black rings~\cite{Emparan:2002:RotatingBlackRing, Emparan:2008:BlackHolesHigher}.
Similarly it may be possible that such techniques could be used in $d = 4$ to derive multicentre solutions (for instance one could imagine adding rotation to both centres successively, changing coordinate system in-between to place the origin of the coordinates at each centre).

\subsection{Myers--Perry black hole}
\label{sec:higher-jna:5d:myers-perry}

In this section we show how to recover the Myers--Perry black hole in five dimensions through the Giampieri prescription.
This is a solution of $5$-dimensional pure Einstein theory which possesses two angular momenta and it generalizes the Kerr black hole.
The importance of this solution lies in the fact that it can be constructed in any dimension.

The seed metric is given by the five-dimensional Schwarzschild--Tangherlini metric
\begin{equation}
	\dd s^2 = - f(r)\, \dd t^2 + f(r)^{-1}\, \dd r^2 + r^2\, \dd \Omega_3^2
\end{equation}
where $\dd \Omega_3^2$ is the metric on $S^3$, which can be expressed in Hopf coordinates (see \cref{app:coord:5d:hopf})
\begin{equation}
	\label{higher-jna:eq:coord-S3-spherical}
	\dd \Omega_3^2 = \dd\theta^2 + \sin^2 \theta\, \dd\phi^2 + \cos^2 \theta\, \dd\psi^2,
\end{equation} 
and the function $f(r)$ is given by
\begin{equation}
	f(r) = 1 - \frac{m}{r^2}.
\end{equation}

An important feature of the JN algorithm is the fact that a given set of transformations in the $(r,\phi)$-plane generates rotation in the latter.
Generating several angular momenta in different 2-planes would then require successive applications of the JN algorithm on different hypersurfaces.
In order to do so, one has to identify what are the 2-planes which will be submitted to the algorithm.
In five dimensions, the two different planes that can be made rotating are the planes $(r,\phi)$ and $(r,\psi)$.
We claim that it is necessary to dissociate the radii of these 2-planes in order to apply separately the JN algorithm on each plane and hence to generate two distinct angular momenta.
In order to dissociate the parts of the metric that correspond to the rotating and non-rotating $2$-planes, one can protect the function $r^2$ to be transformed under complex transformations in the part of the metric defining the plane which will stay static.
We thus introduce the function
\begin{equation}
	R(r) = r
\end{equation} 
such that the metric in null coordinates reads
\begin{equation}
	\label{higher-jna:5d-jna:metric:static:general-ur}
	\dd s^2 = - \dd u\, (\dd u + 2 \dd r)
		+ (1 - f)\, \dd u^2
		+ r^2 (\dd\theta^2 + \sin^2 \theta\, \dd\phi^2) + R^2 \cos^2 \theta\, \dd\psi^2.
\end{equation} 
The first transformation -- hence concerning the $(r,\phi)$-plane -- is
\begin{equation}
	\label{higher-jna:eq:5d-ansatz-hopf-1}
	\begin{gathered}
		u = u' + i a \cos \chi_1, \qquad
		r = r' - i a \cos \chi_1, \\
		i\, \dd \chi_1 = \sin \chi_1\ \dd\phi, \qquad\text{~~with~~}\chi_1 = \theta, \\
		\dd u = \dd u' - a \sin^2 \theta\, \dd\phi, \qquad
		\dd r = \dd r' + a \sin^2 \theta\, \dd\phi,
	\end{gathered}
\end{equation}
and $f$ is replaced by $\tilde f^{\{1\}} = \tilde f^{\{1\}}(r, \theta)$.
Indeed one needs to keep track of the order of the transformation, since the function $f$ will be complexified twice consecutively.
On the other hand $R(r) = \Re(r)$ is transformed\footnotemark{} into $R' = r'$ and one finds (omitting the primes)%
\footnotetext{%
	Note that as a function this corresponds to the rule \eqref{gen:eq:rules:r} but we will see below that $R$ is better interpreted as a coordinate since below it will appear as $\dd R$.
}
\begin{equation}
	\begin{aligned}
	\dd s^2 = - \dd u^2 &- 2\, \dd u \dd r
		+ \big(1 - \tilde f^{\{1\}} \big) (\dd u - a \sin^2 \theta\, \dd \phi)^2
		+ 2 a \sin^2 \theta\, \dd r \dd \phi \\
		&+ (r^2 + a^2 \cos^2 \theta) \dd\theta^2
		+ (r^2 + a^2) \sin^2 \theta\, \dd\phi^2
		+ r^2 \cos^2 \theta\, \dd \psi^2.
	\end{aligned}
\end{equation} 
The function $\tilde f^{\{1\}}$ is
\begin{equation}
	\tilde f^{\{1\}} = 1 - \frac{m}{\abs{r}^2} = 1 - \frac{m}{r^2 + a^2 \cos^2 \theta}.
\end{equation} 
There is a cancellation between the $(u, r)$ and the $(\theta, \phi)$ parts of the metric
\begin{subequations}
\begin{align}
	\dd s_{u,r}^2 &= (1 - \tilde f^{\{1\}})\, (\dd u - a \sin^2 \theta\, \dd \phi)^2
		- \dd u (\dd u + 2 \dd r )
		+ 2 a \sin^2 \theta \, \dd r \dd \phi
		+ a^2 \sin^4 \theta\, \dd \phi^2, \\
	\dd s_{\theta,\phi}^2 &= (r^2 + a^2 \cos^2 \theta) \dd\theta^2
			+ \big(r^2 + a^2 (1 - \sin^2 \theta) \big) \sin^2 \theta\, \dd\phi^2.
\end{align}
\end{subequations}

In addition to the terms present in \eqref{higher-jna:5d-jna:metric:static:general-ur} one obtains new components corresponding to the rotation of the first plane $(r, \phi)$.
Since the structure is very similar one can perform a transformation\footnotemark{} in the second plane $(r, \psi)$%
\footnotetext{%
	The easiest justification for choosing the sinus here is by looking at the transformation in terms of direction cosines, see \cref{sec:higher-jna:examples:myers-perry-5d}.
	Otherwise this term can be guessed by looking at Myers--Perry non-diagonal terms.
}
\begin{equation}
	\label{higher-jna:eq:5d-ansatz-hopf-2}
	\begin{gathered}
		u = u' + i b\, \sin \chi_2, \qquad
		r = r' - i b\, \sin \chi_2, \\
		i\, \dd \chi_2 = - \cos \chi_2\, \dd\psi, \qquad \text{~~with~~}\chi_2 = \theta, \\
		\dd u = \dd u' - b \cos^2 \theta\, \dd\psi, \qquad
		\dd r = \dd r' + b \cos^2 \theta\, \dd\psi,
	\end{gathered}
\end{equation}
can be applied directly to the metric
\begin{equation}
	\begin{aligned}
	\dd s^2 = - \dd u^2 &- 2\, \dd u \dd r
		+ \big(1 - \tilde f^{\{1\}} \big) (\dd u - a \sin^2 \theta\, \dd \phi)^2
		+ 2 a \sin^2 \theta\, \dd R \dd \phi \\
		&+ \rho^2 \dd\theta^2
		+ (R^2 + a^2) \sin^2 \theta\, \dd\phi^2
		+ r^2 \cos^2 \theta\, \dd \psi^2
	\end{aligned}
\end{equation} 
where we introduced once again the function $R(r) = \Re(r)$ to protect the geometry of the first plane to be transformed under complex transformations.

The final result (using again $R = r'$ and omitting the primes) becomes
\begin{equation}
	\begin{aligned}
	\dd s^2 = - \dd u^2 &- 2\, \dd u \dd r
		+ \big(1 - \tilde f^{\{1, 2\}} \big) (\dd u - a \sin^2 \theta\, \dd \phi - b \cos^2 \theta\, \dd \psi)^2
		\\
		&+ 2 a \sin^2 \theta\, \dd r \dd \phi
		+ 2 b \cos^2 \theta\, \dd r\dd \psi \\
		&+ \rho^2 \dd\theta^2
		+ (r^2 + a^2) \sin^2 \theta\, \dd\phi^2
		+ (r^2 + b^2) \cos^2 \theta\, \dd \psi^2
	\end{aligned}
\end{equation} 
where
\begin{equation}
	\rho^2 = r^2 + a^2 \cos^2 \theta + b^2 \sin^2 \theta.
\end{equation} 
Furthermore, the function $\tilde f^{\{1\}}$ has been complexified as
\begin{equation}
 	\tilde f^{\{1,2\}} = 1 - \frac{m}{\abs{r}^2 + a^2 \cos^2 \theta}
		= 1 - \frac{m}{r'^2 + a^2 \cos^2 \theta + b^2 \sin^2 \theta}
		= 1 - \frac{m}{\rho^2}.
\end{equation}

The metric can then be transformed into the Boyer--Lindquist (BL) using
\begin{equation}
	\label{higher:change:5d-bl}
	\dd u = \dd t - g(r)\, \dd r, \qquad
	\dd\phi = \dd\phi' - h_\phi(r)\, \dd r, \qquad
	\dd\psi = \dd\psi' - h_\psi(r)\, \dd r.
\end{equation} 
Defining the parameters\footnotemark{}%
\footnotetext{%
	See \eqref{higher-jna:metric:rotating:result-jna-bl-parameters} for a definition of $\Delta$ in terms of $\tilde f$.
}
\begin{equation}
	\Pi = (r^2 + a^2) (r^2 + b^2), \qquad
	\Delta = r^4 + r^2 (a^2 + b^2- m) + a^2 b^2,
\end{equation}
the functions can be written
\begin{equation}
	\label{higher:change:myers-perry:bl-g-h}
	g(r) = \frac{\Pi}{\Delta}, \qquad
	h_\phi(r) = \frac{\Pi}{\Delta}\, \frac{a}{r^2 + a^2}, \qquad
	h_\psi(r) = \frac{\Pi}{\Delta}\, \frac{b}{r^2 + b^2}.
\end{equation} 
Finally one gets
\begin{equation}
	\label{higher-jna:metric:rotating:5d-2-moments-bl}
	\begin{aligned}
		\dd s^2 = - \dd t^2
			&+ \big(1 - \tilde f^{\{1, 2\}} \big) (\dd t - a \sin^2 \theta\, \dd \phi - b \cos^2 \theta\, \dd \psi)^2
			+ \frac{r^2 \rho^2}{\Delta}\, \dd r^2 \\
			&+ \rho^2 \dd\theta^2
			+ (r^2 + a^2) \sin^2 \theta\, \dd\phi^2
			+ (r^2 + b^2) \cos^2 \theta\, \dd \psi^2.
	\end{aligned}
\end{equation} 
One recovers here the five dimensional Myers--Perry black hole with two angular momenta~\cite{Myers:1986:BlackHolesHigher}.

\subsection{BMPV black hole}
\label{sec:higher-jna:5d:bmpv}

\subsubsection{Few properties and seed metric}

In this section we focus on another example in five dimensions, which is the BMPV black hole~\cite{Breckenridge:1997:DbranesSpinningBlack, Gauntlett:1999:BlackHolesD5}.
This solution possesses many interesting properties, in particular it can be proven that it is the only asymptotically flat rotating BPS black hole in five dimensions with the corresponding near-horizon geometry~\cites[sec.~7.2.2, 8.5]{Emparan:2008:BlackHolesHigher}{Reall:2003:HigherDimensionalBlack}.\footnotemark{}%
\footnotetext{%
	Other possible near-horizon geometries are $S^1 \times S^2$ (for black rings) and $T^3$, even if the latter does not seem really physical.
	BMPV horizon corresponds to the squashed $S^3$.
}
It is interesting to notice that even if this extremal solution is a slowly rotating metric, it is an exact solution (whereas Einstein equations need to be truncated for consistency of usual slow rotation).

For a rotating black hole the BPS and extremal limits do not coincide~\cites[sec.~7.2]{Emparan:2008:BlackHolesHigher}[sec.~1]{Gauntlett:1999:BlackHolesD5}: the first implies that the mass is related to the electric charge,\footnote{It is a consequence from the BPS bound $m \ge \sqrt{3}/2\, \abs{q}$.} while extremality\footnotemark{}%
\footnotetext{%
	Regularity is given by a bound, which is saturated for extremal black holes.
}
implies that one linear combination of the angular momenta vanishes, and for this reason we set $a = b$ from the beginning.\footnotemark{}%
\footnotetext{%
	If we had kept $a \neq b$ we would have discovered later that one cannot transform the metric to Boyer--Lindquist coordinates without setting $a = b$.
}
Thus two independent parameters are left and are taken to be the mass and one angular momentum.

In the non-rotating limit BMPV black hole reduces to the charged extremal Schwarz\-schild--Tangherlini (with equal mass and charge) written in isotropic coordinates.
For non-rotating black hole the extremal and BPS limit are equivalent.

Both the charged extremal Schwarzschild--Tangherlini and BMPV black holes are solutions of minimal ($N = 2$) $d = 5$ supergravity (Einstein--Maxwell plus Chern--Simons) whose bosonic action is~\cites[sec.~1]{Gauntlett:1999:BlackHolesD5}[sec.~2]{Aliev:2014:SuperradianceBlackHole}[sec.~2]{Gauntlett:2003:AllSupersymmetricSolutions}
\begin{equation}
	\label{higher-jna:higher-jna:action:N=2-d=5-sugra}
	S = - \frac{1}{16\pi G} \int \left(R\, \hodge{1} + F \wedge \hodge{F} + \frac{2\lambda}{3 \sqrt{3}}\, F \wedge F \wedge A \right),
\end{equation} 
where supersymmetry imposes $\lambda = 1$.

Since extremal limits are different for static and rotating black holes we can guess that the black hole obtained from the algorithm will not be a solution of the equations of motion and that it will be necessary to take some limit.

The charged extremal Schwarzschild--Tangherlini black hole is taken as a seed metric~\cites[sec.~3.2]{Gauntlett:2003:AllSupersymmetricSolutions}[sec.~4]{Gibbons:1994:SupersymmetricSelfGravitatingSolitons}[sec.~1.3.1]{Puhm:2013:BlackHolesString}
\begin{equation}
	\label{higher-jna:metric:5d-bmpv}
	\dd s^2 = - H^{-2}\, \dd t^2 + H\, (\dd r^2 + r^2\, \dd\Omega_3^2 )
\end{equation} 
where $\dd\Omega_3^2$ is the metric of the $3$-sphere written in
\eqref{higher-jna:eq:coord-S3-spherical}.
The function $H$ is harmonic
\begin{equation}
	H(r) = 1 + \frac{m}{r^2},
\end{equation} 
and the electromagnetic field reads
\begin{equation}
	\label{higher-jna:pot:5d-bmpv}
	A = \frac{\sqrt{3}}{2 \lambda}\, \frac{m}{r^2}\, \dd t
		= (H - 1)\, \dd t.
\end{equation} 

In the next subsections we apply successively the transformations \eqref{higher-jna:eq:5d-ansatz-hopf-1} and \eqref{higher-jna:eq:5d-ansatz-hopf-2} with $a = b$ in the case $\lambda = 1$.

\subsubsection{Transforming the metric}

The transformation to $(u, r)$ coordinates of the seed metric \eqref{higher-jna:metric:5d-bmpv}
\begin{equation}
	\dd t = \dd u + H^{3/2}\, \dd r
\end{equation} 
gives
\begin{subequations}
\begin{align}
	\dd s^2 &= - H^{-2}\, \dd u^2 - 2 H^{-1/2}\, \dd u \dd r + H r^2\, \dd\Omega_3^2 \\
		&= - H^{-2}\, \big(\dd u - 2 H^{3/2}\, \dd r \big)\, \dd u + H r^2\, \dd\Omega_3^2.
\end{align}
\end{subequations}

For transforming the above metric one should follow the recipe of the previous section: the transformations \eqref{higher-jna:eq:5d-ansatz-hopf-1}
\begin{equation}
	u = u' + i a \cos \theta, \qquad
	\dd u = \dd u' - a \sin^2 \theta\, \dd\phi,
\end{equation}
and \eqref{higher-jna:eq:5d-ansatz-hopf-2}
\begin{equation}
	u = u' + i a\, \sin \theta, \qquad
	\dd u = \dd u' - a \cos^2 \theta\, \dd\psi
\end{equation} 
are performed one after another, transforming each time only the terms that pertain to the corresponding rotation plane.\footnotemark{}%
\footnotetext{%
	For another approach see \cref{sec:higher-jna:5d:bmpv-second-approach}.
}
In order to preserve the isotropic form of the metric the function $H$ is complexified everywhere (even when it multiplies terms that belong to the other plane).

Since the procedure is exactly similar to the Myers--Perry case we give only the final result in $(u, r)$ coordinates
\begin{equation}
	\label{higher-jna:metric:5d-bmpv:ur-before-limit}
	\begin{aligned}
		\dd s^2 = &- \tilde H^{-2} \big(\dd u
				- a (1 - \tilde H^{3/2}) (\sin^2 \theta\, \dd\phi + \cos^2 \theta\, \dd\psi) \big)^2 \\
			&- 2 \tilde H^{-1/2} \big(\dd u - a (1 - \tilde H^{3/2})\, (\sin^2 \theta\, \dd\phi + \cos^2 \theta\, \dd\psi) \big)\, \dd r \\
			&+ 2 a \tilde H\, (\sin^2 \theta\, \dd\phi + \cos^2 \theta\, \dd\psi)\, \dd r
			- 2 a^2 \tilde H \cos^2 \theta \sin^2 \theta\, \dd\phi \dd\psi
			\\
			&+ \tilde H\, \Big(
				(r^2 + a^2) (\dd \theta^2 + \sin^2 \theta\, \dd\phi^2 + \cos^2 \theta\, \dd\psi^2)
				+ a^2 (\sin^2 \theta\, \dd\phi + \cos^2 \theta\, \dd\psi)^2 \Big).
	\end{aligned}
\end{equation} 
After both transformations the resulting function $\tilde H$ is
\begin{equation}
	\label{higher-jna:eq:5d-bmpv:tilde-H}
	\tilde H = 1 + \frac{m}{r^2 + a^2 \cos^2\theta + a^2 \sin^2\theta}
		= 1 + \frac{m}{r^2 + a^2}
\end{equation}
which does not depend on $\theta$.

It is easy to check that the Boyer--Lindquist transformation \eqref{higher:change:5d-bl}
\begin{equation}
	\dd u = \dd t - g(r)\, \dd r, \qquad
	\dd\phi = \dd\phi' - h_\phi(r)\, \dd r, \qquad
	\dd\psi = \dd\psi' - h_\psi(r)\, \dd r
\end{equation} 
is ill-defined because the functions depend on $\theta$.
The way out is to take the extremal limit alluded above.

Following the prescription of \cite{Breckenridge:1997:DbranesSpinningBlack, Gauntlett:1999:BlackHolesD5} and taking the extremal limit
\begin{equation}
	\label{higher-jna:eq:5d-bmpv-extremal-limit}
	a, m \longrightarrow 0, \qquad
	\text{imposing} \qquad
	\frac{m}{a^2} = \cst ,
\end{equation}
one gets at leading order
\begin{equation}
	\tilde H(r) = 1 + \frac{m}{r^2} = H(r), \qquad
	a\, (1 - \tilde H^{3/2}) = - \frac{3\, m a}{2\, r^2}
\end{equation} 
which translate into the metric
\begin{equation}
	\begin{aligned}
		\dd s^2 = - H^{-2}\, & \left(\dd u
				+ \frac{3\, m a}{2\, r^2}\, (\sin^2 \theta\, \dd\phi + \cos^2 \theta\, \dd\psi) \right)^2 \\
			&- 2 H^{-1/2} \left( \dd u + \frac{3\, m a}{2\, r^2}\, (\sin^2 \theta\, \dd\phi + \cos^2 \theta\, \dd\psi) \right) \dd r \\
			&+ H\, r^2 (\dd \theta^2 + \sin^2 \theta\, \dd\phi^2 + \cos^2 \theta\, \dd\psi^2).
	\end{aligned}
\end{equation} 
Then Boyer--Lindquist functions are
\begin{equation}
	\label{higher:change:bmpv:g-h}
	g(r) = H(r)^{3/2}, \qquad
	h_\phi(r) = h_\psi(r) = 0
\end{equation} 
and one gets the metric in $(t, r)$ coordinates
\begin{equation}
	\label{higher-jna:metric:5d-bmpv:bmpv-metric}
	\begin{aligned}
		\dd s^2 = &- \tilde H^{-2} \left(\dd t
				+ \frac{3\, m a}{2\, r^2}\, (\sin^2 \theta\, \dd\phi + \cos^2 \theta\, \dd\psi) \right)^2 \\
			&+ \tilde H\, \Big(\dd r^2 + r^2 \big( \dd \theta^2 + \sin^2 \theta\, \dd\phi^2 + \cos^2 \theta\, \dd\psi^2 \big) \Big).
	\end{aligned}
\end{equation} 
One can recognize the BMPV solution~\cites[p.~4]{Breckenridge:1997:DbranesSpinningBlack}[p.~16]{Gauntlett:1999:BlackHolesD5}.
The fact that this solution has only one rotation parameter can be seen more easily in Euler angle coordinates~\cites[sec.~3]{Gauntlett:1999:BlackHolesD5}[sec.~2]{Gibbons:1999:SupersymmetricRotatingBlack} or by looking at the conserved charges in the $\phi$- and $\psi$-planes~\cite[sec.~3]{Breckenridge:1997:DbranesSpinningBlack}.

\subsubsection{Transforming the Maxwell potential}

The seed gauge field \eqref{higher-jna:pot:5d-bmpv} in the $(u, r)$ coordinates is
\begin{equation}
	A = \frac{\sqrt{3}}{2}\, (H - 1)\, \dd u,
\end{equation} 
since the $A_r(r)$ component can be removed by a gauge transformation.
One can apply the two JN transformations \eqref{higher-jna:eq:5d-ansatz-hopf-1} and \eqref{higher-jna:eq:5d-ansatz-hopf-2} with $b = a$ to obtain
\begin{equation}
	A = \frac{\sqrt{3}}{2}\, (\tilde H - 1) \Big( \dd u - a\, (\sin^2 \theta\, \dd\phi + \cos^2 \theta\, \dd\psi) \Big).
\end{equation} 

Then going into BL coordinates with \eqref{higher:change:5d-bl} and \eqref{higher:change:bmpv:g-h} provides
\begin{equation}
	A = \frac{\sqrt{3}}{2}\, (\tilde H - 1) \Big( \dd t - a\, (\sin^2 \theta\, \dd\phi + \cos^2 \theta\, \dd\psi) \Big) + A_r(r)\, \dd r.
\end{equation} 
Again $A_r$ depends only on $r$ and can be removed by a gauge transformation.
Applying the extremal limit \eqref{higher-jna:eq:5d-bmpv-extremal-limit} finally gives
\begin{equation}
	A = \frac{\sqrt{3}}{2}\, \frac{m}{r^2} \Big( \dd t - a\, (\sin^2 \theta\, \dd\phi + \cos^2 \theta\, \dd\psi) \Big),
\end{equation}
which is again the result presented in~\cite[p. 5]{Breckenridge:1997:DbranesSpinningBlack}.

Despite the fact that the seed metric \eqref{higher-jna:metric:5d-bmpv} together with the gauge field \eqref{higher-jna:pot:5d-bmpv} solves the equations of motion for any value of $\lambda$, the resulting rotating metric solves the equations only for $\lambda = 1$ (see~\cite[sec.~7]{Gauntlett:1999:BlackHolesD5} for a discussion).
An explanation in this reduction can be found in the limit \eqref{higher-jna:eq:5d-bmpv-extremal-limit} that was needed for transforming the metric to Boyer--Lindquist coordinates and which gives a supersymmetric black hole -- which necessarily has $\lambda = 1$.

\subsection{Another approach to BMPV}
\label{sec:higher-jna:5d:bmpv-second-approach}

In \cref{sec:higher-jna:5d:bmpv} we applied the same recipe given in \cref{sec:higher-jna:5d:myers-perry} which, according to our claim, is the standard procedure in five dimensions.

There is another way to derive BMPV black hole.
Indeed, by considering that terms quadratic in the angular momentum do not survive in the extremal limit, they can be added to the metric without modifying the final result.
Hence we can decide to transform all the terms of the metric\footnotemark{} since the additional terms will be subleading.%
\footnotetext{%
	In opposition to our initial recipe, but this is done in a controlled way.
}
As a result the BL transformation is directly well defined and overall formulas are simpler, but we need to take the extremal limit before the end (this could be done either in $(u, r)$ or $(t, r)$ coordinates).
This section shows that both approaches give the same result.

Applying the two transformations
\begin{subequations}
\begin{gather}
	u = u' + i a \cos \theta, \qquad
	\dd u = \dd u' - a \sin^2 \theta\, \dd\phi, \\
	u = u' + i a\, \sin \theta, \qquad
	\dd u = \dd u' - a \cos^2 \theta\, \dd\psi
\end{gather}
\end{subequations}
successively on all the terms one obtains the metric
\begin{equation}
	\begin{aligned}
		\dd s^2 = &- \tilde H^{-2} \big(\dd u
				- a (1 - \tilde H^{3/2}) (\sin^2 \theta\, \dd\phi + \cos^2 \theta\, \dd\psi) \big)^2 \\
			&- 2 \tilde H^{-1/2} \big(\dd u - a (\sin^2 \theta\, \dd\phi + \cos^2 \theta\, \dd\psi) \big)\, \dd r \\
			&+ \tilde H\, \Big(
				(r^2 + a^2) (\dd \theta^2 + \sin^2 \theta\, \dd\phi^2 + \cos^2 \theta\, \dd\psi^2)
				+ a^2 (\sin^2 \theta\, \dd\phi + \cos^2 \theta\, \dd\psi)^2 \Big),
	\end{aligned}
\end{equation} 
where again $\tilde H$ is given by \eqref{higher-jna:eq:5d-bmpv:tilde-H}
\begin{equation}
	\tilde H = 1 + \frac{m}{r^2 + a^2}.
\end{equation} 
Only one term is different when comparing with \eqref{higher-jna:metric:5d-bmpv:ur-before-limit}.

The BL transformation \eqref{higher:change:5d-bl} is well-defined and the corresponding functions are
\begin{equation}
	\label{higher-jna:change:bmpv-2:bl-gh}
	g(r) = \frac{a^2 + (r^2 + a^2) \tilde H(r)}{r^2 + 2 a^2}, \qquad
	h_\phi(r) = h_\psi(r) = \frac{a}{r^2 + 2 a^2}
\end{equation} 
which do not depend on $\theta$.
They lead to the metric
\begin{equation}
	\begin{aligned}
		\dd s^2 = &- \tilde H^{-2} \big(\dd t
				- a (1 - \tilde H^{3/2}) (\sin^2 \theta\, \dd\phi + \cos^2 \theta\, \dd\psi) \big)^2 \\
			&+ \tilde H\, \bigg[
				(r^2 + a^2) \left(\frac{\dd r^2}{r^2 + 2 a^2} + \dd \theta^2 + \sin^2 \theta\, \dd\phi^2 + \cos^2 \theta\, \dd\psi^2 \right) \\
				&\qquad\quad+ a^2 (\sin^2 \theta\, \dd\phi + \cos^2 \theta\, \dd\psi)^2 \bigg].
	\end{aligned}
\end{equation} 

At this point it is straightforward to check that this solution does not satisfy Einstein equations and we need to take the extremal limit \eqref{higher-jna:eq:5d-bmpv-extremal-limit}
\begin{equation}
	a, m \longrightarrow 0, \qquad
	\text{imposing} \qquad
	\frac{m}{a^2} = \cst
\end{equation}
in order to get the BMPV solution \eqref{higher-jna:metric:5d-bmpv:bmpv-metric}
\begin{equation}
	\begin{aligned}
		\dd s^2 = &- \tilde H^{-2} \left(\dd t
				+ \frac{3\, m a}{2\, r^2}\, (\sin^2 \theta\, \dd\phi + \cos^2 \theta\, \dd\psi) \right)^2 \\
			&+ \tilde H\, \Big(\dd r^2 + r^2 \big( \dd \theta^2 + \sin^2 \theta\, \dd\phi^2 + \cos^2 \theta\, \dd\psi^2 \big) \Big).
	\end{aligned}
\end{equation} 

It is surprising that the BL transformation is simpler in this case.
Another point that is worth stressing is that we did not need to take the extremal limit at an intermediate stage, whereas in \cref{sec:higher-jna:5d:bmpv} we had to in order to get a well-defined BL transformation.

\subsection{CCLP black hole}
\label{sec:higher-jna:5d:cclp}

The CCLP black hole~\cite{Chong:2005:GeneralNonExtremalRotating} (see also~\cite[sec.~2]{Aliev:2014:SuperradianceBlackHole}) corresponds to the non-extremal generalization of the BMPV solution and it possesses four independent charges: two angular momenta $a$ and $b$, an electric charge $q$ and the mass $m$.
It is a solution of $d = 5$ minimal supergravity \eqref{higher-jna:higher-jna:action:N=2-d=5-sugra}.

The solution reads
\begin{subequations}
\begin{gather}
	\label{higher-jna:higher-jna:metric:cclp}
	\begin{aligned}
		\dd s^2 = - \dd t^2
			&+ (1 - \tilde f) (\dd t - a \sin^2 \theta\, \dd \phi - b \cos^2 \theta\, \dd \psi)^2
			+ \frac{r^2 \rho^2}{\Delta_r}\, \dd r^2 \\
			&+ \rho^2 \dd\theta^2
			+ (r^2 + a^2) \sin^2 \theta\, \dd\phi^2
			+ (r^2 + b^2) \cos^2 \theta\, \dd \psi^2 \\
			&- \frac{2 q}{\rho^2}\, (b \sin^2 \theta\, \dd \phi + a \cos^2 \theta\, \dd \psi) (\dd t - a \sin^2 \theta\, \dd \phi - b \cos^2 \theta\, \dd \psi),
	\end{aligned} \\
	A = \frac{\sqrt{3}}{2}\, \frac{q}{\rho^2} (\dd t - a \sin^2 \theta\, \dd \phi - b \cos^2 \theta\, \dd \psi),
\end{gather}
\end{subequations}
where the function are given by
\begin{subequations}
\begin{align}
	\rho^2 &= r^2 + a^2 \cos^2 \theta + b^2 \sin^2 \theta, \\
	\tilde f &= 1 - \frac{2 m}{\rho^2} + \frac{q^2}{\rho^4}, \\
	\Delta_r &= \Pi + 2 a b q + q^2 - 2 m r^2.
\end{align}
\end{subequations}

Yet, using our prescription, it appears that the metric of this black hole cannot entirely be recovered.
Indeed while the gauge field can be found straightforwardly, all the terms of the metric but one are generated by our algorithm.
The missing term (corresponding to the last one in \eqref{higher-jna:higher-jna:metric:cclp}) is proportional to the electric charge and the current prescription cannot generate it since the latter can only appear in $\tilde f$ (or in the gauge field); moreover the algorithm cannot explain the first term in parenthesis since $a$ and $b$ always appear with $\dd\phi$ and $\dd\psi$ respectively.

This issue may be related to the fact that the CCLP solution cannot be written as a Kerr--Schild metric but rather as an extended Kerr--Schild one~\cite{Aliev:2009:NoteRotatingCharged, Ett:2010:ExtendedKerrSchildAnsatz, Malek:2014:ExtendedKerrSchildSpacetimes}, which includes an additional term proportional to a spacelike vector.
It appears that the missing term corresponds precisely to this additional term in the extended Kerr--Schild metric and it is well-known that the JN algorithm works mostly for Kerr--Schild metrics.
Moreover the $\Delta$ computed from \eqref{higher-jna:metric:rotating:result-jna-bl-parameters} depends on $\theta$ and the BL transformation would not be well-defined if the additional term is not present to modify $\Delta$ to $\Delta_r$.

%% file: sections/higher_dimensions.tex
\section{Algorithm in any dimension}
\label{sec:higher}

Following the same prescription in dimensions higher than five does not lead as nicely to the exact Myers--Perry solution.
Indeed we show in this section that, while the transformation of the metric can be done along the same line, the -- major -- obstacle comes from the function $f$ that cannot be transformed as expected.
Finding the correct complexification seems very challenging and it may be necessary to use a different complex coordinate transformation in order to perform a completely general transformation in any dimension.
It might be possible to gain insight into this problem by computing the transformation within the framework of the tetrad formalism.
One may think that a possible solution would be to replace complex numbers by quaternions, assigning one angular momentum to each complex direction but it is straightforward to check that this approach is not working.

The key element to perform the algorithm on the metric is to parametrize the metric on the sphere by direction cosines since these coordinates are totally symmetric under permutation of angular momenta (at the opposite of the spherical coordinates).
We are able to derive the general form of a rotating metric with the maximal number of angular momenta it can have in $d$ dimensions, but we are not able to apply this result to any specific example for $d \ge 6$, except if all momenta but one are vanishing.
Nonetheless this provides a unified view of the JN algorithm in any $d \ge 3$.
We conclude this section by few examples, including the singly-rotating Myers--Perry solution in any dimension and the rotating BTZ black hole.

It would be very desirable to derive the general $d$-dimensional Myers--Perry solution~\cite{Myers:1986:BlackHolesHigher}, or at least to understand why only the metric can be found, and not the function inside.

\subsection{Metric transformation}
\label{sec:higher-jna:any-dimension}

We consider the JN algorithm applied to a general static $d$-dimension metric and show how the tensor structure can be transformed.
In the following the dimension is taken to be odd in order to simplify the computations but the final result holds also for $d$ even.

\subsubsection{Seed metric and discussion}

Consider the $d$-dimensional static metric (notations are defined in \cref{app:coord:general-d})
\begin{equation}
	\dd s^2 = - f\, \dd t^2 + f^{-1}\, \dd r^2 + r^2\, \dd \Omega_{d-2}^2
\end{equation} 
where $\dd \Omega_{d-2}^2$ is the metric on $S^{d-2}$
\begin{equation}
	\dd \Omega_{d-2}^2 = \dd\theta_{d-2} + \sin^2 \theta_{d-2}\, \dd \Omega_{d-3}^2
		= \sum_{i=1}^n \big( \dd\mu_i^2 + \mu_i^2 \dd\phi_i^2).
\end{equation} 
The number $n = (d-1) / 2$ counts the independent $2$-spheres.

In Eddington--Finkelstein coordinates the metric reads
\begin{equation}
	\label{higher-jna:higher-jna:metric:static-seed}
	\dd s^2 = (1 - f)\, \dd u^2 - \dd u\, (\dd u + 2 \dd r)
			+ r^2 \sum_i \Big(\dd \mu_i^2 + \mu_i^2\, \dd \phi_i^2 \Big).
\end{equation} 

The metric looks like a $2$-dimensional space $(t, r)$ with a certain number of additional $2$-spheres $(\mu_i, \phi_i)$ which are independent from one another.
Then we can consider only the piece $(u, r, \mu_i, \phi_i)$ (for fixed $i$) which will transform like a $4$-dimensional spacetime, while the other part of the metric $(\mu_j, \phi_j)$ for all $j \neq i$ will be unchanged.
After the first transformation we can move to another $2$-sphere.
We can thus imagine to put in rotation only one of these spheres.
Then we will apply again and again the algorithm until all the spheres have angular momentum: the whole complexification will thus be a $n$-steps process.
Moreover if these $2$-spheres are taken to be independent this implies that we should not complexify the functions that are not associated with the plane we are putting in rotation.

To match these demands the metric is rewritten as
\begin{equation}
	\label{higher-jna:higher-jna:higher-jna:metric:static-seed-ur}
	\dd s^2 = (1 - f)\, \dd u^2 - \dd u\, (\dd u + 2 \dd r_{i_1})
		+ r_{i_1}^2 (\dd\mu_{i_1}^2 + \mu_{i_1}^2 \dd\phi_{i_1}^2)
		+ \sum_{i \neq i_1} \Big(r_{i_1}^2 \dd \mu_i^2 + R^2 \mu_i^2\, \dd \phi_i^2 \Big).
\end{equation} 
where we introduced the following two functions of $r$
\begin{equation}
	r_{i_1}(r) = r, \qquad R(r) = r.
\end{equation} 
This allows to choose different complexifications for the different terms in the metric.
It may be surprising to note that the factors in front of $\dd \mu_i^2$ have been chosen to be $r_{i_1}^2$ and not $R^2$, but the reason is that the $\mu_i$ are all linked by the constraint
\begin{equation}
	\sum_i \mu_i^2 = 1
\end{equation} 
and the transformation of one $i_1$-th $2$-sphere will change the corresponding $\mu_{i_1}$, but also all the others, as it is clear from the formula \eqref{coord:eq:spherical-to-oblate-mu} with all the $a_i$ vanishing but one (this can also be observed in $5d$ where both $\mu_i$ are gathered into $\theta$).

\subsubsection{First transformation}

The transformation is chosen to be
\begin{subequations}
\label{higher-jna:higher-jna:change:jna-1}
\begin{equation}
	r_{i_1} = r'_{i_1} - i\, a_{i_1} \sqrt{1 - \mu_{i_1}^2}, \qquad
	u = u' + i\, a_{i_1} \sqrt{1 - \mu_{i_1}^2}
\end{equation} 
which, together with the ansatz
\begin{equation}
	i\, \frac{\dd \mu_{i_1}}{\sqrt{1 - \mu_{i_1}^2}} = \mu_{i_1}\, \dd \phi_{i_1},
\end{equation} 
gives the differentials
\begin{equation}
	\dd r_{i_1} = \dd r'_{i_1} + a_{i_1} \mu_{i_1}^2\, \dd \phi_{i_1}, \qquad
	\dd u = \dd u' - a_{i_1} \mu_{i_1}^2\, \dd \phi_{i_1}.
\end{equation} 
\end{subequations}
It is easy to check that this transformation reproduces the one given in four and five dimensions.
The complexified version of $f$ is written as $\tilde f^{\{i_1\}}$: we need to keep track of the order in which we gave angular momentum since the function $\tilde f$ will be transformed at each step.

We consider separately the transformation of the $(u, r)$ and $\{ \mu_i, \phi_i \}$ parts.
Inserting the transformations \eqref{higher-jna:higher-jna:change:jna-1} in \eqref{higher-jna:higher-jna:metric:static-seed} results in
\begin{subequations}
\begin{align*}
	\dd s_{u,r}^2 &= (1 - \tilde f^{\{i_1\}})\, \Big(\dd u - a_{i_1} \mu_{i_1}^2\, \dd \phi_{i_1} \Big)^2
		- \dd u\, (\dd u + 2 \dd r_{i_1})
		+ 2 a_{i_1} \mu_{i_1}^2\, \dd r_{i_1} \dd \phi_{i_1}
		+ a_{i_1}^2 \mu_{i_1}^4\, \dd \phi_{i_1}^2, \\
	\dd s_{\mu,\phi}^2 &= \big( r_{i_1}^2 + a_{i_1}^2 \big) (\dd\mu_{i_1}^2 + \mu_{i_1}^2 \dd\phi_{i_1}^2)
		+ \sum_{i \neq i_1} \big( r_{i_1}^2 \dd \mu_i^2 + R^2 \mu_i^2\, \dd \phi_i^2 \big) - a_{i_1}^2 \mu_{i_1}^4\, \dd \phi_{i_1}^2 \\
		&\qquad + a_{i_1}^2 \bigg[- \mu_{i_1}^2 \dd \mu_{i_1}^2 + (1 - \mu_{i_1}^2) \sum_{i \neq i_1} \dd \mu_i^2 \bigg].
\end{align*}
\end{subequations}

The term in the last bracket vanishes as can be seen by using the differential of the constraint
\begin{equation}
	\sum_i \mu_i^2 = 1 \Longrightarrow
	\sum_i \mu_i \dd\mu_i = 0.
\end{equation} 
Since this step is very important and non-trivial we expose the details
\begin{align*}
	[\cdots] &= \mu_{i_1}^2 \dd \mu_{i_1}^2 - (1 - \mu_{i_1}^2) \sum_{i \neq i_1} \dd \mu_i^2
		= \left(\sum_{i \neq i_1} \mu_i \dd\mu_i \right)^2 - \sum_{j \neq i_1} \mu_j^2 \sum_{i \neq i_1} \dd \mu_i^2 \\
		&= \sum_{i,j \neq i_1} \big(\mu_i \mu_j \dd\mu_i \dd\mu_j - \mu_j^2 \dd \mu_i^2 \big)
		= \sum_{i,j \neq i_1} \mu_j \big(\mu_i \dd\mu_j - \mu_j \dd \mu_i \big) \dd\mu_i
		= 0
\end{align*}
by antisymmetry.

Setting $r_{i_1} = R = r$ one obtains the metric
\begin{equation}
\begin{aligned}
	\dd s^2 &= (1 - \tilde f^{\{i_1\}})\, \Big(\dd u - a_{i_1} \mu_{i_1}^2\, \dd \phi_{i_1} \Big)^2
		- \dd u\, (\dd u + 2 \dd r)
		+ 2 a_{i_1} \mu_{i_1}^2\, \dd r \dd \phi_{i_1} \\
		&\qquad+ \big( r^2 + a_{i_1}^2 \big) (\dd\mu_{i_1}^2 + \mu_{i_1}^2 \dd\phi_{i_1}^2)
		+ \sum_{i \neq i_1} r^2 \big( \dd \mu_i^2 + \mu_i^2\, \dd \phi_i^2 \big).
\end{aligned}
\end{equation}
It corresponds to Myers--Perry metric in $d$ dimensions with one non-vanishing angular momentum.
We recover the same structure as in \eqref{higher-jna:higher-jna:higher-jna:metric:static-seed-ur} with some extra terms that are specific to the $i_1$-th $2$-sphere.

\subsubsection{Iteration and final result}

We should now split again $r$ in functions $(r_{i_2}, R)$.
Very similarly to the first time we have
\begin{equation}
\begin{aligned}
	\dd s^2 &= (1 - \tilde f^{\{i_1\}})\, \Big(\dd u - a_{i_1} \mu_{i_1}^2\, \dd \phi_{i_1} \Big)^2
		- \dd u\, (\dd u + 2 \dd r_{i_2})
		+ 2 a_{i_1} \mu_{i_1}^2\, \dd R \dd \phi_{i_1} \\
		&\qquad+ \big( r_{i_2}^2 + a_{i_1}^2 \big) \dd\mu_{i_1}^2
		+ \big( R^2 + a_{i_1}^2 \big) \mu_{i_1}^2 \dd\phi_{i_1}^2
		+ r_{i_2}^2 ( \dd\mu_{i_2}^2 + \mu_{i_2}^2 \dd\phi_{i_2}^2 ) \\
		&\qquad+ \sum_{i \neq i_1, i_2} \Big(r_{i_2}^2 \dd \mu_i^2 + R^2 \mu_i^2\, \dd \phi_i^2 \Big).
\end{aligned}
\end{equation}
We can now complexify as
\begin{equation}
	r_{i_2} = r'_{i_2} - i a_{i_2} \sqrt{1 - \mu_{i_2}^2}, \qquad
	u = u' + i\, a_{i_1} \sqrt{1 - \mu_{i_2}^2}.
\end{equation} 
The steps are exactly the same as before, except that we have some inert terms.
The complexified functions is now $\tilde f^{\{i_1, i_2\}}$.

Repeating the procedure $n$ times we arrive at
\begin{equation}
	\label{higher-jna:metric:rotating:result-jna-ur}
	\begin{aligned}
		\dd s^2 = &- \dd u^2 - 2 \dd u \dd r
			+ \sum_i (r^2 + a_i^2) (\dd \mu_i^2 + \mu_i^2 \dd \phi_i^2)
			- 2 \sum_i a_i \mu_i^2 \, \dd r \dd \phi_i \\
			&+ \Big(1 - \tilde f^{\{i_1, \ldots, i_n\}} \Big) \left(\dd u + \sum_i a_i \mu_i^2 \dd \phi_i \right)^2.
	\end{aligned}
\end{equation} 
One recognizes the general form of the $d$-dimensional metric with $n$ angular momenta~\cite{Myers:1986:BlackHolesHigher}.

Let's quote the metric in Boyer--Lindquist coordinates (omitting the indices on $\tilde f$)~\cite{Myers:1986:BlackHolesHigher}
\begin{equation}
	\label{higher-jna:metric:rotating:result-jna-bl}
	\dd s^2 = - \dd t^2
		+ (1 - \tilde f) \left(\dd t - \sum_i a_i \mu_i^2 \dd \phi_i \right)^2
		+ \frac{r^2 \rho^2}{\Delta}\, \dd r^2
		+ \sum_i (r^2 + a_i^2) \Big(\dd \mu_i^2 + \mu_i^2\, \dd \phi_i^2 \Big)
\end{equation} 
which is obtained from the transformation
\begin{equation}
	\dd u = \dd t - g\, \dd r, \qquad
	\dd \phi_i = \dd \phi'_i - h_i\, \dd r
\end{equation} 
with functions
\begin{equation}
\label{higher-jna:change:rotating:higher-dim-func-gh}
	g = \frac{\Pi}{\Delta}
		= \frac{1}{1 - F (1 - \tilde f)}, \qquad
	h_i = \frac{\Pi}{\Delta} \, \frac{a_i}{r^2 + a_i^2},
\end{equation}
and where the various quantities involved are (see \cref{app:coord:general-d:oblate-cosines})
\begin{equation}
	\label{higher-jna:metric:rotating:result-jna-bl-parameters}
	\begin{gathered}
		\Pi = \prod_i (r^2 + a_i^2), \qquad
		F = 1 - \sum_i \frac{a_i^2 \mu_i^2}{r^2 + a_i^2} = r^2 \sum_i \frac{\mu_i^2}{r^2 + a_i^2}, \\
		r^2 \rho^2 = \Pi F, \qquad
		\Delta = \tilde f\, r^2 \rho^2 + \Pi (1 - F).
	\end{gathered}
\end{equation}

Before ending this section, we comment the case of even dimensions: the term $\varepsilon'\, r^2 \dd \alpha^2$ is complexified as $\varepsilon'\, r_{i_1}^2 \dd \alpha^2$, since it contributes to the sum
\begin{equation}
	\sum_i \mu_i^2 + \alpha^2 = 1.
\end{equation} 
This can be seen more clearly by defining $\mu_{n+1} = \alpha$ (we can also define $\phi_{n+1} = 0$), in which case the index $i$ runs from $1$ to $n+\varepsilon$, and all the previous computations are still valid.

\subsection{Examples in various dimensions}
\label{sec:higher-jna:examples}

\subsubsection{Flat space}

A first and trivial example is to take $f = 1$.
In this case one recovers Minkowski metric in spheroidal coordinates with direction cosines (\cref{app:coord:general-d:oblate-cosines})
\begin{equation}
	\dd s^2 = - \dd t^2 + F\, \dd \bar r^2 + \sum_i (\bar r^2 + a_i^2) \Big(\dd \bar \mu_i^2 + \bar \mu_i^2\, \dd \bar \phi_i^2 \Big) + \varepsilon'\, r^2 \dd \alpha^2.
\end{equation}
In this case the JN algorithm is equivalent to a (true) change of coordinates and there is no intrinsic rotation.
The presence of a non-trivial function $f$ then deforms the algorithm.

\subsubsection{Myers--Perry black hole with one angular momentum}

The derivation of the Myers--Perry metric with one non-vanishing angular momentum has been found by Xu~\cite{Xu:1988:ExactSolutionsEinstein}.

The transformation is taken to be in the first plane
\begin{equation}
	r = r' - i a \sqrt{1 - \mu^2}
\end{equation} 
where $\mu \equiv \mu_1$.
The transformation to the mixed spherical--spheroidal system (\cref{app:coord:general-d:oblate-spherical} is obtained by setting
\begin{equation}
	\mu = \sin \theta, \qquad
	\phi_1 = \phi.
\end{equation} 
In these coordinates the transformation reads
\begin{equation}
	r = r' - i a \cos \theta.
\end{equation} 
We will use the quantity
\begin{equation}
	\rho^2 = r^2 + a^2 (1 - \mu^2)
		= r^2 + a^2 \cos^2 \theta.
\end{equation} 

The Schwarzschild--Tangherlini metric is~\cite{Tangherlini:1963:SchwarzschildFieldDimensions}
\begin{equation}
	\dd s^2 = - f\, \dd t^2 + f^{-1}\, \dd r^2 + r^2\, \dd \Omega_{d-2}^2, \qquad
	f = 1 - \frac{m}{r^{d-3}}.
\end{equation} 

Applying the previous transformation results in
\begin{equation}
\begin{aligned}
	\dd s^2 &= (1 - \tilde f)\, \Big(\dd u - a \mu^2\, \dd \phi \Big)^2
		- \dd u\, (\dd u + 2 \dd r)
		+ 2 a \mu^2\, \dd r \dd \phi \\
		&\qquad+ \big( r^2 + a^2 \big) (\dd\mu^2 + \mu^2 \dd\phi^2)
		+ \sum_{i \neq 1} r^2 \big( \dd \mu_i^2 + \mu_i^2\, \dd \phi_i^2 \big).
\end{aligned}
\end{equation}
where $f$ has been complexified as
\begin{equation}
	\tilde f = 1 - \frac{m}{\rho^2 r^{d-5}}.
\end{equation} 

In the mixed coordinate system one has~\cite{Xu:1988:ExactSolutionsEinstein, Aliev:2006:RotatingBlackHoles}
\begin{equation}
	\begin{aligned}
		\dd s^2 = &- \tilde f\, \dd t^2
			+ 2 a (1 - \tilde f) \sin^2 \theta\, \dd t \dd\phi
			+ \frac{r^{d-3} \rho^2}{\Delta}\, \dd r^2 + \rho^2 \dd\theta^2 \\
			&+ \frac{\Sigma^2}{\rho^2}\, \sin^2 \theta\, \dd\phi^2
			+ r^2 \cos^2 \theta^2\, \dd\Omega_{d-4}^2.
	\end{aligned}
\end{equation} 
where we defined as usual
\begin{equation}
	\Delta = \tilde f \rho^2 + a^2 \sin^2 \theta, \qquad
	\frac{\Sigma^2}{\rho^2} = r^2 + a^2 + a g_{t\phi}.
\end{equation} 
This last expression explains why the transformation is straightforward with one angular momentum: the transformation is exactly the one for $d = 4$ and the extraneous dimensions are just spectators.

We have not been able to generalize this result for several non-vanishing momenta for $d \ge 6$, even for the case with equal momenta .

\subsubsection{Five-dimensional Myers--Perry}
\label{sec:higher-jna:examples:myers-perry-5d}

We take a new look at the five-dimensional Myers--Perry solution in order to derive it in spheroidal coordinates because it is instructive.

The function
\begin{equation}
	1 - f = \frac{m}{r^2}
\end{equation} 
is first complexified as
\begin{equation}
	1 - \tilde f^{\{1\}} = \frac{m}{\abs{r_1}^2}
		= \frac{m}{r^2 + a^2 (1 - \mu^2)}
\end{equation}
and then as 
\begin{equation}
	1 - \tilde f^{\{1, 2\}} = \frac{m}{\abs{r_2}^2 + a^2 (1 - \mu^2)}
		= \frac{m}{r^2 + a^2 (1 - \mu^2) + b^2 (1 - \nu^2)}.
\end{equation}
after the two transformations
\begin{equation}
	r_1 = r_1' - i a \sqrt{1 - \mu^2}, \qquad
	r_2 = r_2' - i b \sqrt{1 - \nu^2}.
\end{equation} 
For $\mu = \sin \theta$ and $\nu = \cos \theta$ one recovers the transformations from \cref{sec:higher-jna:5d:myers-perry,sec:higher-jna:5d:bmpv}.

Let's denote the denominator by $\rho^2$ and compute
\begin{align*}
	\frac{\rho^2}{r^2} &= r^2 + a^2 (1 - \mu^2) + b^2 (1 - \nu^2)
		= (\mu^2 + \nu^2) r^2 + \nu^2 a^2 + \mu^2 b^2 \\
		&= \mu^2 (r^2 + b^2) + \nu^2 (r^2 + a^2)
		= (r^2 + b^2) (r^2 + a^2) \left( \frac{\mu^2}{r^2 + a^2} + \frac{\nu^2}{r^2 + b^2} \right).
\end{align*}
and thus
\begin{equation}
	r^2 \rho^2 = \Pi F.
\end{equation} 
Plugging this into $\tilde f^{\{1, 2\}}$ we have~\cite{Myers:1986:BlackHolesHigher}
\begin{equation}
	1 - \tilde f^{\{1, 2\}} = \frac{m r^2}{\Pi F}.
\end{equation}

\subsubsection{Three dimensions: BTZ black hole}
\label{sec:higher-jna:examples:btz}

As another application we show how to derive the $d = 3$ rotating BTZ black hole from its static version~\cite{Banados:1992:BlackHoleThree}
\begin{equation}
	\dd s^2 = - f\, \dd t^2 + f^{-1}\, \dd r^2 + r^2 \dd\phi^2, \qquad
	f(r) = - M + \frac{r^2}{\ell^2}.
\end{equation} 

In three dimensions the metric on $S^1$ in spherical coordinates is given by
\begin{equation}
	\dd\Omega_1^2 = \dd\phi^2.
\end{equation} 
Introducing the coordinate $\mu$ we can write it in oblate spheroidal coordinates
\begin{equation}
	\dd\Omega_1^2 = \dd\mu^2 + \mu^2 \dd\phi^2
\end{equation} 
with the constraint
\begin{equation}
	\mu^2 = 1.
\end{equation} 

Application of the transformation
\begin{equation}
	u = u' + i a \sqrt{1 - \mu^2}, \qquad
	r = r' - i a \sqrt{1 - \mu^2}
\end{equation} 
gives from \eqref{higher-jna:metric:rotating:result-jna-ur}
\begin{equation}
	\begin{aligned}
		\dd s^2 = &- \dd u^2 - 2 \dd u \dd r
			+ (r^2 + a^2) (\dd \mu^2 + \mu^2 \dd \phi^2)
			- 2 a \mu^2 \, \dd r \dd \phi \\
			&+ (1 - \tilde f) (\dd u + a \mu^2 \dd \phi )^2.
	\end{aligned}
\end{equation} 
The transformation of $f$ is
\begin{equation}
	\tilde f = - m + \frac{\rho^2}{\ell^2}, \qquad
	\rho^2 = r^2 + a^2 (1 - \mu^2).
\end{equation} 

The transformation \eqref{higher-jna:change:rotating:higher-dim-func-gh}
\begin{equation}
	g = \frac{\rho^2 (1 - \tilde f)}{\Delta}, \qquad
	h = \frac{a}{\Delta}, \qquad
	\Delta = r^2 + a^2 + (\tilde f - 1) \rho^2
\end{equation}
to Boyer--Lindquist coordinates leads to the metric \eqref{higher-jna:metric:rotating:result-jna-bl}
\begin{equation}
	\dd s^2 = - \dd t^2
		+ (1 - \tilde f) (\dd t + a \mu^2 \dd \phi )^2
		+ \frac{\rho^2}{\Delta}\, \dd r^2
		+ (r^2 + a^2) (\dd \mu^2 + \mu^2\, \dd \phi^2 ).
\end{equation} 

Finally the constraint $\mu^2 = 1$ can be used to remove the $\mu$.
In this case one finds
\begin{equation}
	\rho^2 = r^2, \qquad
	\Delta = a^2 + \tilde f r^2
\end{equation}
and the metric simplifies to
\begin{equation}
	\dd s^2 = - \dd t^2
		+ (1 - \tilde f) (\dd t + a \dd \phi )^2
		+ \frac{r^2}{a^2 + r^2 \tilde f}\, \dd r^2
		+ (r^2 + a^2) \dd \phi^2.
\end{equation} 

We define the function
\begin{equation}
	N^2 = \tilde f + \frac{a^2}{r^2} = - M + \frac{r^2}{\ell^2} + \frac{a^2}{r^2}.
\end{equation} 
Then redefining the time variable as~\cite{Kim:1997:NotesSpinningAdS3, Kim:1999:SpinningBTZBlack}
\begin{equation}
	t = t' - a \phi
\end{equation} 
we get (omitting the prime)
\begin{equation}
	\dd s^2 = - N^2 \dd t^2 + N^{-2}\, \dd r^2 + r^2 (N^\phi \dd t + \dd \phi)^2
\end{equation} 
with the angular shift
\begin{equation}
	N^\phi(r) = \frac{a}{r^2}.
\end{equation} 
This is the solution given in~\cite{Banados:1992:BlackHoleThree} with $J = -2a$.

It has already been showed by Kim that the rotating BTZ black hole can be derived through the JN algorithm in a different settings~\cite{Kim:1997:NotesSpinningAdS3, Kim:1999:SpinningBTZBlack}: he views the $d = 3$ solution as the slice $\theta = \pi/2$ of the $d = 4$ solution.
Obviously this is equivalent to our approach: we have seen that $\mu = \sin \theta$ in $d = 4$ (\cref{app:coord:4d}), and the constraint $\mu^2 = 1$ is solved by $\theta = \pi/2$.
Nonetheless our approach is more direct since the result just follows from a suitable choice of coordinates and there are no need for advanced justification.

Starting from the charged BTZ black hole
\begin{equation}
       f(r) = - M + \frac{r^2}{\ell^2} - Q^2 \ln r^2, \qquad
       A = - \frac{Q}{2}\, \ln r^2,
\end{equation} 
it is not possible to find the charged rotating BTZ black hole from~\cites{Clement:1993:ClassicalSolutionsThreedimensional, Clement:1996:SpinningChargedBTZ}[sec.~4.2]{Martinez:2000:ChargedRotatingBlack}: the solution solves Einstein equations, but not the Maxwell ones.
This has been already remarked using another technique in~\cite[app.~B]{Lambert:2014:ConformalSymmetriesGravity}.
It may be possible that a more general ansatz is necessary, following \cref{sec:general} but in $d = 3$.

%% file: sections/coordinates.tex
\section{Coordinate systems}
\label{app:coord}

This appendix is partly based on~\cite{Tangherlini:1963:SchwarzschildFieldDimensions, Myers:1986:BlackHolesHigher, Gibbons:2005:GeneralKerrdeSitter}.
We present formulas for any dimension before summarizing them for $4$ and $5$ dimensions.

\subsection{\texorpdfstring{$d$}{d}-dimensional}
\label{app:coord:general-d}

Let's consider $d = N + 1$ dimensional Minkowski space whose metric is denoted by
\begin{equation}
	\dd s^2 = \eta_{\mu\nu}\; \dd x^\mu \dd x^\nu, \qquad
	\mu = 0, \ldots, N.
\end{equation} 
In all the following coordinates systems the time direction can separated from the spatial (positive definite) metric as
\begin{equation}
	\dd s^2 = - \dd t^2 + \dd \Sigma^2, \qquad
	\dd \Sigma^2 = \gamma_{ab}\; \dd x^a \dd x^b, \qquad
	a = 1, \ldots, N,
\end{equation} 
where $x^0 = t$.

One defines by $n$ the number of independent $2$-planes of rotation
\begin{equation}
	n = \floor{\frac{N}{2}}
\end{equation} 
such that
\begin{equation}
	\label{coord:eq:d-dim-epsilon}
	d + \varepsilon = 2n + 2, \qquad
	N + \varepsilon = 2n + 1, \qquad
	\varepsilon' = 1 - \varepsilon
\end{equation} 
where
\begin{equation}
	\varepsilon = \frac{1}{2} (1 - (-1)^d ) =
	\begin{cases}
		0 & \text{$d$ even (or $N$ odd)} \\
		1 & \text{$d$ odd (or $N$ even)},
	\end{cases}
\end{equation} 
and conversely for $\varepsilon'$.

\subsubsection{Cartesian system}

The usual Cartesian metric is
\begin{equation}
	\dd \Sigma^2 = \delta_{ab} \dd x^a \dd x^b
		= \dd x^a \dd x^a
		= \dd \vec x^2.
\end{equation}

\subsubsection{Spherical}

Introducing a radial coordinate $r$, the flat space metric can be written as a $(N-1)$-sphere of radius $r$
\begin{equation}
	\label{coord:metric:flat-d:spherical}
	\dd \Sigma^2 = \dd r^2 + r^2 \dd \Omega_{N-1}^2.
\end{equation} 
The term $\dd \Omega_{N-1}^2$ corresponds the metric on the unit $(N-1)$-sphere $S^{N-1}$, which is parame\-trized by $(N-1)$ angles $\theta_i$ and is defined recursively as
\begin{equation}
	\dd \Omega_{N-1}^2 = \dd \theta_{N-1}^2 + \sin^2 \theta_{N-1} \; \dd \Omega_{N-2}^2.
\end{equation} 

This surface can be embedded in $N$-dimensional flat space with coordinates $X^a$ constrained by
\begin{equation}
	\label{coord:eq:spherical-embedding}
	X^a X^a = 1.
\end{equation}

\subsubsection{Spherical with direction cosines}

In $d$-dimensions there are $n$ orthogonal $2$-planes,\footnotemark{} thus we can pair $2n$ of the embedding coordinates $X^a$ \eqref{coord:eq:spherical-embedding} as $(X_i, Y_i)$ which are parametrized as%
\footnotetext{%
	Note that this is linked to the fact that the little group of massive representation in $D$ dimension is $\group{SO}(N)$, which possess $n$ Casimir invariants~\cite{Myers:1986:BlackHolesHigher}.
}
\begin{equation}
	X_i + i Y_i = \mu_i \e^{i\phi_i}, \qquad
	i = 1, \ldots n.
\end{equation} 
For $d$ even there is an extra unpaired coordinate that is taken to be
\begin{equation}
	X^N = \alpha.
\end{equation}

Each pair parametrizes a $2$-sphere of radius $\mu_i$.
The $\mu_i$ are called the \emph{direction cosines} and satisfy
\begin{equation}
	\sum_i \mu_i^2 + \varepsilon' \alpha^2 = 1
\end{equation} 
since there is one superfluous coordinate from the embedding.
Finally the metric is
\begin{equation}
	\dd \Omega_{N-1}^2 = \sum_i \Big(\dd \mu_i^2 + \mu_i^2\; \dd \phi_i^2 \Big) + \varepsilon'\, \dd \alpha^2.
\end{equation} 

The interest of these coordinates is that all rotational directions are symmetric.

\subsubsection{Spheroidal with direction cosines}
\label{app:coord:general-d:oblate-cosines}

From the previous system we can define the spheroidal $(\bar r, \bar\mu_i, \bar\phi_i)$ system – adapted when some of the $2$-spheres are deformed to ellipses – by introducing parameters $a_i$ such that (for $d$ odd)
\begin{equation}
	\label{coord:eq:spherical-to-oblate-mu}
	r^2 \mu_i^2 = (\bar r^2 + a_i^2) \bar \mu_i^2, \qquad
	\sum_i \bar \mu_i^2 = 1.
\end{equation} 
This last condition implies that
\begin{equation}
	r^2 = \sum_i (\bar r^2 + a_i^2) \bar \mu_i^2
		= \bar r^2 + \sum_i a_i^2 \bar \mu_i^2.
\end{equation} 

In these coordinates the metric reads
\begin{equation}
	\label{coord:metric:flat-d:spheroidal}
	\dd \Sigma^2 = F\; \dd \bar r^2 + \sum_i (\bar r^2 + a_i^2) \Big(\dd \bar \mu_i^2 + \bar \mu_i^2\; \dd \bar \phi_i^2 \Big) + \varepsilon'\, r^2 \dd \alpha^2
\end{equation} 
and we defined
\begin{equation}
	\label{coord:eq:flat-d:spheroidal:F}
	F = 1 - \sum_i \frac{a_i^2 \bar \mu_i^2}{\bar r^2 + a_i^2} = \sum_i \frac{\bar r^2 \bar \mu_i^2}{\bar r^2 + a_i^2}.
\end{equation} 

Here the $a_i$ are just introduced as parameters in the transformation, but in the main text they are interpreted as "true" rotation parameters, i.e.
angular momenta (per unit of mass) of a black hole.
They all appear on the same footing.

Another quantity of interest is
\begin{equation}
	\label{coord:eq:flat-d:spheroidal:Pi}
	\Pi = \prod_i (\bar r^2 + a_i^2).
\end{equation}

\subsubsection{Mixed spherical–spheroidal}
\label{app:coord:general-d:oblate-spherical}

We consider the deformation of the spherical metric where one of the $2$-sphere is replaced by an ellipse~\cite[sec.~3]{Aliev:2006:RotatingBlackHoles}.

To shorten the notation let's define
\begin{equation}
	\theta = \theta_{N-1}, \qquad
	\phi = \theta_{N-2}.
\end{equation} 
Doing the change of coordinates
\begin{equation}
	\sin^2 \theta \sin^2 \phi = \cos^2 \theta.
\end{equation}
the metric becomes
\begin{equation}
	\dd \Sigma^2 = \frac{\rho^2}{r^2 + a^2}\, \dd r^2
		+ \rho^2 \dd\theta^2 \\
		+ (r^2 + a^2)\, \sin^2 \theta\, \dd\phi^2
		+ r^2 \cos^2 \theta^2\, \dd\Omega_{d-4}^2
\end{equation} 
where as usual
\begin{equation}
	\rho^2 = r^2 + a^2 \cos^2 \theta.
\end{equation} 
Except for the last term one recognizes $4$-dimensional oblate spheroidal coordinates \eqref{coord:metric:4d:spheroidal}.

\subsection{4-dimensional}
\label{app:coord:4d}

In this section one considers
\begin{equation}
	d = 4, \quad
	N = 3, \quad
	n = 1.
\end{equation}

\subsubsection{Cartesian system}

\begin{equation}
	\dd \Sigma^2 = \dd x^2 + \dd y^2 + \dd z^2.
\end{equation}

\subsubsection{Spherical}

\begin{subequations}
\begin{gather}
	\dd \Sigma^2 = \dd r^2 + r^2 \dd \Omega^2, \\
	\dd \Omega^2 = \dd \theta^2 + \sin^2 \theta\; \dd \phi^2,
\end{gather}
\end{subequations}
where $\dd \Omega^2 \equiv \dd \Omega_2^2$.

\subsubsection{Spherical with direction cosines}

\begin{subequations}
\begin{gather}
	\dd \Omega^2 = \dd \mu^2 + \mu^2\; \dd \phi^2 + \dd \alpha^2, \\
	\mu^2 + \alpha^2 = 1,
\end{gather}
\end{subequations}
where
\begin{equation}
	x + iy = r \mu\, \e^{i\phi}, \qquad
	z = r \alpha,
\end{equation} 

Using the constraint one can rewrite
\begin{equation}
	\dd \Omega^2 = \frac{1}{1 - \mu^2}\; \dd \mu^2 + \mu^2\; \dd \phi^2.
\end{equation} 
Finally the change of coordinates
\begin{equation}
	\alpha = \cos \theta, \qquad
	\mu = \sin \theta.
\end{equation} 
solves the constraint and gives back the spherical coordinates.

\subsubsection{Spheroidal with direction cosines}

The oblate spheroidal coordinates from the Cartesian ones are~\cite[p.~15]{Visser:2009:KerrSpacetimeBrief}
\begin{equation}
	x + i y = \sqrt{r^2 + a^2}\, \sin \theta\, \e^{i\phi}, \qquad
	z = r \cos\theta,
\end{equation} 
and the metric is
\begin{equation}
	\label{coord:metric:4d:spheroidal}
	\dd \Sigma^2 = \frac{\rho^2}{r^2 + a^2}\; \dd r^2 + \rho^2 \dd\theta^2 + (r^2 + a^2) \sin^2 \theta\; \dd \phi^2, \qquad
	\rho^2 = r^2 + a^2 \cos^2 \theta.
\end{equation} 

In terms of direction cosines one has
\begin{equation}
	\dd \Sigma^2 = \left(1 - \frac{r^2 \mu^2}{r^2 + a^2} \right)\; \dd r^2 + (r^2 + a^2) \Big(\dd \mu^2 + \mu^2\; \dd \phi^2 \Big) + r^2 \dd \alpha^2.
\end{equation}

\subsection{5-dimensional}
\label{app:coord:5d}

In this section one considers
\begin{equation}
	d = 4, \quad
	N = 3, \quad
	n = 1.
\end{equation}

\subsubsection{Spherical with direction cosines}

\begin{equation}
	\label{coord:metric:5d:spherical}
	\dd\Omega_3^2 = \dd \mu^2 + \mu^2\, \dd\phi^2 + \dd \nu^2 + \nu^2\, \dd\psi^2, \qquad
	\mu^2 + \nu^2 = 1
\end{equation} 
where for simplicity
\begin{equation}
	\mu = \mu_1, \qquad
	\mu = \mu_2, \qquad
	\phi = \phi_1, \qquad
	\psi = \phi_2.
\end{equation}

\subsubsection{Hopf coordinates}
\label{app:coord:5d:hopf}

The constraint \eqref{coord:metric:5d:spherical} can be solved by
\begin{equation}
	\mu = \sin \theta, \qquad
	\nu = \cos \theta
\end{equation} 
and this gives the metric in Hopf coordinates
\begin{equation}
	\label{coord:metric:5d:hopf}
	\dd \Omega_3^2 = \dd\theta^2 + \sin^2 \theta\, \dd\phi^2 + \cos^2 \theta\, \dd\psi^2.
\end{equation} 

%% file: sections/sugra.tex
\section{Review of \texorpdfstring{$N=2$}{N = 2} ungauged supergravity}
\label{app:N=2-sugra}

In order for this review to be self-contained we recall the basic elements of $N = 2$ supergravity without hypermultiplets -- we refer the reader to the standard references for more details~\cite{Freedman:2012:Supergravity, Andrianopoli:1996:GeneralMatterCoupled, Andrianopoli:1997:N2SupergravityN2}.

The gravity multiplet contains the metric and the graviphoton
\begin{equation}
	\{ g_{\mu\nu}, A^0 \}
\end{equation} 
while each of the vector multiplets contains a gauge field and a complex scalar field
\begin{equation}
	\{ A^i, \tau^i \}, \qquad i = 1, \ldots, n_v.
\end{equation} 
The scalar fields $\tau^i$ (the conjugate fields $\conj{(\tau^i)}$ are denoted by $\bar\tau^{\bar\imath}$) parametrize a special Kähler manifold with metric $g_{i\bar\jmath}$.
This manifold is uniquely determined by an holomorphic function called the prepotential $F$.
The latter is better defined using the homogeneous (or projective) coordinates $X^\Lambda$ such that
\begin{equation}
	\tau^i = \frac{X^i}{X^0}.
\end{equation} 
The first derivative of the prepotential with respect to $X^\Lambda$ is denoted by
\begin{equation}
	F_\Lambda = \frac{\pd F}{\pd X^\Lambda}.
\end{equation} 
Finally it makes sense to regroup the gauge fields into one single vector
\begin{equation}
	A^\Lambda = (A^0, A^i).
\end{equation} 

One needs to introduce two more quantities, respectively the Kähler potential and the Kähler connection
\begin{equation}
	K = - \ln i (\bar X^\Lambda F_\Lambda - X^\Lambda \bar F^\Lambda), \qquad
	\mc A_\mu = - \frac{i}{2} (\pd_i K\, \pd_\mu \tau^i - \pd_{\bar\imath} K\, \pd_\mu \bar\tau^{\bar\imath}).
\end{equation} 

The Lagrangian for the theory without gauge group is given by
\begin{equation}
	\mc L = - \frac{R}{2}
		+ g_{i\bar\jmath}(\tau, \bar \tau)\, \pd_\mu \tau^i \pd^\nu \bar\tau^{\bar\imath}
		+ \mc I_{\Lambda\Sigma}(\tau, \bar \tau)\, F^\Lambda_{\mu\nu} F^{\Sigma\,\mu\nu}
		- \mc R_{\Lambda\Sigma}(\tau, \bar \tau)\, F^\Lambda_{\mu\nu} \hodge{F}^{\Sigma\,\mu\nu}
\end{equation} 
where $R$ is the Ricci scalar and $\hodge{F}^\Lambda$ is the Hodge dual of $F^\Lambda$.
The matrix
\begin{equation}
	\mc N = \mc R + i\, \mc I
\end{equation} 
can be expressed in terms of $F$.
From this Lagrangian one can introduce the symplectic dual of $F^\Lambda$
\begin{equation}
	G_\Lambda = \frac{\var \mc L}{\var F^\Lambda} = \mc R_{\Lambda\Sigma} F^\Sigma - \mc I_{\Lambda\Sigma} \hodge{F}^\Sigma.
\end{equation} 

%% file: sections/properties.tex
\section{Technical properties}
\label{app:technical-properties}

In this chapter we describe few technical properties of the algorithm.
We comment on the group properties that some of the JN transformations possess~\cite{Erbin:2016:DecipheringGeneralizingDemianskiJanisNewman}.
Another useful property of Giampieri's prescription is to allow to chain all coordinate transformation, making computations easier~\cite{Erbin:2015:JanisNewmanAlgorithmSimplifications}.
Then finally we discuss the fact that not all the rules \eqref{gen:eq:rules} are independent and several choices of complexification are equivalent~\cite{Erbin:2015:JanisNewmanAlgorithmSimplifications}, contrary to what is commonly believed.

\subsection{Group properties}
\label{app:group-properties}

We want to study the JN transformations that form a group: the main motivation is to state clearly the effect of chaining several transformations.
This observation can be useful for chaining several transformations, therefore adding charges to a solution that is already non-static (for example adding rotation to a solution that already contains a NUT charge).
More importantly this provides a setting where the algorithm has good chances to preserve Einstein equations.

We will make the assumptions that the functions $F(\theta)$ and $G(\theta)$ are linear in some parameters $\pi^A$ (implicit sum over $A$)
\begin{equation}
	F(\theta) = \pi^A F_A(\theta), \qquad
	G(\theta) = \pi^A G_A(\theta),
\end{equation} 
where $\{ F_A(\theta) \}$ and $\{ G_A(\theta) \}$ are the functions associated to the parameters and $A$ runs over the dimension of this space.
Mathematically the functions are member of an additive group $\mc G$ with elements in\footnotemark{} $\mc F \times \mc F$ ($\mc F$ being the space of functions with second derivatives) with generators $\big( F_A(\theta), G_A(\theta) \big)$, $A = 1, \ldots, \dim \mc V$ since there is an identity element $0$ and each element with coefficients $\pi^A$ possesses an inverse given by $- \pi^A$.%
\footnotetext{%
	For simplicity we consider the case where $F$ and $G$ are expanded over the same parameters, but this is not necessarily the case.
}
Adding the multiplication by a scalar turns this group into a vector space but we do not need this extra structure.
As a consequence the sum of two functions $F_1 = \pi_1^A F_A$ and $F_2 = \pi_2^A F_A$ gives another function $F_3 = \pi_3^A F_A$ with $\pi_3^A = \pi_1^A + \pi_2^A$.
These assumptions are motivated by the results of \cref{sec:derivation} where $F$ and $G$ were solutions of (non-homogeneous) second order linear differential equations where the $\pi^A$ are the integration constants.

After a first transformation
\begin{equation}
	r = r' + i\, F_1, \qquad
	u = u' + i\, G_1
\end{equation} 
one obtains the metric (omitting the primes)
\begin{equation}
	\begin{aligned}
		\dd s^2 = &- \tilde f^{\{1\}}_t (\dd u + H G_1'\, \dd\phi)^2
			+ \tilde f^{\{1\}}_\Omega (\dd\theta^2 + H^2 \dd\phi^2) \\
			&- 2 \sqrt{\tilde f^{\{1\}}_t \tilde f^{\{1\}}_r} (\dd u + G_1' H\, \dd\phi) (\dd r + F_1' H\, \dd\phi)
	\end{aligned}
\end{equation} 
where
\begin{equation}
	\tilde f^{\{1\}}_i = \tilde f^{\{1\}}_i(r, F_1).
\end{equation} 
Performing a second transformation
\begin{equation}
	r = r' + i\, F_2, \qquad
	u = u' + i\, G_2
\end{equation} 
the previous metric becomes (omitting the primes)
\begin{equation}
	\label{topdown:eq:metric-two-transf}
	\begin{aligned}
		\dd s^2 = &- \tilde f^{\{1,2\}}_t \big( \dd u + H (G_1' + G_2')\, \dd\phi \big)^2
			+ \tilde f^{\{1,2\}}_\Omega (\dd\theta^2 + H^2 \dd\phi^2) \\
			&- 2 \sqrt{\tilde f^{\{1,2\}}_t \tilde f^{\{1,2\}}_r} \big( \dd u + (G_1' + G_2') H\, \dd\phi \big) \big( \dd r + (F_1' + F_2') H\, \dd\phi \big)
	\end{aligned}
\end{equation} 
where
\begin{equation}
	\tilde f^{\{1,2\}}_i = \tilde f^{\{1,2\}}_i(r, F_1, F_2).
\end{equation}
This function is required to satisfy the following conditions (omitting the primes)
\begin{equation}
	\tilde f^{\{1,2\}}_i(r, F_1, 0) = \tilde f^{\{1\}}_i(r, F_1), \qquad
	\tilde f^{\{1,2\}}_i(r, F_1, F_2) = \tilde f^{\{2,1\}}_i(r, F_2, F_1).
\end{equation} 
The second condition means that the order of the transformations should not matter because we want to obtain the same solution given identical seed metric and parameters.

The metric \eqref{topdown:eq:metric-two-transf} is obviously equivalent to the one we would get with a unique transformation\footnotemark{}%
\footnotetext{%
	This breaks down when the metric is transformed with more complicated rules, such as in higher dimensions~\cite{Erbin:2015:FivedimensionalJanisNewmanAlgorithm}.
}
\begin{equation}
	r = r' + i\, (F_1 + F_2), \qquad
	u = u' + i\, (G_1 + G_2).
\end{equation} 
Then, for the transformations which are such that
\begin{equation}
	\label{topdown:eq:fi-sum-F}
	\tilde f^{\{1,2\}}_i(r, F_1, F_2) = \tilde f^{\{1\}}_i(r, F_1 + F_2),
\end{equation} 
the DJN transformations form an Abelian group thanks to the group properties of the function space.
This structure implies that we can first add one parameter, and later another one (say first the NUT charge, and then an angular momentum).
Said another way this group \emph{preserves Einstein equations} when the seed metric is a known (stationary) solution.
But note that it may be very difficult to do it as soon as one begins to replace the $F$ in the functions by their expression, because it obscures the original function – in one word we can not find $\tilde f_i(r, F)$ from $\tilde f_i(r, \theta)$.

Another point worth to mention is that not all DJN transformation are in this group since the condition \eqref{topdown:eq:fi-sum-F} may not satisfied: we recall that imposing or not the latter is a choice that one is doing when performing the algorithm.
A simple example is provided by
\begin{equation}
	f(r) = r^2,
\end{equation} 
which can be transformed under the two successive transformations
\begin{equation}
	r = r' + i F_1, \qquad
	r' = r'' + i F_2
\end{equation} 
in two ways:
\begin{subequations}
\begin{align}
	1.& \qquad
		\tilde f^{\{1\}} = \abs{r}^2
			= r'^2 + F_1^2, \qquad
		\tilde f^{\{1,2\}} = \abs{r'}^2 + F_1^2
			= r''^2 + F_1^2 + F_2^2, \\
	2.& \qquad
		\tilde f^{\{1\}} = \abs{r}^2
			= \abs{r' + i F_1}^2, \qquad
		\begin{aligned}
			\tilde f^{\{1,2\}} &= \abs{r'' + i (F_1 + F_2)}^2 \\
				&= r''^2 + F_1^2 + F_2^2 + 2 F_1 F_2.
		\end{aligned}
\end{align}
\end{subequations}
Only the second option satisfy the property \eqref{topdown:eq:fi-sum-F} that leads to a group.
Such an example is provided in $5d$ where the function $f_\Omega(r) = r^2$ is successively transformed as~\cite{Erbin:2015:FivedimensionalJanisNewmanAlgorithm}
\begin{equation}
	r^2 \longrightarrow \abs{r}^2 = r^2 + a^2 \cos^2 \theta \longrightarrow \abs{r}^2 + a^2 \cos^2 \theta = r^2 + a^2 \cos^2 \theta + b^2 \sin^2 \theta,
\end{equation} 
with the functions
\begin{equation}
	F_1 = a \cos \theta, \qquad
	F_2 = b \sin \theta.
\end{equation} 
The condition \eqref{topdown:eq:fi-sum-F} is clearly not satisfied.

\subsection{Chaining transformations}
\label{sec:jna-prop:chaining}

The JN algorithm is summarized by the following table
\begin{equation}
	\begin{array}{cccccccccc}
 		t             & \to & u & \to & u \in \C    & \to & u' & \to & t'    \\
		r             &     &   & \to & r \in \C    & \to & r' &     &      \\
		\phi          &     &   &     &             &     &    & \to & \phi' \\
		f             &     &   & \to & \tilde f    &     &    &     &
	\end{array}
\end{equation}
where the arrows correspond to the different steps of the algorithm.

A major advantage of Giampieri's prescription is that one can chain all these transformations since it involves only substitutions and no tensor operations.
For this reason it is much easier to implement on a computer algebra system such as Mathematica.
It is then possible to perform a unique change of variables that leads directly from the static metric to the rotating metric in any system defined by the function $(g, h)$.
For example in the case of rotation for a metric with a single function one finds
\begin{subequations}
\begin{align}
	\dd t &= \dd t' + \big(a h \sin^2 \theta\, (1 - \tilde f^{-1}) - g + \tilde f^{-1} \big)\, \dd r'
		+ a \sin^2 \theta\, (\tilde f^{-1} - 1)\, \dd \phi', \\
	\dd r &= (1 - a h \sin^2 \theta)\, \dd r' + a \sin^2 \theta\, \dd \phi', \\
	\dd \phi &= \dd \phi' - h\, \dd r',
\end{align}
\end{subequations}
where the complexification of the metric function $f$ can be made at the end.
It is impressive that the algorithm from section~\ref{sec:algo} can be written in such a compact way.

\subsection{Arbitrariness of the transformation}
\label{sec:jna-prop:arbitrariness}

We provide a short comment on the arbitrariness of the complexification rules \eqref{gen:eq:rules}.
In particular let's consider the functions
\begin{equation}
	f_1(r) = \frac{1}{r}, \qquad
	f_2(r) = \frac{1}{r^2}.
\end{equation} 

The usual rule is to complexify these two functions as
\begin{equation}
	\label{prop:eq:arb-usual-rules}
	\tilde f_1(r) = \frac{\Re r}{\abs{r}^2}, \qquad
	\tilde f_2(r) = \frac{1}{\abs{r}^2}
\end{equation} 
using respectively the rules \eqref{gen:eq:rules:1/r} and \eqref{gen:eq:rules:r2} (in the denominator).

But it is possible to arrive at the same result with a different combinations of rules.
In fact the functions can be rewritten as
\begin{equation}
	f_1(r) = \frac{r}{r^2}, \qquad
	f_2(r) = \frac{1}{r}\, \frac{1}{r}.
\end{equation} 
The following set of rules results again in \eqref{prop:eq:arb-usual-rules}:
\begin{itemize}
	\item $f_1$: \eqref{gen:eq:rules:r} (numerator) and \eqref{gen:eq:rules:r2} (denominator);
	\item $f_2$: \eqref{gen:eq:rules:r} (first fraction) and \eqref{gen:eq:rules:1/r} (second fraction).
\end{itemize}